\documentclass{statsoc}
\usepackage[a4paper]{geometry}
\usepackage{amsmath, amssymb,mathtools}
\usepackage{amsfonts, multirow, epsfig, floatrow}
\usepackage{graphicx, pdflscape, verbatim, enumerate, colortbl, setspace}
\usepackage{setspace, color,bm}
\usepackage[normalem]{ulem}
\usepackage{cite}
\usepackage{url}
\usepackage{multirow} 
\usepackage{booktabs,array}
\usepackage{hyperref}
\usepackage{float}
\usepackage{caption}
\usepackage{subcaption} 
\usepackage{tikz}
\usepackage{bibunits}
\usepackage{algorithm, algorithmicx, algpseudocode}
\newtheorem{Assumption}{Assumption}
\newtheorem{Lemma}{Lemma}

\newtheorem{Theorem}{Theorem}
\newtheorem{Example}{Example} 
\newtheorem{Remark}{Remark}

\newcommand{\PP}{{\mathbf{P}}}

\newcommand{\RR}{\mathbb{R}}
\newcommand{\R}{\mathbb{R}}
\newcommand{\EE}{\mathbf{E}}

\newcommand{\cip}{\overset{p}{\to}}
\newcommand{\cid}{\overset{d}{\to}}

\newcommand{\MC}{\mathcal{MC}}

\newcommand{\m}{[m]}

\newcommand{\Data}{ \mathcal{O}}

\newcommand{\E}{ {\mathbf E}}

\newcommand{\err}{ {\rm err}}
\newcommand{\bga}{\boldsymbol{\alpha}}
\newcommand{\bgb}{\boldsymbol{\gamma}}
\newcommand{\bge}{\boldsymbol{\eta}}
\newcommand{\T}{{g}}
\newcommand{\DD}{\mathcal{D}}
\newcommand{\LL}{{\mathcal{L}}}
\newcommand{\sig}{\nu}
\newcommand{\Thre}{T}

\DeclareMathOperator*{\argmin}{arg\,min}

\def\VV{\mathcal{V}}
\def\sigmahat{\widehat{\sigma}}
\def\betahat{\widehat{\beta}}
\def\thetahat{\widehat{\theta}}
\def\supl{^{(l)}}
\def\supk{^{(k)}}

\def\sublk{_{l,k}}
\def\SEhat{\widehat{\mbox{SE}}}

\def\Hhat{\widehat{H}}
\def\VVhat{\widehat{\VV}}

\definecolor{purple}{RGB}{250,000,180}

\def\trans{^{\scriptscriptstyle \sf T}}

\def\betahatstar{\betahat^*}
\def\tX{\widetilde{X}}

\title[Inference for Meta-Learning]{Robust Inference for Federated Meta-Learning}

\author{Zijian Guo}
\address{Rutgers University, Piscataway, USA}
\author{Xiudi Li}
\address{Harvard University, Boston, USA}
\author{Larry Han}
\address{Harvard University, Boston, USA}
\author[Guo, Li, Han $\&$ Cai]{Tianxi Cai}
\address{Harvard University, Boston, USA}

\begin{document}
\maketitle 

\begin{abstract}
Synthesizing information from multiple data sources is critical to ensure knowledge generalizability. Integrative analysis of multi-source data is challenging due to the heterogeneity across sources and data-sharing constraints due to privacy concerns. In this paper, we consider a general robust inference framework for federated meta-learning of data from multiple sites, enabling statistical inference for the prevailing model, defined as the one matching the majority of the sites. Statistical inference for the prevailing model is challenging since it requires a data-adaptive mechanism to select eligible sites and subsequently account for the selection uncertainty. We propose a novel sampling method to address the additional variation arising from the selection. Our devised CI construction does not require sites to share individual-level data and is shown to be valid without requiring the selection of eligible sites to be error-free. The proposed robust inference for federated meta-learning (RIFL) methodology is broadly applicable and illustrated with three inference problems: aggregation of parametric models, high-dimensional prediction models, and inference for average treatment effects. We use RIFL to perform federated learning of mortality risk for patients hospitalized with COVID-19 using real-world EHR data from 16 healthcare centers representing 275 hospitals across four countries. 
\end{abstract}
\keywords{Post-selection Inference; Heterogeneous Data; Multi-source Data; Privacy Preserving; High-dimensional Inference.}

%%%%%%%%%%%%%%%%%%%%%%%%%%%%%%%%%%%%%%%%%%%%%%%%%%%%%%%%%%%%%%%%%%%%%%%%%%%%%%%%%%%%%%
\section{Introduction}
%%%%%%%%%%%%%%%%%%%%%%%%%%%%%%%%%%%%%%%%%%%%%%%%%%%%%%%%%%%%%%%%%%%%%%%%%%%%%%%%%%%%%%

Crowdsourcing, or the process of aggregating crowd wisdom to solve problems, is a useful community-based method to improve decision-making in disciplines ranging from education \citep{heffernan2014assistments} to public health \citep{han2018crowdsourcing, wang2020crowdsourcing}. Compared to traditional expert-driven solutions made by a single group, incorporating the opinions of multiple diverse groups can improve the quality of the final decision \citep{surowiecki2005wisdom}. In health research, crowdsourcing has led to the discovery of new drugs during pandemics \citep{chodera2020crowdsourcing}, the design of patient-centered mammography reports \citep{short2017patient}, and the development of machine learning algorithms to classify tumors for radiation therapy \citep{mak2019use}. 

Underlying the phenomenon of the ``wisdom of the crowds" is the statistical and philosophical notion that learning from multiple data sources is desirable. Incorporating information from diverse data sources can increase the generalizability and transportability of findings compared to learning from a single data source. Findings from a single data source may not be generalizable to a new target population of interest due to poor data quality or heterogeneity in the underlying data generating processes. %It may also be underpowered to draw accurate predictions or valid causal inferences. The issue of limited data from a single data source is particularly relevant for underrepresented populations or individuals with rare diseases. Regardless of its size, findings from a single data source may not be generalizable due to poor data quality or biases associated with a specific health system. 

Integrative analysis of data from multiple sources can be a valuable alternative to using a single data source alone. However, directly pooling multiple data sources into a single dataset for analysis is often unsatisfactory or even infeasible. Heterogeneity between different data sources can severely bias predictions or inferences made by such a pooled analysis strategy \citep{leek2010tackling, ling2022batch}. As an alternative to pooled analysis, meta-analysis has frequently synthesized information from multiple studies. Standard meta-analysis methods aggregate quantitative summary of evidence from multiple studies. Variations of meta-analysis, such as random effects meta-analysis, have been adopted to explore between-study heterogeneity, potential biases such as publication bias, and small-study effects. However, most existing meta-analysis tools that account for heterogeneity require strong modeling assumptions and do not consider the validity of inference when data from certain sites have substantially different distributions from other sites.  

Another challenge of particular importance is the issue of data privacy pertaining to biomedical studies. Regulations in the United States, such as the Health Insurance Portability and Accountability Act (HIPAA) Privacy Rule, and those in the European Union, such as the General Data Protection Regulation (GDPR) and the European Medicines Agency (EMA) Privacy Statement, protect the personal information of patients and preclude the transfer of patient-level data between sites. These regulations make the promise of integrative data analysis more difficult to attain, highlighting the need for federated integrative analysis methods that do not require sharing of individual-level data.

When cross-study heterogeneity is substantial and outliers exist, a desirable strategy of integrative analysis is to identify a prevailing model to achieve consensus learning. The prevailing model is defined as the  model satisfied by the majority of the sites. 
%The prevailing model here is taken as a generalizable concept with a wide range of applications in practice, including aggregation of parametric models, construction of high-dimensional prediction models, and inference for the average treatment effect (ATE). 
Identifying the prevailing model can  be intuitively achieved via the majority rule  \citep{sorkin1998group,kerr2004group,hastie2005robust}, which chooses the alternative that more than half of individuals agree upon. The majority rule is widespread in modern liberal democracies and is deployed in various streams of research. For example, %legislative bodies often consider multiple policy proposals over many rounds of deliberation until one achieves the majority rule. The majority rule is also employed in various streams of research. 
genomics data is usually separated into batches, %due to computational constraints, 
but heterogeneity across batches can lead to undesirable variation in the data \citep{leek2010tackling, ling2022batch}. This setting aims to identify batches that show low levels of concordance with the majority of the batches and adjust for such differences in downstream analyses \citep{trippa2015bayesian}. As another example, in the design of clinical trials, it is often infeasible or unethical to enroll patients in the control arm. In such cases, it is possible to use data from historical trials or observational studies to construct an external control arm \citep{jahanshahi2021use, ventz2019design, davi2020informing}. However, when many such historical data sources exist, it is crucial to carefully select data sources that show high levels of similarity with the majority of the other data sources. The last example is Mendelian Randomization, where multiple genetic markers are used as instrumental variables (IVs) to account for potential unmeasured confounders. Every single IV will have its causal effect estimator, and the goal is to identify the causal effect matching the majority of the estimated effects  \citep{burgess2017review, bowden2016consistent, kang2016instrumental}. 

Without prior knowledge of the prevailing model, it is critical to employ data-adaptive approaches to select appropriate sites for inferring the prevailing model. In addition, confidence intervals (CIs) for the target parameter of the prevailing model need to appropriately adjust for the site selection variability. Most existing statistical inference methods rely on perfectly separating eligible and ineligible sites, which may be unrealistic for practical applications. There is a paucity of statistical inference methods for the prevailing model that can achieve efficient and robust inference while being applicable to a broad set of scenarios without restrictive assumptions such as a perfect separation.

%There is a lack of  achieve inference under the prevailing model framework for specific scenarios, no existing methods can be achieve highly efficient and robust inference while being applicable to a broad set of scenarios without restrictive assumptions. \tcomm{zijian, you need a high level summary of limitations of existing methods}

%\vspace{-3.5mm}
%%%%%%%%%%%%%%%%%%%%%%%%%%%%%%%%%%%%%%%%%%%%%%%%%%%%%%%%%%%
%\subsection{Our results and contributions}
%%%%%%%%%%%%%%%%%%%%%%%%%%%%%%%%%%%%%%%%%%%%%%%%%%%%%%%%%%%%%%%%%%%%%%%%%%%%%%
In this paper, we fill this gap by developing a broad theoretically justified framework for making robust inferences for federated meta-learning (RIFL) of an unknown prevailing model using multi-source data. The RIFL method selects the eligible sites to infer the prevailing model by assessing dissimilarities between sites with regard to the parameter of interest. We employ a novel resampling method to construct uniformly valid CIs. %%
The RIFL inference method is robust to the errors in separating the sites belonging to the majority group and the remaining sites; see Theorem \ref{thm: coverage general}. {We also show in Theorem \ref{thm: efficiency} that our proposed sampling CI can be as short as the oracle CI with the prior knowledge of the eligible sites.} Our general sampling algorithm is privacy-preserving in that it is implemented using site-specific summary statistics and without requiring sharing individual-level data across different sites. Our proposed RIFL methodology is demonstrated with three inference problems: aggregation of low-dimensional parametric models, construction of high-dimensional prediction models, and inference for the average treatment effect (ATE). 

To the best of our knowledge, our proposed RIFL method is the first CI guaranteeing uniform coverage of the prevailing model under the majority rule. We have further compared via simulation studies with three other inference procedures that can potentially be used under the majority rule, including the majority voting estimator, the median estimator \citep[e.g.,][]{bowden2016consistent}, and the m-out-of-n bootstrap \citep[e.g.,][]{chakraborty2013inference,andrews2000inconsistency}. Numerical results demonstrate that these three CIs fail to achieve the desired coverage property, while our RIFL method leads to a uniformly valid CI. We provide  the reasoning for under-coverage for these existing methods in Sections \ref{sec: post-selection} and \ref{sec: median}.

%and de 
% Through numerical studies, we demonstrate that our proposed CI is nearly as efficient as an oracle bias-aware confidence interval; and substantially more efficient than which does not efficiently combine information from the valid sources. In addition, we illustrate that  will suffer under-coverage due to the selection error. 
 
\vspace{-3.5mm}

%%%%%%%%%%%%%%%%%%%%%%%%%%%%%%%%%%%%%%%%%%%%%%%%%%%%%%%%%%%%%%%%%%%%%%%%%%%%%%
\subsection{Related literature}
%%%%%%%%%%%%%%%%%%%%%%%%%%%%%%%%%%%%%%%%%%%%%%%%%%%%%%%%%%%%%%%%%%%%%%%%%%%%%%

The RIFL method is related to multiple streams of literature, including post-selection inference, mendelian randomization, integrative analysis of multi-source data, transfer learning, and federated learning. We next detail how RIFL differs from the existing literature and highlight its contributions. 

A wide range of novel methods and theories  have been established to address the post-selection inference problem \citep[e.g.,][]{berk2013valid,lee2016exact,leeb2005model,zhang2014confidence,javanmard2014confidence,van2014asymptotically,chernozhukov2015post,belloni2014inference,cai2017confidence,xie2022repro}. However, most post-selection inference literature focuses on inferences after selecting a small number of important variables under high-dimensional regression models. The selection problem under the RIFL framework is fundamentally different: the selection error comes from comparing different sites, and there is no outcome variable to supervise the selection process. Additionally, RIFL only requires the majority rule, while the variable selection methods typically require a small proportion of variables to affect the outcome. 

%\Zijian{Tianxi, please check this paragraph.} 
In Mendelian Randomization, various methods have been developed to leverage the majority rule and make inferences for the one-dimensional causal effect \citep{bowden2016consistent,windmeijer2019use,kang2016instrumental,guo2018confidence,windmeijer2021confidence}. A recent work by \citet{guo2021causal} demonstrated the post-selection problem due to IV selection errors. %\citet{guo2021causal} addressed the post-selection problem by searching for the casual effect satisfying the majority rule and further improved the confidence interval precision by a sampling method. We shall emphasize that the method in \citet{guo2021causal} is only useful when the model parameter is one-dimensional. 
However, the uniformly valid inference method in \citet{guo2021causal} relies on searching the one-dimensional space of the causal effect and cannot be easily generalized to multivariate settings, not to mention high-dimensional settings. In contrast, RIFL is distinct from the existing searching method and is useful in addressing a much broader collection of post-selection problems as detailed in Section \ref{sec: applications}. %Our methodology is applied to make inferences for the prevailing models in multiple and high dimensions; see Sections \ref{sec: app 1} and \ref{sec: app 2} for more details.  

The integrative analysis of multi-source data has been investigated in different directions. \citet{wang2021robust} studied the data fusion problem with robustness to biased sources. The identification condition in \citet{wang2021robust} differs from the majority rule, and the validity of their proposal requires correctly identifying unbiased sources. {\citet{maity2022meta} studied meta-analysis in high-dimensional settings where the data sources
are similar but non-identical and require the majority rule to be satisfied as well as a large separation between majority and outlier sources to perfectly identify eligible sites.} In contrast, the RIFL CI is valid without requiring the selection step to perfectly identify eligible sites.    \citet{meinshausen2015maximin, buhlmann2015magging,rothenhausler2016confidence,guo2020inference} made inference for the maximin effect, which is defined as a robust prediction model across heterogeneous datasets. \citet{cai2021individual,liu2021integrative,zhao2016partially} imposed certain similar structures across different sources and made inferences for the shared component of regression models. \citet{peters2016causal,arjovsky2019invariant} studied the multi-source data problem and identified the causal effect by invariance principles. Unlike existing methods, the RIFL framework only assumes that a majority of the sites have similar models but allows non-eligible sites to differ  arbitrarily from the majority group. 

RIFL relates to the existing literature on federated learning and transfer learning. 
%, aiming to perform integrative analysis via joint analyses of summary-level data. 
Privacy-preserving and communication-efficient algorithms have been recently developed to learn from multiple sources of electronic health records (EHR) \citep{rasmy2018study, tong2022distributed} and multiple sources of diverse genetic data \citep{kraft2009replication, keys2020cross}. Federated regression and predictive modeling \citep{chen2006regression, li2013statistical,chen2014split,lee2017communication, lian2017divide,wang2019distributed, duan2020learning} and causal modeling  \citep{xiong2021federated, vo2021federated, han2021federated} have been developed. However, none of these federated learning methods study inference for the prevailing model when some sites may not be valid, which is the main focus of RIFL. The RIFL framework also differs from the recently developed transfer learning algorithms  \citep[e.g.,][]{li2020transfer,tian2022transfer,han2021federated}. These algorithms require pre-specification of an anchor model to which the models obtained from source data sets can be compared. In contrast, RIFL targets a more challenging scenario: we do not assume the availability of such an anchor model but leverage the majority rule to identify the unknown prevailing model.

\vspace{-3.5mm}

\subsection{Paper organization and notations}
The paper proceeds as follows. Section \ref{sec: formulation} describes the  multi-source data setting and highlights the challenge of inferring the prevailing model. Section \ref{sec: general principle} proposes the RIFL methodology,  and Section \ref{sec: theory} establishes its related theory. In Section \ref{sec: applications}, we illustrate our proposal in three applications. In Section \ref{sec: simulation}, we provide extensive simulation results comparing our method to existing methods. Section \ref{sec: real data} illustrates our method using real-world international EHR data from 16 participating healthcare centers representing 275 hospitals  across four countries as part of the multi-institutional Consortium for the Clinical Characterization of COVID-19 by EHR (4CE) \citep{brat2020international}. %Due to data privacy constraints, patient-level data could not be shared across the sites. Using RIFL, we find that the majority rule is satisfied and are able to identify and estimate a prevailing mortality risk prediction model for patients hospitalized with COVID-19. We find that advanced age, black or Hispanic race, low serum albumin levels, and high C-reactive protein (CRP) and aspartate aminotransferase (AST) levels are associated with a significantly higher 14-day mortality risk. Our results have implications for clinical practice, e.g., suggesting that measuring serum albumin, CRP, and AST levels may serve as useful biomarkers for the early identification of patients at high risk of disease progression and mortality. 

We introduce the notations used throughout the paper. For a set $A$, $|A|$ denotes the cardinality of the set. For a vector $x$, we define its $\ell_q$ norm as $\|x\|_{q}=\left(\sum_{l=1}^{p}|x_l|^q\right)^{\frac{1}{q}}$ for $q \geq 0$ with $\|x\|_0=\left|\{1\leq l\leq p: x_l \neq 0\}\right|$ and $\|x\|_{\infty}=\max_{1\leq l \leq p}|x_l|$. For a matrix $X$, $X_{i,\cdot}$ and $X_{\cdot, j}$ denote its $i$-th row and $j$-th column, respectively. For two positive sequences $a_n$ and $b_n$, $a_n \ll b_n$ if $\limsup_{n\rightarrow\infty} {a_n}/{b_n}=0$. For a matrix $A$, we use $\|A\|_{F}$, $\|A\|_2$ and $\|A\|_{\infty}$ to denote its Frobenius norm, spectral norm, and element-wise maximum norm, respectively.

%%%%%%%%%%%%%%%%%%%%%%%%%%%%%%%%%%%%%%%%%%%%%%%%%%%%%%%%%%%%%%%%%%%%%%%%%%%
\section{Formulation and Statistical Inference Challenges}
\label{sec: formulation}
%%%%%%%%%%%%%%%%%%%%%%%%%%%%%%%%%%%%%%%%%%%%%%%%%%%%%%%%%%%%%%%%%%%%%%%%%%%%
%%%%%%%%%%%%%%%%%%%%%%%%%%%%%%%%%%%%%%%%%%%%%%%%%%%%%%%%%%%%%%%%%%%%%%%%%%%%
%We investigate how to efficiently aggregate information from $L$ possibly heterogeneous source sites. Throughout, we use the terms `sites', `populations', and `data sets' interchangeably. The goal is to make inference for the prevailing model, defined as the model that matches with more than half of the source sites. Such a target model is uniquely defined and generalizable since it captures the structures shared by the majority of all sites. 

%extract the information from these $L$ source populations to make inference for a target population, which is unseen and possibly different from the source populations. 

%%%%%%%%%%%%%%%%%%%%%%%%%%%%%%%%%%%%%%%%%%%%%%%%%%%%%%%%%%%%%%%%%%%%%%%%%%%%
\subsection{Model assumptions and overview of RIFL}
%%%%%%%%%%%%%%%%%%%%%%%%%%%%%%%%%%%%%%%%%%%%%%%%%%%%%%%%%%%%%%%%%%%%%%%%%%%%
Throughout the paper, we consider that we have access to $L$ independent training data sets drawn from $L$ source populations. 
For $1\leq l\leq L$, we use $\mathbb{P}^{(l)}$ to denote the distribution of the $l$-th source population and use $\theta^{(l)}=\theta(\mathbb{P}^{(l)})\in \R^{d}$ to denote the associated model parameter. For any $\theta\in \R^{d}$, we define the index set $\mathcal{V}(\theta)\subset \{1,\cdots,L\}$ as  \begin{equation}
\mathcal{V}(\theta) \coloneqq \{1\leq l\leq L: \theta^{(l)}=\theta\},
\label{eq: equal in distribution}
\end{equation}
which contains the indexes of all sites having the same model parameter as $\theta.$ We now introduce the majority rule.
 
\begin{Assumption}[Majority Rule] There exists $\theta^*\in \R^d$ such that $\left|\mathcal{V}(\theta^*)\right|>L/2.$
\label{assump: majority rule}
\end{Assumption}
We shall refer to $\theta^*$ as the {\em prevailing model} that matches with more than half of $\{\theta^{(l)}\}_{1\leq l\leq L}$, and the corresponding index set $\mathcal{V}(\theta^*)$ as the prevailing set. %When there is no confusion, we write $\mathcal{V}$ for $\mathcal{V}(\theta^*).$ 
Our goal is to construct a confidence region for a low dimensional functional of $\theta^*$, denoted as $\beta^*=\T(\theta^*)\in \R^{q}$, for some $q\geq 1,$ where $\T(\cdot)\in \R^{q}$ is a prespecified low-dimensional transformation. Examples of $\beta^*=\T(\theta^*)$ include
\begin{enumerate} 
\item Single coefficient or sub-vector: $\beta^*=\theta^*_j$ for $1\leq j\leq d$ or $\beta^*=\theta^*_{G}$ with $G\subset\{1,\cdots,d\}$; 
\item Linear transformation: $\beta^*=x^{\intercal}\theta^*$ for any $x\in \R^{d}$;
\item Quadratic form: $\beta^*=\|\theta^*\|_2^2$. 
\end{enumerate}
For notational ease, we focus on $q=1$ primarily and discuss the extension to the setting with $q\geq 2$ in Section \ref{sec: multivariate}.

If the prevailing set $\mathcal{V}(\theta^*)$ were known, standard meta and federated learning methods could be used to make inferences about $\theta^*$ using data from sites belonging to $\mathcal{V}(\theta^*).$ However, as highlighted in  Section \ref{sec: post-selection}, inference for $\theta^*$ without prior knowledge of  $\mathcal{V}(\theta^*)$ except for the majority rule is substantially more challenging due to the need to estimate $\mathcal{V}(\theta^*)$. Our proposed RIFL procedure involves several key steps: (i) for $l=1, ..., L$, construct local estimates of $\theta\supl$ and $\beta\supl=g(\theta\supl)$, denoted by $\widehat{\theta}\supl$ and $\widehat{\beta}\supl$, respectively; (ii) for $1 \le l < k \le L$, estimate pairwise dissimilarity measure $\DD_{l,k}=D(\theta\supl, \theta\supk)$ and $\LL\sublk = \beta\supl-\beta\supk$ as $\widehat\DD\sublk$ and $\widehat\LL\sublk$ along with their standard errors $\SEhat(\widehat\DD\sublk)$ and $\SEhat(\widehat\LL\sublk)$; (iii) construct a robust estimate for the prevailing set $\VV(\theta^*)$; (iv) derive robust resampling-based  confidence set for $\beta^*$ accounting for post-selection uncertainty. 
The construction of $\widehat{\theta}\supl$ and $\widehat{\beta}\supl$ follows standard procedures for the specific problems of interest.  We next detail (ii) the construction of the dissimilarity measures and (iii) the prevailing set estimator. The most challenging step of RIFL is the resampling-based inference, which is described in Section \ref{sec: general principle}.

%%%%%%%%%%%%%%%%%%%%%%%%%%%%%%%%%%%%%%%%%%%%%%%%%%%%%%%%%%%%
\subsection{Dissimilarity measures}
\label{sec: dissimilarity}
%%%%%%%%%%%%%%%%%%%%%%%%%%%%%%%%%%%%%%%%%%%%%%%%%%%%%%%%%%%%
A critical step of applying the majority rule is to evaluate the (dis)similarity between any pair of parameters $\theta^{(l)}$ and $\theta^{(k)}$ for $1\leq l,k\leq L.$ We form two sets of dissimilarity measures, the local dissimilarity between $\beta\supl$ and $\beta\supk$,  $\LL_{l,k}=\beta\supl-\beta\supk$, and a global dissimilarity $\DD_{l,k}=\DD(\theta^{(l)},\theta^{(k)}) = \|\theta^{(l)}-\theta^{(k)}\|_2^2$. Although other vector norms can be considered for $\DD(\cdot, \cdot)$, we focus on the quadratic norm due to its smoothness and ease of inference, especially in the high-dimensional setting.

We assume that 
 $\{\widehat{\beta}\supl,\widehat{\sigma}_l\}_{1\leq l\leq L}$ satisfy 
\begin{equation}
\frac{1}{{\sigma}_l}(\widehat{\beta}\supl-\beta\supl)\cid N(0,1) \quad \text{and}\quad \frac{\widehat{\sigma}_l}{{\sigma}_l}\cip 1,
\label{eq: limiting distribution}
\end{equation}
with $\sigma_l$ denoting the standard error of $\betahat\supl.$ In the low-dimensional setting, most existing estimators satisfy \eqref{eq: limiting distribution} under standard regularity conditions. In the high-dimensional setting, various asymptotically normal de-biased estimators have recently been proposed and shown to satisfy \eqref{eq: limiting distribution}; see more discussions at the end of Section \ref{sec: app 2}. 

Let $\widehat\LL\sublk = \betahat\supl-\betahat\supk$ and $\widehat\DD\sublk$ be the point estimators for $\LL\sublk$ and $\DD\sublk$, respectively. We estimate their standard errors as $\SEhat(\widehat\LL\sublk) = \sqrt{\widehat\sigma_l^2+\widehat\sigma_k^2}$ and $\SEhat(\widehat\DD\sublk)$, with $\sigmahat_l^2$ denoting the estimated variance of  $\betahat\supl$. For the global dissimilarity measure, we assume that  $\widehat{\DD}_{l,k}$ and $\widehat{\rm SE}(\widehat{\DD}_{l,k})$ satisfy
\begin{equation}
\limsup_{n\rightarrow \infty}\PP\left({\left|\widehat{\DD}_{l,k}-{\DD}_{l,k}\right|}/{\widehat{\rm SE}(\widehat{\DD}_{l,k})}\geq z_{\alpha}\right)\leq \alpha \quad \text{for}\quad 0<\alpha<1, 
\label{eq: key assumption}
\end{equation}
where $z_{\alpha}$ denotes the $\alpha$ upper quantile of a standard normal distribution. Although $\widehat{\DD}_{l,k}$ can be constructed as $\|\thetahat\supl - \thetahat\supk\|_2^2$ in the low-dimensional setting, deriving 
$\{\widehat{\DD}_{l,k}, \widehat{\rm SE}(\widehat{\DD}_{l,k})\}$ that satisfies (\ref{eq: key assumption}) is much more challenging in the high-dimensional setting due to the inherent bias in  regularized estimators. 
In Sections \ref{sec: app 1} and \ref{sec: app 2}, we demonstrate that our proposed estimators of ${\DD}_{l,k}$ satisfy \eqref{eq: key assumption} for a broad class of applications in both low and high dimensions.

Based on both sets of dissimilarity measures, we determine the concordance between sites $k$ and $l$ with respect to inference for $\beta^*=g(\theta^*)$ based on the following test statistic
\begin{equation}
\widehat{S}_{l,k}\coloneqq \max\left\{\left|\widehat{\DD}_{l,k}/{\widehat{\rm SE}(\widehat{\DD}_{l,k})}\right|,\left|\widehat{\LL}_{l,k}/{\widehat{\rm SE}(\widehat{\LL}_{l,k})}\right|\right\}.
\label{eq: global difference}
\end{equation} 
For $1\leq l< k\leq L$, we can then implement the following significance test of whether the $k$-th and $l$-th sites share the same parameters, 
\begin{equation}
\widehat{H}_{l,k}={\bf 1}\left(\widehat{S}_{l,k}\leq z_{0.05/[2L(L-1)]}\right), 
\label{eq: voting est}
\end{equation}
where $0.05$ is a pre-selected significance level for testing the similarity among different sites and $z_{0.05/[2L(L-1)]}$ denotes the $0.05/[2L(L-1)]$ upper quantile of the standard normal distribution. The statistic $\widehat{S}_{l,k}$ measures the level of evidence that the two sites differ from each other based on observed data. The binary decision $\widehat{H}_{l,k}$ in \eqref{eq: voting est} essentially estimates $H_{l,k}={\bf 1}\{\theta^{(l)}=\theta^{(k)}\}.$ We specify the threshold as $z_{0.05/[2L(L-1)]}$ to adjust for the multiplicity of hypothesis testing.  Since the matrix $H$ is symmetric and $H_{l,l}=1$, we construct an estimate for the full voting matrix $\Hhat = [\Hhat\sublk]_{l=1,...,L}^{k=1,...,L}$ by setting $\Hhat_{k,l}=\Hhat\sublk$ and $\Hhat_{l,l}=1$. 
The estimated voting matrix $\widehat{H}$ summarizes all cross-site similarities, which is then used to estimate the prevailing set.  

\begin{Remark}[Univariate case] \rm For the special setting with a univariate $\theta^*,$ we may simplify the construction of the test statistics in \eqref{eq: global difference}  and the vote in \eqref{eq: voting est} as 
\begin{equation}
\widehat{H}_{l,k}={\bf 1}\left(\widehat{S}_{l,k}\leq z_{0.05/[L(L-1)]}\right) \quad \text{with}\quad \widehat{S}_{l,k}=\left|\widehat{\LL}_{l,k}/{\widehat{\rm SE}(\widehat{\LL}_{l,k})}\right| \quad \text{for}\quad 1\leq l<k\leq L.
\label{eq: univariate}
\end{equation}
\label{rem: univariate}
\end{Remark}

%The term $\widehat{H}_{l,k}$ in \eqref{eq: voting est} is a data-dependent estimator of $H_{l,k}={\bf 1}(\theta^{(l)}=\theta^{(k)}).$ Note that $H_{l,k}=1$ represents that the $l$-th site and $k$-th site `vote' for each other. Hence, we refer to $H_{l,k}$ as `vote' and the corresponding matrix $H=\{H_{l,k}\}_{1\leq l,k\leq L}$ as `voting matrix'. 

%%%%%%%%%%%%%%%%%%%%%%%%%%%%%%%
\subsection{Prevailing set estimation and post-selection problem}
\label{sec: post-selection}
\begin{figure}[H]
\centering
\begin{tikzpicture}[node distance={15mm}, main/.style = {draw, circle}] 
\node[main] (1) {$1$}; 
\node[main] (2) [above right of=1] {$2$};
\node[main] (3) [below right of=2] {$3$}; 
\node[main] (4) [right of=3] {$4$};
%\node[main] (5) [above right of=4] {$5$}; 
\node[main] (5) [above left of=4] {$5$};
\node[main] (6) [right of=4] {$6$};
\draw (1) -- (2);
\draw (1) -- (3);
\draw (2) -- (3);
\draw (3) -- (4);
\draw (4) -- (5);
\draw (4) -- (6);
\draw (5) -- (1);
\draw (5) -- (2);
\draw (5) -- (3);
\end{tikzpicture} 
\caption{ The graph $\mathcal{G}([L],\widehat{H})$ with $L=6.$ $\widetilde{\mathcal{V}}=\{1,2,3,4,5\}$ and $\widehat{\mathcal{V}}=\{1,2,3,5\}$.}
\label{fig: MC}
\end{figure}
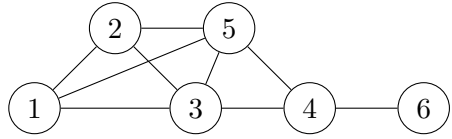

%%%%%%%%%%%%%%%%%%%%%%%%%%%%%%%%%%%%%%%%%%%%%%%%%%%%%%%%%
In the following, we construct two estimators of the prevailing set $\mathcal{V}(\theta^*)$, which are used to make inference for $\beta^*.$ We construct the first estimator as 
\begin{equation}
\widetilde{\mathcal{V}}\coloneqq\{1\leq l\leq L: \|\widehat{H}_{l,\cdot}\|_0>L/2\}.
\label{eq: voting}
\end{equation}
The set $\widetilde{\mathcal{V}}$ contains all sites receiving `majority votes'. The second estimator is constructed by utilizing the maximum clique from graph theory \citep{carraghan1990exact}. Specifically, we define the graph $\mathcal{G}([L],\widehat{H})$ with vertices $[L]\coloneqq\{1,2,\cdots,L\}$ and the adjacency matrix $\widehat{H}$ with $\widehat{H}_{l,k}=1$ and $\widehat{H}_{l,k}=0$ denoting that the $l$-th and $k$-th vertexes are connected and disconnected, respectively. 
The maximum clique of the graph $\mathcal{G}([L],\widehat{H})$ is defined as the largest fully connected sub-graph. We use the term {\it maximum clique set} to denote the corresponding vertex set in the maximum clique, denoted as $\MC([L],\widehat{H}).$ We construct $\widehat{\mathcal{V}}$ by identifying the maximum clique set of $\mathcal{G}([L],\widehat{H})$, that is, 
\begin{equation}
\widehat{\mathcal{V}}\coloneqq \MC([L],\widehat{H}).
\label{eq: voting MC}
\end{equation}
If the majority rule holds, the prevailing set $\mathcal{V}(\theta^*)$ is the maximum clique set of $\mathcal{G}([L], H)$, that is, $\mathcal{V}(\theta^*)=\MC([L],{H})$. 
If $\widehat{H}$ is a sufficiently accurate estimator of $H$, both set estimators $\widetilde{\mathcal{V}}$ and  $\widehat{\mathcal{V}}$ exactly recover the prevailing set $\mathcal{V}(\theta^*).$ However, since $\widehat{H}$ might be different from the true $H$ due to the limited sample size in practice, $\widetilde{\mathcal{V}}$ and  $\widehat{\mathcal{V}}$ 
 may be different from $\mathcal{V}(\theta^*)$.  When the maximum clique set has the cardinality above $L/2,$ we have $\widehat{\mathcal{V}}\subset\widetilde{\mathcal{V}}$, that is,  $\widehat{\mathcal{V}}$ may be a more restrictive set estimator than $\widetilde{\mathcal{V}}.$ We illustrate the definitions of $\widetilde{\mathcal{V}}$ in \eqref{eq: voting} and $\widehat{\mathcal{V}}$ in  \eqref{eq: voting MC} in Figure \ref{fig: MC}.

In the following, we demonstrate the subsequent analysis after obtaining $\widehat{\mathcal{V}}.$ The argument is easily extended to the estimated set $\widetilde{\mathcal{V}}.$ One may aggregate $\{\widehat{\beta}\supl,\widehat{\sigma}_{l}\}_{l\in \widehat{\mathcal{V}}}$ to estimate $\beta^*$ as the following inverse variance weighted estimator, 
\begin{equation}
\betahatstar=\frac{\sum_{l\in \widehat{\mathcal{V}}}{\widehat{\beta}\supl
}/{\widehat{\sigma}_{l}^2}}{\sum_{l\in \widehat{\mathcal{V}}}1/{\widehat{\sigma}_{l}^2}}.
\label{eq: post-selection est}
\end{equation}
A naive $1-\alpha$ confidence interval for $\beta^*$ can be constructed as
\begin{equation}
{\rm CI}_{\rm post}=\left(\betahatstar-z_{\alpha/2}\frac{1}{\sqrt{\sum_{l\in \widehat{\mathcal{V}}}1/{\widehat{\sigma}_{l}^2}}}, \betahatstar+z_{\alpha/2}\frac{1}{\sqrt{\sum_{l\in \widehat{\mathcal{V}}}1/{\widehat{\sigma}_{l}^2}}}\right),
\label{eq: post CI}
\end{equation}
where $z_{\alpha/2}$ denotes the $\alpha/2$ upper quantile of the standard normal distribution.

Unfortunately, similar to other settings in the `post-selection' literature, such naive construction can lead to bias in the point estimation and under-coverage in the confidence interval due to ignoring the variability in the selection of $\VVhat$.  We illustrate the post-selection problem of the naive confidence interval in  \eqref{eq: post CI} with the following example.

\begin{Example}\rm
We construct the confidence interval for a target population's average treatment effect (ATE) in the multi-source causal inference setting detailed in Section \ref{sec: app 3}. We have $L = 10$ source sites with $n_l = 1000$, $1 \leq l \leq 10$. In each source site, we observe the data $\{X^{(l)}_{i}, A^{(l)}_i, Y^{(l)}_i\}_{1\leq i\leq n_l},$ where $X^{(l)}_{i} \in \mathbb{R}^{10}$ denotes a 10-dimensional vector of baseline covariates, $A^{(l)}_{i} \in \{0,1\}$ denotes the treatment assignment (treatment or control) and $Y^{(l)}_{i}\in \R$ denotes the outcome. 
The first six source sites are generated such that the target ATE has a value of $-1$, while the remaining four source sites are generated such that the target ATE has values $-1.2$, $-1.2$, $-1.1$, and $-1.1$, respectively. In this case, the first six source sites form the majority group. The confidence intervals relying on $\widehat{\mathcal{V}}$ and $\widetilde{\mathcal{V}}$ suffer from the under-coverage due to wrongly selected sites being included.
Based on 500 simulations, the confidence interval in  \eqref{eq: post CI} has an empirical coverage of only $43.2\%$. If we replace $\widehat{\mathcal{V}}$ in \eqref{eq: post CI} with $\widetilde{\mathcal{V}}$ in \eqref{eq: voting}, the empirical coverage drops to $27.4\%$.

\end{Example}

%\begin{figure}[H]
%    \centering
%    \includegraphics[scale=0.25]{graphics/Cov2.500.png}
%    \caption{Illustration of under-coverage of the confidence interval in \eqref{eq: post CI}, with only $219$ of the $500$ CI's covering the true ATE of $3.0$ (light blue).}
%    \label{fig: undercover2}
%\end{figure}

\subsection{Challenge for the median-based confidence interval}
\label{sec: median}
%\Zijian{Tianxi and Xiudi, I moved Xiudi's discussion about median estimator to here and rename the title of Section 2 by replacing ``post-selection challenge" with ``Inference Challenges". Please double check.}

A commonly used consistent estimator of the prevailing model parameter under the majority rule is the median estimator \citep[e.g.,][]{bowden2016consistent}. We construct the median estimator as the median of $\{\widehat{\beta}\supl\}_{1\leq l\leq L}$ and estimate its standard error by parametric bootstrap. It is worth noting that although the median estimator is consistent under the majority rule as the sample size in each site approaches infinity, it may not be suitable for the purpose of statistical inference due to its bias. Consequently, the CI based on the median estimator does not achieve the desired coverage property.  To illustrate this, let us consider a special case where $L$ is odd, and there are $(L+1)/2$ sites in the prevailing set. Without loss of generality, we assume that $\mathcal{V}(\theta^*)=\{1,\ldots, (L+1)/2\}$. Furthermore, suppose that $\beta\supl < \beta^*$ for $l \notin \mathcal{V}$. In this scenario, when the parameter values in the non-majority sites are well-separated from the parameter value in the prevailing set, with high probability, the median estimator will coincide with $\min\{\widehat\beta_1, \ldots, \widehat\beta_{(L+1)/2}\}$. That is, the median estimator is the smallest order statistics of $(\widehat\beta_1,\ldots,\widehat\beta_{(L+1)/2})$. Even when the site-specific estimator $\widehat\beta\supl$ is unbiased for $\beta^*$ and normally distributed for $l \in \mathcal{V}$, the smallest order statistics typically has a non-normal distribution with a mean value below $\beta^*$. More generally, the limiting distribution of the median estimator is that of an order statistics and has an asymptotic bias that is not negligible for the purpose of statistical inference. This same issue has been discussed in more detail in \citet{windmeijer2019use} in the context of invalid instrumental variables. Our numerical results in Section \ref{sec: simulation} show that the CI based on the median estimator fails to achieve the desired coverage.

%%%%%%%%%%%%%%%%%%%%%%%%%%%%%%%%%%%%%%%%%%%%%%%%%%%%%%%%%%%%
\section{RIFL Inference}
\label{sec: general principle}
%%%%%%%%%%%%%%%%%%%%%%%%%%%%%%%%%%%%%%%%%%%%%%%%%%%%%%%%%%%%
In this section, we devise resampling-based methods for  deriving a valid confidence interval for $\beta^*$, addressing the post-selection issue in aggregating multi-source data. 
%We provide theoretical justifications for the proposed inference procedure in Section \ref{sec: theory}. 

%%%%%%%%%%%%%%%%%%%%%%%%%%%%%%%%%%%%%%%%%%%%%%%%%%%%%%%%%%%%
\subsection{RIFL: resampling-based inference}
\label{sec: RIFL method}
%%%%%%%%%%%%%%%%%%%%%%%%%%%%%%%%%%%%%%%%%%%%%%%%%%%%%%%%%%%%
The RIFL interval construction consists of two steps. In the first step, we resample the dissimilarity measures and screen out the inaccurate resampled measures. In the second step, we use the resampled dissimilarity measures to estimate the prevailing set, which is further used to generate a sampled confidence interval. 
\vspace{1.5mm}

\noindent {\bf Step 1: resampling and screening.} Conditioning on the observed data, for $1\leq l<k\leq L,$ we generate $\{\widehat{\DD}^{[m]}_{l,k}\}_{1\leq m\leq M}$ and $\{\widehat{\LL}^{[m]}_{l,k}\}_{1\leq m\leq M}$ following 
\begin{equation}
\widehat{\DD}^{[m]}_{l,k}\stackrel{i.i.d}{\sim} N\left(\widehat{\DD}_{l,k},\widehat{\rm SE}^2(\widehat{\DD}_{l,k})\right), \quad  \widehat{\LL}^{[m]}_{l,k} \stackrel{i.i.d}{\sim} N\left(\widehat{\LL}_{l,k},\widehat{\rm SE}^2(\widehat{\LL}_{l,k})\right) \quad \text{for}\quad 1\leq m\leq M .
\label{eq: resampling}
\end{equation}
The above generating mechanism in \eqref{eq: resampling} guarantees that the distributions of
$\widehat{\DD}^{[m]}_{l,k}-\widehat{\DD}_{l,k}$ and $\widehat{\LL}^{[m]}_{l,k}-\widehat{\LL}_{l,k}$ approximate those of $\widehat{\DD}_{l,k}-{\DD}_{l,k}$ and $\widehat{\LL}_{l,k}-{\LL}_{l,k}$, respectively. 

%\Larry{Zijian, add a remark that we only sample the upper right triangle and the distances are 0 along the diagonal.}

\begin{Remark}\rm
The random variables $\{\widehat{\DD}^{[m]}_{l,k}\}_{1\leq l<k\leq L}$ and $\{\widehat{\LL}^{[m]}_{l,k}\}_{1\leq l<k\leq L}$ are independently generated. Such a resampling method is effective even though  $\{\widehat{\DD}_{l,k}\}_{1\leq l<k\leq L}$ and $\{\widehat{\LL}_{l,k}\}_{1\leq l<k\leq L}$ are correlated. Our proposal can be generalized to capture the correlation structure among  $\{\widehat{\DD}_{l,k}\}_{1\leq l<k\leq L}$ and $\{\widehat{\LL}_{l,k}\}_{1\leq l<k\leq L}$. However, our focus is on the resampling in \eqref{eq: resampling} since it does not require all sites to provide the correlation structure of all dissimilarity measures. 
\end{Remark}

With the resampled data, we mimic \eqref{eq: global difference}
 and define the resampled test statistics 
 \begin{equation}
\widehat{S}^{[m]}_{l,k}\coloneqq \max\left\{\left|\widehat{\DD}^{[m]}_{l,k}/{\widehat{\rm SE}(\widehat{\DD}_{l,k})}\right|,\left|\widehat{\LL}^{[m]}_{l,k}/{\widehat{\rm SE}(\widehat{\LL}_{l,k})}\right|\right\} \quad \text{for}\quad 1\leq m\leq M.
\label{eq: global difference sampling}
\end{equation}
To properly account for the uncertainty of the test $\Hhat\sublk$ defined in \eqref{eq: voting est}, 
we show that there exists resampled dissimilarity measures $\{\widehat{\DD}^{[m^*]}_{l,k}\}_{1\leq l<k\leq L}$ and $\{\widehat{\LL}^{[m^*]}_{l,k}\}_{1\leq l<k\leq L}$ that are nearly the same as the corresponding true dissimilarity measures $\{{\DD}_{l,k}\}_{1\leq l<k\leq L}$ and $\{{\LL}_{l,k}\}_{1\leq l<k\leq L}$. In particular, the following Theorem \ref{thm: sampling} establishes that with probability larger than $1-\nu$, there exists $1\leq m^*\leq M$ and $\rho(M)=c_*(\nu)\left({\log n }/{M}\right)^{1/[L(L-1)]}$ such that 
\begin{equation}
\max_{1\leq l<k\leq L}\max\left\{\left|\frac{\widehat{\DD}^{[m^*]}_{l,k}-\DD_{l,k}}{\widehat{\rm SE}(\widehat{\DD}_{l,k})}\right|,\left|\frac{\widehat{\LL}^{[m^*]}_{l,k}-\LL_{l,k}}{\widehat{\rm SE}(\widehat{\LL}_{l,k})}\right|\right\}\leq {\rho(M)}\cdot \Thre, \quad \text{with}\quad \Thre= z_{\sig/[2L(L-1)]},
\label{eq: shrinking}
\end{equation}
where $n=\min_{1\leq l\leq L} n_l$, and $c_*(\nu)$ is a positive constant dependent on the probability $\nu$ but independent of $n$ and $d$. With a large resampling size $M$, we have $\rho(M)\rightarrow 0$, indicating that $\widehat{\DD}^{[m^*]}$ and $\widehat{\LL}^{[m^*]}$ are nearly the same as $\DD$ and $\LL$, respectively. Similar to  \eqref{eq: voting est}, the threshold level $\Thre$ in \eqref{eq: shrinking} can be interpreted as the Bonferroni correction
threshold level, which adjusts for the multiplicity of testing the similarity between all sites. 

Following \eqref{eq: shrinking}, we adjust the threshold $\Thre$ by a factor of $\rho(M)$ when constructing the resampled voting matrix. Specifically, we define $\widehat{H}^{[m]}=\{\widehat{H}^{[m]}_{l,k}\}_{1\leq l<k\leq L}$ as 
\begin{equation}
\widehat{H}^{[m]}_{l,k}={\bf 1}\left(\widehat{S}^{[m]}_{l,k}\leq {\rho(M)}\cdot \Thre\right) \quad \text{for}\quad 1\leq l<k\leq L.
\label{eq: voting sampling}
\end{equation}
We define $\widehat{H}^{[m]}_{l,l}=1$ for $1\leq l\leq L,$ and $\widehat{H}^{[m]}_{l,k}=\widehat{H}^{[m]}_{k,l}$ for $1\leq k<l\leq L.$  The empirical guidance of choosing the tuning parameter $\rho(M)\in (0,1)$ is provided in the following Section \ref{sec: tuning selection}.

%\tcomm{zijian: not sure if you have more comments on the role of $\rho(M)$} \Zijian{Tianxi, I have added a bit more discussion here.}

The shrinkage parameter $\rho(M)$ reduces the critical values of all pairs of similarity tests in \eqref{eq: voting sampling} and leads to a stricter rule of claiming any two sites to be similar. 
The probabilistic statement in (\ref{eq: shrinking}) suggests that at least one of the resampled voting matrices can accurately reflect the true voting matrix $H$ with the significantly reduced threshold $\rho(M)\cdot \Thre$.  On the other hand, some of the resampled statistics $\widehat{S}^{[m]}_{l,k}$ may deviate from $0$, leading to a very sparse $\widehat{H}^{[m]}$ that may be different from the true $H$. To synthesize information from $M$ resamples, we first infer the subset of the resamples representative of the underlying structure and subsequently take a union of the confidence intervals constructed based on those plausible subsets of sites. To this end, we first compute the maximum clique of the graph $\mathcal{G}([L],\widehat{H}^{[m]}),$ denoted as
\begin{equation}
\widehat{\mathcal{V}}^{[m]}=\MC([L],\widehat{H}^{[m]}) \quad \text{for}\quad 1\leq m\leq M.
\label{eq: Vhat sampling}
\end{equation}
We use   $|\widehat{\mathcal{V}}^{[m]}|$ to determine whether $\widehat{H}^{[m]}$ is similar to $H$. If  $|\widehat{\mathcal{V}}^{[m]}|$ is less than or equal to $L/2$, we view $\widehat{H}^{[m]}$ as being far from $H$ and discard the $m$-th resampled dissimilarity measures. We shall only retain the resampled voting matrices $H^{[m]}$ belonging to the index set $\mathcal{M}$ defined as
\begin{equation}
\mathcal{M}\coloneqq \left\{1\leq m\leq M: |\widehat{\mathcal{V}}^{[m]}|>L/2\right\} .
\label{eq: index set}
\end{equation}

\vspace{1.5mm}
 
\noindent {\bf Step 2: aggregation.}
For $m\in \mathcal{M}$, we construct the estimated prevailing set $\widetilde{\mathcal{V}}^{[m]}$ as, 
\begin{equation}
\widetilde{\mathcal{V}}^{[m]}=\{1\leq l\leq L: \|\widehat{H}^{[m]}_{l,\cdot}\|_0>L/2\}.
\label{eq: enlarged resampling}
\end{equation}
The set $\widetilde{\mathcal{V}}^{[m]}$ contains all indexes receiving more than half of the votes.  With $\widetilde{\mathcal{V}}^{[m]}$ in \eqref{eq: enlarged resampling}, we apply the inverse variance weighted estimator 
\begin{equation*}
\widehat{\beta}^{[m]}=\frac{\sum_{l\in \widetilde{\mathcal{V}}^{[m]}}{\widehat{\beta}\supl
}/{\widehat{\sigma}_{l}^2}}{\sum_{l\in \widetilde{\mathcal{V}}^{[m]}}1/{\widehat{\sigma}_{l}^2}}.
\end{equation*}
For the significance level $\alpha,$ we construct the $1-\alpha$ confidence interval as 
\begin{equation}
{\rm CI}^{[m]}=\left(\widehat{\beta}^{[m]}-z_{\alpha_1/2}\frac{1}{\sqrt{{\sum_{l\in \widetilde{\mathcal{V}}^{[m]}}1/{\widehat{\sigma}_{l}^2}}}},\widehat{\beta}^{[m]}+z_{\alpha_1/2}\frac{1}{\sqrt{{\sum_{l\in \widetilde{\mathcal{V}}^{[m]}}1/{\widehat{\sigma}_{l}^2}}}}\right),
\label{eq: CI sampling}
\end{equation} 
{where $\alpha_1=\alpha-\nu$ with $\nu$ denoting a pre-specified small probability used to guarantee the sampling property in \eqref{eq: shrinking}. We choose the default value of $\nu$ as $\nu=\alpha/20$ throughout the paper.}

%\Zijian{with $\alpha_1=0.95 \alpha.$ In \eqref{eq: CI sampling}, instead of $\alpha,$ we use $\alpha_1<\alpha$ to construct the confidence interval, where the difference $\alpha-\alpha_1$ is used to control the sampling uncertainty. Any $\alpha_1<\alpha$ is sufficient for our proposal, and we set $\alpha_1=0.95\alpha$ as the default choice.}  \tcomm{this is  not right, you need $\alpha_1$ to depend on $\nu$ based on your previous statement}
Finally, we construct the CI for $\beta^*$ as 
\begin{equation}
{\rm CI}=\cup_{m\in \mathcal{M}}{\rm CI}^{[m]},
\label{eq: CI union}
\end{equation}
with $\mathcal{M}$ and ${\rm CI}^{[m]}$ defined in \eqref{eq: index set} and \eqref{eq: CI sampling}, respectively.
We refer to ${\rm CI}$ as a confidence interval although $\cup_{m\in \mathcal{M}}{\rm CI}^{[m]}$ may not be an interval. The RIFL algorithm for constructing CI is also summarized in Algorithm \ref{algo: multi-source sampling} in Section \ref{sec: algorithm} of the supplement.

{The average of the maximum and minimum value of the confidence interval defined in \eqref{eq: CI union} serves as a point estimator of $\beta^*.$} Additionally, we may use $\widehat{p}_l={\sum_{m\in \mathcal{M}} {\bf 1}(l\in \widetilde{\mathcal{V}}^{[m]})}/{|\mathcal{M}|}$, the proportion of times site $l$ being included in the majority group,  as a generalizability measure for the $l$-th site.

\begin{Remark}[Difference between $\widehat{\mathcal{V}}^{[m]}$ and $\widetilde{\mathcal{V}}^{[m]}$]\rm 
For $m\in \mathcal{M}$, we have $\widehat{\mathcal{V}}^{[m]}\subset \widetilde{\mathcal{V}}^{[m]},$ that is, the set $\widetilde{\mathcal{V}}^{[m]}$ in \eqref{eq: enlarged resampling} is less restrictive compared to the maximum clique set $\widehat{\mathcal{V}}^{[m]}$ defined in \eqref{eq: index set}. This relationship helps explain why two different set  estimators $\widehat{\mathcal{V}}^{[m]}$ and  $\widetilde{\mathcal{V}}^{[m]}$ are used in our construction. Firstly, the maximum clique set $\widehat{\mathcal{V}}^{[m]}$ imposes a stricter rule than  $\widetilde{\mathcal{V}}^{[m]}$ in \eqref{eq: enlarged resampling} and the maximum clique set $\widehat{\mathcal{V}}^{[m]}$ is likely to screen out more inaccurate resamples. Secondly, our theory shows that there exists $m^*\in \mathcal{M}$ such that   $\widetilde{\mathcal{V}}^{[m^*]}$ is guaranteed to recover the true prevailing set $\mathcal{V}(\theta^*)$. For $m\in \mathcal{M}$, the set $\widetilde{\mathcal{V}}^{[m]}$ tends to contain more sites than $\widehat{\mathcal{V}}^{[m]}$, leading to shorter ${\rm CI}^{[m]}.$  The use of $\widetilde{\mathcal{V}}^{[m]}$ in \eqref{eq: enlarged resampling} enhances the precision of the resulting RIFL confidence interval.
\end{Remark}

%\begin{Remark}[Leveraging additional prior information]\rm  
%\end{Remark}

%%%%%%%%%%%%%%%%%%%%%%%%%%%%%%%%%%%%%%%%%%%%%%%%%%%%%%%%%%%%
%\subsection{Leverage of additional prior information}
%%%%%%%%%%%%%%%%%%%%%%%%%%%%%%%%%%%%%%%%%%%%%%%%%%%%%%%%%%%%

%%%%%%%%%%%%%%%%%%%%%%%%%%%%%%%%%%%%%%%%%%
\subsection{Extension to multivariate target parameters}
\label{sec: multivariate}
%%%%%%%%%%%%%%%%%%%%%%%%%%%%%%%%%%%%%%%
%%%%%%%%%%%%%%%%%%%%%%%%%%%%%%%%%%%%%%%%%%%%%%%%%%%%%%%%%%%%%%%%%%%%%%%%%%%%%%%%%
Our method can be easily extended to construct confidence regions for a multi-dimensional target parameter $\beta^*\in \R^{q}$.  As a generalization of \eqref{eq: limiting distribution}, we assume the site-specific estimators $\{\widehat{\beta}\supl,\widehat{\Omega}_l\}_{1\leq l\leq L}$ satisfy 
$
\widehat{\Omega}_l^{-1/2}(\widehat{\beta}\supl-\beta\supl)\cid N(0,{\rm I}_{q\times q}),
$
where $\betahat\supl\in \R^{q}$,  $\widehat{\Omega}_l\in \R^{q\times q}$ denotes the estimated covariance matrix of $\betahat\supl,$ and ${\rm I}_{q\times q}$ is the $q\times q$ identity matrix.

We shall highlight two main adjustments to the RIFL algorithm described above for the univariate $\beta^*$. Firstly, we modify the computation of 
$\widehat{\LL}_{l,k}$ and $\widehat{\rm SE}(\widehat{\LL}_{l,k})$  as 
$\widehat{\LL}_{l,k}=\|\betahat\supl-\widehat{\beta}\supk\|_2^2$
and 
$$\widehat{\rm SE}(\widehat{\LL}_{l,k})=\sqrt{4(\betahat\supl-\widehat{\beta}\supk)^{\intercal}(\widehat{\Omega}_l+\widehat{\Omega}_k) (\betahat\supl-\widehat{\beta}\supk)+1/\min\{n_l,n_k\}},$$
where $1/\min\{n_l,n_k\}$ is used to control higher order approximation error of $\|\betahat\supl-\widehat{\beta}\supk\|_2^2.$
Secondly, we generalize the construction of ${\rm CI}^{[m]}$ in equation \eqref{eq: CI sampling} as 
\begin{equation}
{\rm CS}^{[m]}=\left\{\beta\in \R^{q}: (\beta-\widehat{\beta}^{[m]})^{\intercal}\left(\sum_{l\in \widetilde{\mathcal{V}}^{[m]}} \widehat{\Omega}_l^{-1}\right)(\beta-\widehat{\beta}^{[m]})\leq \chi^2_q(\alpha)\right\}
\end{equation}
where $\widehat{\beta}^{[m]}=\left(\sum_{l\in \widetilde{\mathcal{V}}^{[m]}} \widehat{\Omega}_l^{-1}\right)^{-1}\left(\sum_{l\in \widetilde{\mathcal{V}}^{[m]}} \widehat{\Omega}_l^{-1}\betahat\supl\right)$
and $\chi^2_q(\alpha)$ denotes the upper $\alpha$ quantile of the $\chi^2$ distribution with $q$ degrees of freedom. Our final confidence set is ${\rm CS}=\cup_{m\in \mathcal{M}}{\rm CS}^{[m]}.$
%%%%%%%%%%%%%%%%%%%%%%%%%%%%%%%%%%%%%%%%%%%%%%%%%%%%%%%%%%%%%%%%%%%%%%%%%%%%%%%%%
\subsection{Tuning parameter selection}
\label{sec: tuning selection}
%%%%%%%%%%%%%%%%%%%%%%%%%%%%%%%%%%%%%%%%%%%%%%%%%%%%%%%%%%%%%%%%%%%%%%%%%%%%%%%%%
{Implementing the RIFL CI requires the specification of the resampling size $M$ and the shrinkage parameter $\rho(M)\in (0,1)$ used in \eqref{eq: voting sampling}. The proposed method is not sensitive to the choice of the size $M$ as long as it is sufficiently large (e.g., $M\geq 500$). We set $M = 500$ as the default value. Our theoretical results in Section \ref{sec: theory} suggest the choice of $\rho(M)$ as $\rho(M)=c_*\left({\log n }/{M}\right)^{1/[L(L-1)]}$ for some positive constant $c_*$; see the following equation \eqref{eq: choice of rho}.  If the constant $c_*$ is chosen to be too small, most resampled dissimilarity measures will not produce a maximum clique set satisfying the majority rule, resulting in a very small $|\mathcal{M}|$. Consequently, we can use $|\mathcal{M}|$ to determine whether the tuning parameter $\rho(M)$ is chosen to be sufficiently large. In practice, we start with a small value of $c_*$ (e.g., $c_*=1/12$) and increase the value of $c_*$ until a pre-specified proportion, say ${\rm prop}=10\%$, of resampled dissimilarity measures produce maximum clique sets satisfying the majority rule. The RIFL CI is nearly invariant to the original threshold $\Thre$ in \eqref{eq: shrinking}. Our selected value of $\rho(M)$ will ensure a large enough $\rho(M)\cdot \Thre$ such that more than ${\rm prop}=10\%$ of the resampled sets satisfy the majority rule. 
Even if the threshold $\Thre$ may be conservative due to the Bonferroni correction, this choice of $\Thre$ does not affect the performance of our RIFL CI since the choice of $\rho(M)$ will be adaptive to the specification of the threshold $\Thre.$ We demonstrate the robustness to the tuning parameters over the numerical studies in Section \ref{sec: tuning parameter}. 

%{\blue In Section 6, we further discuss alternative strategies for choosing the tuning parameter $\rho$.} \tcomm{xiudi: plz check}}

%\Zijian{Xiudi, move this to supplement (how about Section \ref{sec: algorithm}?) and also check my discussion in the discussion section \ref{sec: discussion}.}
%\Xiudi{Another rule suggested by Tianxi: start with $K = \textnormal{floor}(L+1)/2 + 1$, find $\rho$ such that 10\% of resampled data estimated a prevailing set of size greater than or equal to $K$. Let $\mathcal{M}_K$ denote the index of these resamples. Now for $j \in \{1,\ldots,L\}$, compute $f_j = \sum_{m \in \mathcal{M}_K} I\{ j \in \hat{\mathcal{V}}^m\}/|\mathcal{M}_K|$ and define $\mathcal{U}_K = \{j: f_j \geq K/L\}$. If $|\mathcal{U}_K| \leq K$, stop; otherwise set $K = K + 1$, repeat.}\Zijian{Xiudi, I have defined $\widehat{p}_l={\sum_{m\in \mathcal{M}} {\bf 1}\left(l\in \widetilde{\mathcal{V}}^{[m]}\right)}/{|\mathcal{M}|}$ as the generalizability measure and please adopt the same notation.}
%\tcomm{you have a rule above but now you have a different suggestion??}\Zijian{Please double check.}

% \Zijian{Larry, please confirm that the default value is $M=1000$ or $M=500.$} 
%Our theoretical and numerical results suggest that the RIFL performance is nearly invariant to the choice of $M$ provided that it is sufficiently large (e.g., $M\geq 500$); see Figure \ref{fig:rho} for the numerical results. 

%%%%%%%%%%%%%%%%%%%%%%%%%%%%%%%%%%%%%%%%%%%%%%%%%%%%%%%%%%%%
\section{Theoretical justification for RIFL inference}
\label{sec: theory}
%%%%%%%%%%%%%%%%%%%%%%%%%%%%%%%%%%%%%%%%%%%%%%%%%%%%%%%%%%%%
We next provide theoretical justifications for the validity of the RIFL CI. Let $n=\min_{1\leq l\leq L} n_l$ and define 
\begin{equation}
\err_n(M,\nu) =c_*(\nu)\left[\frac{ \log n}{ M}\right]^{\frac{1}{L(L-1)}} \; \mbox{with} \; \;
c_*(\nu)=2^{\frac{1}{L(L-1)}-\frac{1}{2}}{\sqrt{\pi}} \exp\left(\frac{1}{2}z^2_{\nu/[2L(L-1)]}\right),
\label{eq: sampling accuracy}
\end{equation}
where $L\geq 2$ and $0<\nu<1/2.$ The term $\err_n(M,\nu)$ quantifies the sampling accuracy, which denotes the smallest difference between the true dissimilarity measures and the resampled ones after resampling $M$ times. For a constant $\nu\in (0,1/2)$ and fixed $L,$ $c_*(\nu)$ is a constant only depending on the pre-specified probability $\nu$ and $\err_n(M,\nu)$ is of order $(\log n /M)^{\frac{1}{L(L-1)}}$, which tends to 0 with a sufficiently large $M$. 

The following theorem establishes the critical sampling property, which provides the theoretical support for the threshold reduction in \eqref{eq: voting sampling}.
\begin{Theorem} Suppose that the site-specific estimators
$\{\widehat{\beta}\supl,\widehat{\sigma}_l\}_{1\leq l\leq L}$  satisfy \eqref{eq: limiting distribution} and the dissimilarity measures $\{\widehat{\DD}_{l,k},\widehat{\rm SE}(\widehat{\DD}_{l,k})\}_{1\leq l<k \leq L}$ satisfy \eqref{eq: key assumption}. Then the resampled dissimilarity measures $\{\widehat{\DD}^{[m]}_{l,k}\}_{1\leq m\leq M}$ and $\{\widehat{\LL}^{[m]}_{l,k}\}_{1\leq m\leq M}$ defined in \eqref{eq: resampling} satisfy
\begin{equation*}
\liminf_{n\rightarrow\infty}\liminf_{M\rightarrow\infty}\PP\left(\min_{1\leq m\leq M}\left[\max_{1\leq l<k\leq L}\max\left\{\left|\frac{\widehat{\DD}^{[m]}_{l,k}-\DD_{l,k}}{\widehat{\rm SE}(\widehat{\DD}_{l,k})}\right|,\left|\frac{\widehat{\LL}^{[m]}_{l,k}-\LL_{l,k}}{\widehat{\rm SE}(\widehat{\LL}_{l,k})}\right|\right\}\right] \leq \err_n(M,\nu)\right)\geq 1-\nu,
\end{equation*}
for any positive constant $0<\nu<1/2.$
\label{thm: sampling}
\end{Theorem}
The above theorem shows that, with a high probability, there exists $1\leq m^*\leq M$ such that 
$$\max_{1\leq l<k\leq L}\max\left\{\left|\frac{\widehat{\DD}^{[m^*]}_{l,k}-\DD_{l,k}}{\widehat{\rm SE}(\widehat{\DD}_{l,k})}\right|,\left|\frac{\widehat{\LL}^{[m^*]}_{l,k}-\LL_{l,k}}{\widehat{\rm SE}(\widehat{\LL}_{l,k})}\right|\right\}  \leq \err_n(M,\nu)\asymp \left(\frac{\log n }{M}\right)^{\frac{1}{L(L-1)}}.$$
This provides the theoretical founding for \eqref{eq: shrinking}.  
%where the threshold is shrunk by a factor of $\rho(M)=c_*\left({\log n }/{M}\right)^{1/L(L-1)}$. 

The following theorem establishes the coverage property of the proposed RIFL CI. 
\begin{Theorem}
Suppose that the assumptions of Theorem \ref{thm: sampling} hold and the thresholding level $\rho(M)\cdot \Thre$ used in \eqref{eq: shrinking} satisfies 
\begin{equation}
{\rho(M)} \cdot \Thre \geq \err_n(M,\nu)\quad \text{and}\quad  \lim_{M\rightarrow \infty}{\rho(M)} \cdot \Thre = 0.
\label{eq: tuning parameter condition}
\end{equation}
Then the confidence interval defined in \eqref{eq: CI union} satisfies 
\begin{equation*}
\liminf_{n\rightarrow\infty}\liminf_{M\rightarrow \infty}\PP\left(\beta^*\in {\rm CI}\right)\geq 1-\alpha,
\end{equation*}
with $\beta^*=\T(\theta^*)$ and $0<\alpha<1$ denoting the significance level used in \eqref{eq: CI sampling}. 
\label{thm: coverage general}
\end{Theorem}

{Equation \eqref{eq: tuning parameter condition} is a condition on the shrinkage parameter $\rho(M).$ We can choose $\rho(M)$ as 
\begin{equation}
 {\rho(M)}= \err_n(M,\nu)/T=c_* (\log n /M)^{\frac{1}{L(L-1)}}, \quad \text{with} \quad c_*=c_*(\nu)/\Thre,
 \label{eq: choice of rho}
\end{equation}
where $c_*$ is a constant independent of $n$ and $d$.} With a sufficiently large $M$, the choice of $\rho(M)$ in \eqref{eq: choice of rho} automatically satisfies the condition \eqref{eq: tuning parameter condition}.

%\Zijian{Tianxi, check the following Theorem 3. Our method can detect all strongly violated sites.} 
{ 
In the following, we show that the RIFL method can detect all sites whose model parameter is well separated from the prevailing model. Consequently, when all sites from the non-majority group are well separated from those from the majority group, the RIFL CI matches with the oracle CI assuming the prior knowledge $\mathcal{V}(\theta^*)$. 
With the prior knowledge $\mathcal{V}^*=\mathcal{V}(\theta^*)$, we can construct the oracle $1-\alpha$ confidence interval as 
\begin{equation}
{\rm CI}_{\rm ora}=\left(\widehat{\beta}^{\rm ora}-z_{\alpha/2}\frac{1}{\sqrt{\sum_{l\in \mathcal{V}^*}1/{\widehat{\sigma}_{l}^2}}}, \widehat{\beta}^{\rm ora}+z_{\alpha/2}\frac{1}{\sqrt{\sum_{l\in \mathcal{V}^*}1/{\widehat{\sigma}_{l}^2}}}\right), \; \text{with}\; \widehat{\beta}^{\rm ora}=\frac{\sum_{l\in {\mathcal{V}^*}}{\widehat{\beta}\supl
}/{\widehat{\sigma}_{l}^2}}{\sum_{l\in {\mathcal{V}^*}}1/{\widehat{\sigma}_{l}^2}}.
\label{eq: ora CI}
\end{equation}

 \begin{Theorem}
 Suppose that Conditions of Theorem \ref{thm: coverage general} hold. For $k\in \mathcal{V}^c$, if  \begin{equation} |\beta\supk-\beta^*|\geq (2\sqrt{2\log n+ 2\log M}+\rho(M)\cdot \Thre)\cdot \max_{l\in \mathcal{V}}\widehat{\rm SE}(\widehat{\LL}_{l,k}),
 \label{eq: well separation}
 \end{equation} 
 then $\lim_{n\rightarrow \infty}\lim_{M\rightarrow \infty}\PP\left(k\not\in \cup_{m\in \mathcal{M}} \widetilde{\mathcal{V}}^{[m]}\right)=1.$
 Additionally, if  $|\mathcal{V}|=\lfloor L/2\rfloor+1$ and any $k\in \mathcal{V}^c$ satisfies \eqref{eq: well separation}, then the confidence interval defined in \eqref{eq: CI union} satisfies 
$$\lim_{n\rightarrow \infty}\lim_{M\rightarrow \infty}\PP\left({\rm CI}={\rm CI}_{\rm ora}\right)=1.$$
\label{thm: efficiency}
 \end{Theorem}

The condition \eqref{eq: well separation} %\Xiudi{in (25), $\widehat{\rm SE}(\widehat\LL_{l,k})$? or ${\rm SE}(\widehat\LL_{l,k})$?}
requires the  non-majority site $k$ to be significantly different from the majority sites. Such well-separated non-majority site $k$ will not be included in any of the resampled prevailing set $\widetilde{\mathcal{V}}^{[m]}$ defined in \eqref{eq: enlarged resampling}. When all non-majority sites are easy to detect and there are just more than half of the sites belonging to $\mathcal{V}$, the above theorem shows that our proposed CI can be the same as the oracle CI knowing $\mathcal{V}^*$. 
}

%Empirically, we shall choose the smallest constant $c_*>0$ such that a proportion of the resampled $\{\widehat{V}^{[m]}\}_{1\leq m\leq M}$ satisfy the majority rule; see Remark \ref{rem: tuning parameter sel} for the details. \tcomm{comment on $\rho(M)$ and $c_*$ seems repetitive since you alread discussed about} 

%\tcomm{tuning parameters are important but i find it a bit hard to follow when these comments are scattered. you can just have a subsection on tuning. $\nu$ in a way is also a hyper parameter} \Zijian{Tianxi, I have moved all discussion about tuning parameters to section \ref{sec: tuning selection} in the numerical studies. Our procedure does not require this $\nu$, which is only used in the assumption and conclusion of Theorem 1.}

%%%%%%%%%%%%%%%%%%%%%%%%%%%%%%%%%%%%%%%%%%%%%%%%%%%%%%%%%%%%%%%%%%%%%%%%%%%%%%%%%%%%%%
\section{Applications to Multi-source Inference Problems}
\label{sec: applications}
%%%%%%%%%%%%%%%%%%%%%%%%%%%%%%%%%%%%%%%%%%%%%%%%%%%%%%%%%%%%%%%%%%%%%%%%%%%%%%%%%%%%%%

In this section, we detail the RIFL method for three inference problems that arise frequently in practice. We investigate (i) robust inference under a general parametric modeling framework in Section \ref{sec: app 1}, (ii) the statistical inference problem of a prevailing high-dimensional prediction model in Section \ref{sec: app 2}; (iii) estimation of ATEs from multiple sites in Section \ref{sec: app 3}. Across this wide range of applications, the prevailing model is of great interest since it is a generalizable model matching the majority of the observed data sets. 
The application of RIFL method requires the construction of the site-specific estimators
$\{\widehat{\beta}\supl,\widehat{\sigma}_l\}_{1\leq l\leq L}$ together with the dissimilarity measures 
 $\{\widehat{\DD}_{l,k}, \widehat{\rm SE}(\widehat{\DD}_{l,k})\}_{1\leq l<k\leq L},$ which will be detailed in the following for each application. 

%%%%%%%%%%%%%%%%%%%%%%%%%%%%%%%%%%%%%%%%%%%%%%%%%%%%%%%%%%%%%%%%%%%%%%%%%%%%%%%%%%%%%%
\subsection{Parametric model inference}
\label{sec: app 1}
%%%%%%%%%%%%%%%%%%%%%%%%%%%%%%%%%%%%%%%%%%%
%%%%%%%%%%%%%%%%%%%%%%%%%%%%%%%%%%%%%%%%%%%

We consider that we have access to $L$ sites, where the $l$-th site (with $1\leq l\leq L$) is a parametric model with the associated model parameter $\theta^{(l)}\in \R^{d}.$ We focus on the low-dimensional setting in the current subsection and will move to the high-dimensional model in Section \ref{sec: app 2}. For $1\leq l\leq L,$ the $l$-th site outputs an estimator $\widehat{\theta}^{(l)}$ satisfying 
\begin{equation}
\sqrt{n_l}(\widehat{\theta}^{(l)}-\theta^{(l)})\cid N(0, C^{(l)}),
\label{eq: limiting ex1 low}
\end{equation}
where the covariance matrix $C^{(l)}\in \R^{d\times d}$ is consistently estimated by $\widehat{C}\supl$. This generic setting covers many parametric and semi-parametric models, including but not limited to generalized linear models, survival models, and quantile regression. Many existing estimators, including M-estimators, satisfy the condition \eqref{eq: limiting ex1 low} under regularity conditions.

Our goal is to make inferences for the pre-specified transformation $g(\theta^*),$ where the prevailing model $\theta^*$ is defined as the one agreeing with more than half of $\{\theta^{(l)}\}_{1\leq l\leq L}.$ We estimate $\beta^{(l)}$ by the plug-in estimator  $\widehat{\beta}^{(l)}=g(\widehat{\theta}^{(l)})$. For a twice differentiable function $g$, we apply the delta method and construct the standard error estimator of  $\widehat{\beta}^{(l)}$ as $\widehat{\sigma}_l=\sqrt{\left[\triangledown g(\widehat{\theta}^{(l)})\right]^{\intercal}\widehat{C}^{(l)}\triangledown g(\widehat{\theta}^{(l)})}$ with $\triangledown g(\cdot)$ denoting the gradient of the function $g$. 
We measure the dissimilarity between two sites with the quadratic norm $$\DD_{l,k}=\|\theta^{(l)}-\theta^{(k)}\|_2^2 \quad \text{for} \quad 1\leq l<k\leq L.$$ In the following, we specify the estimator $\widehat{\DD}_{l,k}$ and its standard error estimators $\widehat{\rm SE}(\widehat{\DD}_{l,k})$ and then apply RIFL to construct confidence intervals for $\T(\theta^*)$. 

For a given $l$ and $k$, define $\gamma\sublk=\theta^{(l)}-\theta^{(k)}$ and $\widehat{\gamma}\sublk=\widehat{\theta}^{(l)}-\widehat{\theta}^{(k)}.$ With a slight abuse of notation, we drop the subscript $\sublk$ in $\gamma$ and $\widehat\gamma$ next for ease of presentation. We decompose the error of $\widehat{\DD}_{l,k}=\|\widehat{\gamma}\|_2^2$ as follows,
\begin{equation}
\widehat{\DD}_{l,k}-\DD_{l,k}=\|\widehat{\gamma}\|_2^2-\|{\gamma}\|_2^2=2\langle \widehat{\gamma}-\gamma, \gamma\rangle+\| \widehat{\gamma}-\gamma\|_2^2.
\label{eq: error decomposition}
\end{equation}
The first term $2\langle \widehat{\gamma}-\gamma, \gamma\rangle$ on the right-hand side satisfies 
\begin{equation*}
\frac{2\langle \widehat{\gamma}-\gamma, \gamma\rangle}{\sqrt{4\gamma^{\intercal}C^{(l)}\gamma/n_l+4\gamma^{\intercal}C^{(k)}\gamma/n_k}}\cid N(0,1) .
\end{equation*}
Together with the decomposition in \eqref{eq: error decomposition}, we estimate the standard error of $\widehat{\DD}_{l,k}$ by 
\begin{equation}
\widehat{\rm SE}(\widehat{\DD}_{l,k})=\sqrt{4\widehat{\gamma}^{\intercal}\widehat{C}^{(l)}\widehat{\gamma}/n_l+4\widehat{\gamma}^{\intercal}\widehat{C}^{(k)}\widehat{\gamma}/n_k+1/\min\{n_l,n_k\}} .
\label{eq: SE low-dim}
\end{equation}
The extra term $1/\min\{n_l,n_k\}$ in the  definition of $\widehat{\rm SE}(\widehat{\DD}_{l,k})$ is used to control the uncertainty of
the second term $\| \widehat{\gamma}-\gamma\|_2^2$ on the right-hand side of \eqref{eq: error decomposition}. 
The construction of $\widehat{\beta}\supl,\widehat{\sigma}_l,\widehat{\DD}_{l,k}$ and $\widehat{\rm SE}(\widehat{\DD}_{l,k})$ only depends on the summary statistics $\{\widehat{\theta}^{(l)}\}_{1\leq l\leq L}$ and $\{\widehat{C}^{(l)}\}_{1\leq l\leq L}.$ RIFL can be implemented without requiring sharing individual-level data. The following Theorem \ref{thm: low-dim dissimilarity} justifies that $\widehat{\DD}_{l,k}=\|\widehat{\gamma}\|_2^2$  and ${\widehat{\rm SE}(\widehat{\DD}_{l,k})}$ in \eqref{eq: SE low-dim} satisfy \eqref{eq: key assumption} for $1\leq l<k\leq L$.

\begin{Theorem}
Suppose that $\{\widehat{\theta}^{(l)}\}_{1\leq l\leq L}$ satisfy \eqref{eq: limiting ex1 low} and the largest eigenvalue of $C^{(l)}$, denoted as $\lambda_{\max}(C^{(l)})$, is bounded for $1\leq l\leq L.$ If $\widehat{C}^{(l)}$ is consistent estimator of $C^{(l)}$, the point estimator $\widehat{\DD}_{l,k}=\|\widehat{\gamma}\|_2^2$  and its standard error estimator ${\widehat{\rm SE}(\widehat{\DD}_{l,k})}$ in \eqref{eq: SE low-dim} satisfy \eqref{eq: key assumption} for $1\leq l<k\leq L$.
\label{thm: low-dim dissimilarity}
\end{Theorem}

%%%%%%%%%%%%%%%%%%%%%%%%%%%%
%%%%%%%%%%%%%%%%%%%%%%%%%%%%
\subsection{High-dimensional prediction model}
\label{sec: app 2}
%%%%%%%%%%%%%%%%%%%%%%%%%%%%
%%%%%%%%%%%%%%%%%%%%%%%%%%%%
We next consider the inference problem of a prevailing high-dimensional prediction model by aggregating the information from $L$ sites. For the $l$-th site, we consider the following generalized linear model (GLM) for the data $\{X^{(l)}_{i}, Y^{(l)}_{i}\}_{1\leq i\leq n_l},$  
\begin{equation*}
\EE(Y_i^{(l)}\mid X_i^{(l)})=h(\mu_l+[X_i^{(l)}]^{\intercal} \theta^{(l)}) \quad \text{for}\quad 1\leq i\leq n_l,
\label{eq: high-dim GLM}
\end{equation*}
where $h(\cdot)$ is a known link function, $\mu_l$ is the intercept, and $\theta^{(l)}\in \R^{d}$ is the regression vector corresponding to the covariates. We assume $\theta^{(l)}$ to be shared within the prevailing set but allow $\mu_l$ to differ across sites. For illustration purposes, we take the identity link function $h(x)=x$ for continuous outcomes or the logit link function $h(x)=1/[1+\exp(-x)]$ for binary outcomes, but our proposal can be extended to other link functions $h$.

\def\Pbbhat{\widehat{\mathbb{P}}}
\def\subSone{_{\mathcal{S}^{(l)}_1}}
\def\subStwo{_{\mathcal{S}^{(l)}_2}}
\def\subS{_{\mathcal{S}}}
\def\Ssc{\mathcal{S}}
We adopt the same quadratic dissimilarity measure  from the low-dimensional setting in Section \ref{sec: app 1}.  Compared to the low-dimensional setting, it is much more challenging to construct an accurate estimator of {$\DD_{l,k}=\|\theta^{(l)}-\theta^{(k)}\|_2^2$ in high dimensions.} Sample splitting is needed to establish the theoretical properties of the dissimilarity estimators in high dimensions. For $1\leq l\leq L$, we randomly split the index set $\{1,2,\cdots,n_l\}$ into disjoint subsets $\mathcal{S}^{(l)}_1$ and $\mathcal{S}^{(l)}_2$ with $|\mathcal{S}^{(l)}_1|=\lceil n_l/2 \rceil$ and $|\mathcal{S}^{(l)}_2|=n_l-|\mathcal{S}^{(l)}_1|,$ where $\lceil n_l/2 \rceil$ denotes the smallest integer above $n_l/2.$ For any set $\Ssc \subseteq \{1, ..., n_l\}$, let $\Pbbhat\subS\supl f(Y,X)= |\Ssc|^{-1}\sum_{i \in \Ssc} f(Y_i\supl, X_i\supl)$.

Our proposal consists of three steps: in the first step, we construct initial estimators $\{\widetilde{\mu}_l,\widetilde{\theta}^{(l)}\}_{1\leq l\leq L}$ together with the debiased estimators $\{\widehat{\beta}^{(l)}\}_{1\leq l\leq L}$ and the corresponding standard errors $\{\widehat{\sigma}_l\}_{1\leq l\leq L}$ and broadcast these initial estimators to all $L$ sites; in the second step, we estimate the error components of the plug-in estimators; in the third step, we construct the debiased estimators of $\DD_{l,k}$ for $1\leq l<k\leq L.$ 
Our method is designed to protect privacy since it does not require passing individual-level data.

\noindent {\bf Step 1: broadcasting the initial estimators.} For site $l$ with $1\leq l\leq L,$ we construct the penalized maximum likelihood estimator \citep{buhlmann2011statistics} using the data belonging to $\mathcal{S}^{(l)}_1$. Particularly, the initial estimator of $\theta^{(l)}$ is defined as 
\begin{equation}
\left\{\widetilde{\mu}_l,\widetilde{\theta}^{(l)}\right\}=\argmin_{\mu\in \R, \theta\in \R^d}\left[\Pbbhat\subSone\supl\left\{\ell_h(\mu+\theta^{\intercal} X, Y)  \right\}+\lambda \|\theta\|_1\right],
\label{eq: penalized estimator}
\end{equation}
{where $\ell_h(x, y)$ corresponds to the log-likelihood function, which takes the form of $(y-x)^2$ for a linear model with $h(x)=x$ and $\log(1+e^{x})-yx$ for a logistic model with $h(x) = e^x/(1+e^x)$.}
The tuning parameter $0 < \lambda\asymp \sqrt{{\log d}/{|\mathcal{S}^{(l)}_1|}}$ is chosen via cross-validation in practice.  

{
Since our final goal is to make an inference for  $\beta^* = g(\theta^*)$ of the prevailing model $\theta^*$, we require all sites to additionally output a debiased asymptotically normal estimator $\widehat{\beta}\supl$ of $\beta\supl = g(\theta^{(l)})$. In high dimensions, the plug-in estimator $g(\widetilde{\theta}^{(l)})$ is generally a biased estimator of $\beta\supl$ and requires bias correction. %which requires correcting the bias of the plug-in estimator. 
Such bias-correction  procedures have been widely studied in recent years \citep{zhang2014confidence,javanmard2014confidence,van2014asymptotically,chernozhukov2015post,ning2017general,cai2021statistical}, with examples including linear functionals \citep{zhu2018linear,cai2021optimal,guo2021inference}, and quadratic forms \citep{guo2021group,cai2020semisupervised}. 
We shall illustrate our method by considering the example $\beta\supl=\langle e_j,\theta\supl\rangle =\theta\supl_j$ for a specific coordinate $j$ in the high-dimensional regression model. Following \citet{zhang2014confidence,javanmard2014confidence,van2014asymptotically}, a  debiased estimator for $\beta\supl$ has been shown to satisfy the following asymptotic normality property: 
\begin{equation}
(\widehat{\beta}^{(l)}-\beta^{(l)})/{\rm SE}(\widehat{\beta}^{(l)})\cid N (0,1) 
\label{eq: debiasing}
\end{equation}
and a consistent estimator for ${\rm SE}(\widehat{\beta}^{(l)})$, denoted by  $\widehat{\rm SE}(\widehat{\beta}^{(l)})$ was also provided.
} %\tcomm{plz make sure that we do not confuse readers about the debiased estimator and the initial estimator.}

Subsequently, we broadcast $\{\widetilde{\mu}_l,\widetilde{\theta}^{(l)}, \widehat{\beta}\supl, \widehat{\rm SE}(\widehat{\beta}\supl)\}_{1\leq l\leq L}$ to all $L$ sites. 

%  We can obtain the same error decomposition \eqref{eq: error decomposition} as the low-dimensional setting, but the main difference is that the first term $2\langle \widehat{\gamma}-\gamma, \gamma\rangle$ of \eqref{eq: error decomposition} is biased due to the sparsity penalty. 

\noindent {\bf Step 2: estimating the bias components of plug-in estimators.} In Step 2, we obtain additional summary data needed for constructing debiased estimators for $\{\DD_{l,k}\}_{1\leq l<k\leq L}$. To calculate the distance metrics and assess their sampling errors in the high-dimensional setting, we need to correct the bias of the plug-in estimator $\|\widetilde{\theta}\supl - \widetilde{\theta}\supk\|_2^2$ for $1\leq l<k\leq L$. The bias correction for the distance metric estimates requires the construction of a projection for each of the pairwise distances. {To this end, for $1\leq l\leq L$, we define $\eta^{(l)}=(\mu_l,[\theta^{(l)}]^{\intercal})^{\intercal}\in \R^{d+1}$ and $\widetilde{\eta}^{(l)}=(\widetilde{\mu}_l,[\widetilde{\theta}^{(l)}]^{\intercal})^{\intercal}\in \R^{d+1}$ and $\tX^{(l)}_{i}=(1,(X^{(l)}_{i})^{\intercal})^{\intercal}$ for  $1\leq i\leq n_l.$ We fix the indexes $1\leq l<k\leq L$ and let $\widehat{\gamma}_{l,k}\coloneqq(0,[\widetilde{\theta}^{(l)}-\widetilde{\theta}^{(k)}]^{\intercal})^{\intercal}$ and $\gamma_{l,k}\coloneqq (0,[\theta\supl-\theta\supk]^{\intercal})^{\intercal}$. When it is clear from the context, we shall write $\widetilde{\gamma}$ and $\gamma$ for $\widetilde{\gamma}_{l,k}$ and $\gamma_{l,k},$ respectively.}

We analyze the error decomposition of the plug-in estimator $\|\widetilde{\gamma}\|_2^2=\|\widetilde{\theta}^{(l)}-\widetilde{\theta}^{(k)}\|_2^2.$ The main difference from the low-dimensional setting is that $\widetilde{\theta}^{(l)}$ is a biased estimator of $\theta^{(l)}$ due to the penalty term in \eqref{eq: penalized estimator}. Consequently, the plug-in estimator $\|\widetilde{\gamma}\|_2^2$ can be severely biased for $\|\gamma\|_2^2$. {To mitigate this excessive bias, we propose a bias correction procedure following the  decomposition
\begin{equation}
\begin{aligned}
%\|\widetilde{\theta}^{(l)}-\widetilde{\theta}^{(k)}\|_2^2-\|\theta^{(l)}-\theta^{(k)}\|_2^2
\|\widetilde{\gamma}\|_2^2-\|{\gamma}\|_2^2=2\langle \widetilde{\eta}^{(l)}-\eta^{(l)}, \widetilde{\gamma}\rangle-2\langle \widetilde{\eta}^{(k)}-\eta^{(k)}, \widetilde{\gamma}\rangle-\| \widetilde{\gamma}-\gamma\|_2^2.
\end{aligned}
\label{eq: error decomposition high}
\end{equation}
In contrast to the low-dimensional setting, the two terms $2\langle \widetilde{\eta}^{(l)}-\eta^{(l)}, \widetilde{\gamma}\rangle$ and $2\langle \widetilde{\eta}^{(k)}-\eta^{(k)}, \widetilde{\gamma}\rangle$ suffer from excessive bias due to $\widetilde{\eta}^{(l)}$ and $\widetilde{\eta}^{(k)}$ and do not attain the asymptotic normal limiting distribution as in the low-dimensional setting.

Following 
\citet{guo2021inference}, we perform debiasing by estimating the error components $2\langle \widetilde{\eta}^{(l)}-\eta^{(l)}, \widetilde{\gamma}\rangle$ and $2\langle \widetilde{\eta}^{(k)}-\eta^{(k)}, \widetilde{\gamma}\rangle$ using the data belonging to $\mathcal{S}^{(l)}_2$ and $\mathcal{S}^{(k)}_2.$ For $1\leq i\leq n_l$, define  
$$\epsilon^{(l)}_i=Y^{(l)}_i-h([\tX^{(l)}_{i}]^{\intercal} \eta^{(l)}), \
\widehat\epsilon^{(l)}_i=Y^{(l)}_i-h([\tX^{(l)}_{i}]^{\intercal} \widetilde{\eta}^{(l)}), \  W_i^{(l)}=1/h'([\tX^{(l)}_{i}]^{\intercal} \widetilde{\eta}^{(l)}),$$ 
where $h'(x)=dh(x)/dx$. Note that 
 \begin{equation}
\begin{aligned}
u^{\intercal}\Pbbhat\subStwo\supl( W\tX\widehat\epsilon)
\approx u^{\intercal}\Pbbhat\subStwo\supl (W\tX\epsilon)+u^{\intercal}\widehat{\Sigma}^{(l)}(\eta^{(l)}-\widetilde{\eta}^{(l)})%+u^{\intercal}, %\Pbbhat\subStwo\supl(\Delta X),
\end{aligned}
\label{eq: high-dim bias est main}
\end{equation}
with $\widehat{\Sigma}^{(l)}= \Pbbhat\subStwo\supl\{ \tX(\tX)\trans\}$. 
%We aim to identify a projection direction $u \in \R^{d+1}$ such that $u^{\intercal}\Pbbhat\subStwo\supl( W\tX\widehat\epsilon)$ approximates $\langle \widetilde{\eta}^{(l)}-\eta^{(l)}, \widetilde{\gamma}\rangle$. 
Based on \eqref{eq: high-dim bias est main}, we aim to construct the projection direction ${u}\in \R^{d+1}$ such that $\widehat{\Sigma}^{(l)}{u}\approx \widetilde{\gamma}\sublk$, which ensures that $u^{\intercal}\Pbbhat\subStwo\supl( W\tX\widehat\epsilon)$ is an accurate estimator of $\langle \widetilde{\eta}^{(l)}-\eta^{(l)}, \widetilde{\gamma}\rangle$. In particular,  we write $\widehat\gamma\sublk=(0,[\widetilde{\theta}^{(l)}-\widetilde{\theta}^{(k)}]^{\intercal})^{\intercal}$ and construct $\widehat{u}_k\supl$ as follows,
\begin{equation}
\begin{aligned}
\widehat{u}_k\supl=\;\argmin_{u\in \RR^{d}} u^{\intercal}\widehat{\Sigma}^{(l)}u \quad
\text{subject to}&\; \|\widehat{\Sigma}^{(l)}u-\widetilde{\gamma}\sublk\|_{\infty}\leq  \|\widetilde{\gamma}\sublk\|_2 \lambda \\
&\;|\widetilde{\gamma}\sublk^{\intercal}\widehat{\Sigma}^{(l)}u-\|\widetilde{\gamma}\sublk\|_2^2 |\leq \|\widetilde{\gamma}\sublk\|_2^2\lambda 
\\
&\;\max_{i\in \mathcal{S}^{(l)}_2}|u^{\intercal}\tX^{(l)}_i|\leq \|\widetilde{\gamma}\sublk\|_2^2\tau 
\end{aligned}
\label{eq: projection direction}
\end{equation}
where $\lambda\asymp \sqrt{{\log d}/{|\mathcal{S}^{(l)}_2|}}$ and $\tau \asymp \sqrt{\log |\mathcal{S}^{(l)}_2|}.$ Then we obtain the debiasing term 
\begin{equation}
\widehat{\delta}^{(l)}_{k}=\left(\widehat{u}_k\supl\right)^{\intercal}\Pbbhat\subStwo\supl (W\tX \widehat{\epsilon} ), \quad \text{for}\quad 1\leq k \leq L, \; k\neq l, 
\label{eq: error estimate}
\end{equation} 
and the corresponding variance measure 
\begin{equation}
\widehat{\rm V}^{(l)}_k=\left(\widehat{u}_k\supl\right)^{\intercal}\left[ \Pbbhat\subStwo\supl (W\tX\tX^{\intercal})\right]\widehat{u}_k\supl \quad \text{for}\quad 1\leq k \leq L, \; k\neq l.
\label{eq: var component}
\end{equation}
Similarly, we may obtain $\widehat{u}_l\supk$, $\widehat{ {\delta}}_l\supk$, and $\widehat{\rm V}_l\supk$ for estimating $\langle \widetilde{\eta}^{(k)}-\eta^{(k)}, \widetilde{\gamma}\sublk\rangle$. %The above projection direction is obtained by applying the R package \texttt{SIHR} \citep{rakshit2021sihr} with the built-in tuning parameter selection.  More discussions about the projection direction construction can be found in Section 2.3 of \citet{guo2021inference}. 
}

\noindent{\bf Step 3: Constructing the debiased estimators.}
In the last step, the center combines all summary statistics from all sites to construct the following bias-corrected estimator of $\DD_{l,k}$, 
\begin{equation}
\widehat{\DD}_{l,k}=\max\left\{\|\widetilde{\theta}^{(l)}-\widetilde{\theta}^{(k)}\|_2^2+2\widehat{\delta}^{(l)}_{k}-2\widehat{\delta}^{(k)}_{l}, 0\right\}, 
\label{eq: dist high-dim}
\end{equation}
and estimate its standard error by 
\begin{equation}
\widehat{\rm SE}(\widehat{\DD}_{l,k})=\sqrt{4\widehat{\rm V}^{(l)}_{k}+4\widehat{\rm V}^{(k)}_{l}+\frac{1}{\min\{n_l,n_k\}}}.
\label{eq: SE high-dim}
\end{equation}
We have summarized our proposed method in Algorithm \ref{algo: high-dim} in the supplement.
In Theorem \ref{thm: high-dim dissimilarity} of the supplement, we show that $\widehat{\DD}_{l,k}$ in \eqref{eq: dist high-dim}  and ${\widehat{\rm SE}(\widehat{\DD}_{l,k})}$ in \eqref{eq: SE high-dim} satisfy \eqref{eq: key assumption} for $1\leq l<k\leq L$. %Our above construction does not require access to the individual-level data. Only the initial estimators and the estimators of the error components need to be broadcasted or passed to the center. 
With $\{\widehat{\beta}^{(l)},\widehat{\rm SE}(\widehat{\beta}^{(l)}),\widehat{\DD}_{l,k}, \widehat{\rm SE}(\widehat{\DD}_{l,k})\}_{1\leq l<k\leq L}$ defined in \eqref{eq: dist high-dim} and  \eqref{eq: SE high-dim}, we apply RIFL to construct confidence intervals for $\beta^*$.

%%%%%%%%%%%%%%%%%%%%%%%%%%%%%%%%%%%%%%%%%%%%%%%%%%%%%%%%%%%%
\subsection{Inference for average treatment effects}
\label{sec: app 3}
%%%%%%%%%%%%%%%%%%%%%%%%%%%%%%%%%%%%%%%%%%%%%%%%%%%%%%%%%%%%
Multi-center causal modeling is of great value for generating reliable real-world evidence of approved drugs on a new target population, which can be much broader than those patients recruited to clinical trials. Data from a single site may not be sufficient to reliably estimate the causal effect due to limited sample size and may cause potential bias due to insufficient confounding adjustment. We aim to use RIFL to estimate an average treatment effect for a target population with a specified covariate distribution, $f_G(\cdot)$, based on $L$ source sites. Specifically, for the site $1\leq l\leq L$, we observe the data $\{X^{(l)}_{i}, A^{(l)}_i, Y^{(l)}_i\}_{1\leq i\leq n_l},$ where $X^{(l)}_{i} \in \R^p$ denotes the $p$-dimensional baseline covariate vector, $A^{(l)}_{i} \in \{0,1\}$ denotes whether the subject receives an experimental treatment or control and $Y^{(l)}_{i}\in \R$ denotes the outcome. Moreover, we observe $\{X_i^\mathcal{T}\}_{1\leq i \leq N}$ drawn from $f_G(\cdot)$, the covariate distribution in the target population, for some large $N$. Let $f_l(\cdot)$ denote the density function of the covariate distribution in the $l$-th source site. We use $Y^{(l,a)}$ to denote the potential outcome of patients under treatment $A=a$ in site $l$. We define the ATE of the target population projected from the $l$-th source site as 
$$\theta^{(l)}= \int \left\{\E(Y^{(l,1)} \mid X\supl = x) - \E(Y^{(l,0)}\mid X\supl = x) \right\} f_G(x) dx .$$
We use $\beta^*$ to denote the majority of the ATEs $\{\theta^{(l)}\}_{1\leq l\leq L}$ and define $\mathcal{V}(\beta^*)=\{1\leq l\leq L: \theta^{(l)}=\beta^*\}$.
We aim to make inferences for $\beta^*$ using RIFL. 

%\Zijian{Xiudi (and Tianxi), this is the only place that we use bold letters? shall we just use the regular letters as the remaining of the paper?} \Xiudi{I think all bold letters refer to the parameters in the nuisance model, which only appear for the TATE example. I have updated section 6.3 to use bold letters when appropriate to match 5.3.}
To estimate $\beta^*$, we first obtain a standard augmented doubly robust estimator \citep{bang2005doubly,tsiatis2006semiparametric,kang2007demystifying,tao2019doubly} for $\theta\supl$ using data from the $l$th site, which requires the propensity score and outcome model 
\begin{equation*}
\PP(A^{(l)}_i=a \mid X^{(l)}_i)   = \pi_{l} (a,X^{(l)}_i; \bga^{(l)}), 
 \quad \EE(Y^{(l)}_i \mid A^{(l)}_i=a, X^{(l)}_i)  = m(a,X^{(l)}_i; \bgb\supl_{a}),
 %\label{model-PS} 
\end{equation*}
% and an outcome model 
% \begin{equation}
%\EE(Y^{(l)}_i \mid A^{(l)}_i=a, X^{(l)}_i)  = m(a,X^{(l)}_i; \bgb\supl_{a}) . \label{model-outcome}
%\end{equation}  
where $\bga\supl$ and $\{\bgb_0\supl, \bgb_1\supl\}$ are the unknown model parameters and $\pi_l(\cdot)$ and $m(\cdot)$ are specified link functions. In order to infer the target site ATE, one also needs to specify a density ratio model, $f_G(X_i\supl)/f_l(X_i^{(l)}) = \omega_l(X_i^{(l)}; \bge^{(l)})$, which accounts for the covariate shift between the source and the target populations. An example of $\omega_l(X_i^{(l)}; \bge^{(l)})$ can be chosen as $\exp\{\Psi(X_i\supl)\trans\bge\supl\}$, where $\Psi(\cdot)$ denotes a vector of specified basis functions. 
%Notably, under the multi-causal framework, the outcome regression, propensity score, and density ratio models are nuisance models. The ATE $\theta^{(l)}\in \R$ is the parameter of interest.

Let $\widehat{\bga}^{(l)}$, $\widehat{\bgb}^{(l)}_{a}$, and $\widehat{\bge}^{(l)}$ denote consistent estimators of $\bga^{(l)}$, $\bgb_{a}^{(l)}$, and $\bge^{(l)}$, respectively. For example, these finite-dimensional parameters can be estimated locally in each site $l$ by fitting generalized linear models.
For site $l$ with $1\leq l\leq L$, we compute the following doubly robust estimator $\widehat{\theta}^{(l)}=\widehat{M}^{(l)} + \widehat{\delta}^{(l)}$ with  
$\widehat{M}^{(l)} = \frac{1}{N} \sum_{i=1}^{N}
    \left\{m(1,X_i^{\mathcal{T}}; \widehat{\bgb}^{(l)}_{1})  - m(0,X_i^{\mathcal{T}}; \widehat{\bgb}^{(l)}_{0})\right\}$ and 
% %$\widehat{M}^{(l)} = \frac{1}{n_l} \sum_{i=1}^{n_l}
%     \left\{m(1,X_i^{(l)}; \widehat{\bgb}^{(l)}_{1})  - m(0,X_i^{(l)}; \widehat{\bgb}^{(l)}_{0})\right\}$ and 
\begin{equation*}
% \begin{aligned}
%     \widehat{\delta}^{(l)} &= \frac{1}{n_l}\sum_{i=1}^{n_l} \omega_l(X_i^{(l)};\widehat{\bge}^{(l)})
%     \left\{\frac{I(A_i\supl = 1)}{\pi_{l} (1,X^{(l)}_i; \widehat{\bga}^{(l)})} - \frac{I(A_i\supl = 0)}{\pi_{l} (0,X^{(l)}_i; \widehat{\bga}^{(l)})} \right\} \{Y^{(l)}_i - m(A_i\supl,X^{(l)}_i; \widehat{\bgb}^{(l)}_{a})\}. \label{aug-target}
%     \end{aligned}
\begin{aligned}
    \widehat{\delta}^{(l)} &= \frac{1}{n_l}\sum_{i=1}^{n_l} \omega_l(X_i^{(l)};\widehat{\bge}^{(l)})
    \left\{\sum_{a=0}^1\frac{(-1)^{a+1}\mathbf{1}(A_i\supl = a)}{\pi_{l} (a,X^{(l)}_i; \widehat{\bga}^{(l)})} \{Y^{(l)}_i - m(A_i\supl,X^{(l)}_i; \widehat{\bgb}^{(l)}_{a})\} \right\} . \label{aug-target}
    \end{aligned}
 \end{equation*}
It has been shown that the doubly robust estimator  $\widehat{\theta}^{(l)}$ is asymptotically normal with $n_l^{1/2}(\widehat{\theta}\supl - \theta\supl) \cid N(0, V\supl)$ under the corresponding regularity conditions when either the outcome model is correctly specified or both the density ratio model and the propensity score model are correctly specified.
%the density ratio model is correctly specified and either the outcome model or the propensity score model is correctly specified \citep{bang2005doubly,tsiatis2006semiparametric,kang2007demystifying,tao2019doubly}. %; see Theorem 1 and Corollary S1 in \citet{han2021federated} for an example.
%\Zijian{Larry, I agree with Tianxi's previous comments. Can we give a formula for the variance term and can we give more references about the doubly robust estimator?}
We estimate the variance $V\supl$ by $\widehat{V}^{(l)}$ using influence function expansions. We provide the influence functions for calculating $\widehat{V}^{(l)}$ in Appendix~\ref{sec:TATEinfluence} of the supplementary materials for a specific set of model choices. 

\def\SE{\mbox{SE}}
Since $\theta\supl$ is one-dimensional, we only need to estimate the dissimilarity between sites by $
\widehat{\LL}_{l,k} = \widehat{\theta}^{(l)} - \widehat{\theta}^{(k)},
$ whose standard error can be estimated as $\widehat{\SE}(\widehat{\LL}_{l,k}) = \sqrt{\widehat{V}\supl+\widehat{V}\supk}$,  for $1\leq l,k\leq L.$ 
As discussed in Remark \ref{rem: univariate}, the RIFL estimator can be simplified and constructed based on $\{(\widehat{\LL}_{l,k}, \widehat{\SE}(\widehat{\LL}_{l,k})), 1 \le l<k\leq L\}$ along with $\{(\widehat{\theta}\supl, \widehat{V}\supl), l = 1, ..., L\}$. 

%For $1\leq l\leq L$ and $1\leq k\leq L$, we can define the dissimilarity measure as
%with the site-specific ATEs $\theta^{(l)}$ and $\theta^{(k)}$ defined in \eqref{eq: site ATE}. Thus $\DD_{l,k}$ measures the difference in the ATEs for source $l$ and source $k$. We can estimate $\DD_{l,k}$ by $\widehat{\DD}_{l,k} = \widetilde{\theta}^{(l)} - \widetilde{\theta}^{(k)}.$ 

%We estimate the standard error of $\widehat{\DD}_{l,k}$ by calculating the variance of the influence functions of the site-specific ATEs. \Zijian{Larry, can you give the exact formula for variance?} In the propensity score regression model, suppose that $\widehat{\bga}^{(l)} - \bga^{*(l)} = o_p(1)$. In the outcome regression model, suppose also that $\widehat{\bgb_a}^{(l)} - \bgb_a^{*(l)} = o_p(1)$. Then the influence function for $\widetilde{\mu}_{a,l}$ can be written as the summand in the below expression

%\begin{align*}
%    \sqrt{n_l}\left(\widetilde{\mu}_{a,l} - \mu_{a,l}\right) &= \frac{1}{\sqrt{n_l}}\sum_{i=1}^{n_l} \left[\frac{I(A_i=a)Y_i}{\pi_l(a,X_i,\widehat{\bga}^{(l)})} - \left(\frac{I(A_i=a)}{\pi_l(a,X_i,\widehat{\bga}^{(l)})} - 1 \right) m(l,X_i,\widehat{\bgb}_a^{(l)}) - \mu_{a,l} \right]
%\end{align*}

%%%%%%%%%%%%%%%%%%%%%%%%%%%%%%%%%%%%%%%%%%%%%%%%%%%%%%%%%%%%
\section{Simulation Studies}
\label{sec: simulation}
%%%%%%%%%%%%%%%%%%%%%%%%%%%%%%%%%%%%%%%%%%%%%%%%%%%%%%%%%%%%

We showcase RIFL via the three inference problems discussed in sections \ref{sec: app 1},  \ref{sec: app 2}, and \ref{sec: app 3}. The code for implementing RIFL for each of the three problems can be accessed  at \url{https://github.com/celehs/RIFL}.
%We shall implement the general RIFL algorithm and provide the tuning parameter selection in Section \ref{sec: tuning selection} in the supplementary material. \tcomm{put tuning in the supplementary since the flow is not the best here} 
%%
%%
We implement three other estimators for comparison: the confidence interval based on the median estimator, the m-out-of-n bootstrap (MNB) confidence interval, and the voting with maximum clique (VMC) estimator in \eqref{eq: post CI}. We run $500$ simulations for each scenario to examine the empirical coverage and average length of the $95\%$ CIs. We consider 5 different levels of separation between the prevailing set and the remaining sites to examine how the degree of separation impacts the performance of different methods. For simplicity, we set the sample sizes to be the same in all sites, $n_l=n$ for $1\leq l\leq L$, and consider $n=500$, $1000$, and $2000$. Throughout, we let $L = 10$ and focus on the case with $|\mathcal{V}(\theta^*)| = 6$.

%with $B = 500$ since we assume $\widehat\beta\supl$ to be asymptotically normal with estimated standard error $\widehat{\sigma}_l$ under \ref{eq: limiting distribution}. 
We construct the median estimator as the median of $\{\widehat{\beta}\supl\}_{1\leq l\leq L}$ and estimate its standard error by parametric bootstrap. 
For the m-out-of-n bootstrap (MNB), we compute the point estimator for the original data, say $\widehat{\beta}^*$, by VMC. For each site, we subsample $m$ observations with replacement from the  $n$ observations and repeat the process 500 times to construct CIs. According to \citet{bickel2008choice}, the subsample size $m$ should be small relative to $n$, and we set $m = n^\upsilon$, with $\upsilon = 0.8$. For $1\leq j\leq 500$, we compute a point estimator $\widehat{\beta}^{m,j}$ by applying the VMC to the $j$-th resampled data. 
%Since we subsample $500$ times, we obtain $\{\widehat{\beta}^{m,j}\}_{1 \leq j \leq 500}$.
We define $L_n(t) = \frac{1}{500}\sum_{j=1}^{500} \mathbf{1}\{\sqrt{m}(\widehat{\beta}^{m,j} - \widehat{\beta}^*) \leq t\}$ with $\mathbf{1}$ denoting the indicator function. The MNB CI is given by \begin{equation}
\left[\widehat{\beta}^*-\frac{\hat{t}_{1-\alpha/2}}{\sqrt{n}}, \widehat{\beta}^* - \frac{\hat{t}_{\alpha/2}}{\sqrt{n}}\right],
\label{eq: mnb}
\end{equation}
where $\hat{t}_{\alpha/2}$ and $\hat{t}_{1-\alpha/2}$ are the smallest $t$ such that $L_n(t) \geq \alpha/2$ and $L_n(t) \geq 1-\alpha/2$, respectively.  %As explained in Section \ref{sec: median}, the CI based on the median estimator does not achieve the desired coverage due to the bias of the median estimator.

{We additionally include an oracle bias-aware (OBA) confidence interval as a benchmark to provide a fair comparison of methods that account for selection variability. For the estimator $\betahatstar$ defined in \eqref{eq: post-selection est}, we assume $(\betahatstar-\beta^*)/{\rm SE}(\betahatstar)\cid N(b,1)$ with ${\rm SE}(\betahatstar)$ denoting the standard error of $\betahatstar$ and $b$ denoting the corresponding bias.  Following (7) in \citet{armstrong2020bias}, we use the oracle knowledge of $|\E\betahatstar-\beta^*|$ and form the oracle bias-aware CI as 
\begin{equation}
\left(\betahatstar-\chi,\betahatstar+\chi\right) \quad \text{with}\quad \chi=\widehat{\rm SE}(\betahatstar)\cdot \sqrt{{\rm cv}_{\alpha}\left(|\E\betahatstar-\beta^*|^2/\widehat{\rm SE}^2(\betahatstar)\right)},
\label{eq: bias aware}
\end{equation}
where ${\rm cv}_{\alpha}(B^2)$ is the $1-\alpha$ quantile of the $\chi^2$ distribution with 1 degree of freedom and non-centrality parameter $B^2.$ We implement 500 simulations and now specify an oracle method of constructing the estimated standard error $\widehat{\rm SE}(\betahatstar)$ for the $j$-th simulation with $1\leq j\leq 500.$ For the $j$-th simulation with $1\leq j\leq 500,$ we obtain the point estimator $\widehat{\beta}^{*,j}$ as defined in \eqref{eq: post-selection est} together with ${\rm SE}_{j}=\frac{1}{\sqrt{\sum_{l\in \widehat{\mathcal{V}}}1/{\widehat{\sigma}_{l}^2}}}.$ With these 500 estimators, we compute the empirical standard error of $\{\widehat{\beta}^{*,j}\}_{1\leq j\leq 500}$ and define it as the ESE. We then compute the ratio between the ESE and $\sum_{j=1}^{500}{\rm SE}_{j}/500$ and define $\widehat{\rm SE}(\betahatstar)$ in \eqref{eq: bias aware} for the $j$-th simulation as 
$\widehat{\rm SE}(\betahatstar)=\frac{\rm ESE}{\sum_{j=1}^{500}{\rm SE}_{j}/500}\cdot {\rm SE}_{j}.$} We argue that the OBA CI serves as a better benchmark than the oracle CI in \eqref{eq: ora CI} assuming knowledge of the majority group since it accounts for the post-selection error. 

\vspace{-3.5mm}

\subsection{Multi-source low-dimensional prediction simulation}
\label{sims: app1}
We first examine the performance of RIFL for the low-dimensional multi-source prediction problem described in subsection \ref{sec: app 1} with $d=10$ covariates and $L = 10$ sites. The covariates $\{X_i\supl\}_{1\leq i\leq n}$ are i.i.d. generated as multivariate normals with zero mean and covariance $\Sigma \in \mathbb{R}^{d \times d}$ where $\Sigma_{jk} = 0.6^{|j-l|}$ for $1 \leq j,k\leq d$. For the $l$-th site with $1\leq l\leq L$, we generate the binary outcome $Y_i^{(l)}$ as
$
    \PP(Y_i^{(l)}=1|X_i^{(l)}) = 1/[1+\exp(- \mu_l -[X_i^{(l)}]^\top\theta^{(l)})]$, for $1\leq i\leq n_l,$
with site-specific intercept $\mu_l$. We set $(\mu_1,\ldots,\mu_{10})$ to $(0.05,-0.05,0.1,-0.1,0.05,-0.05,0.1,-0.1,0,0)$. For the site in the prevailing set, we set $\theta^{(l)} = \theta^* =  (0.5,0.5,0.5,0.5,0.5,0.1,0.1,0.1,0,0)$ for $l \in \{1,\ldots, 6\}$. For $\theta^{(7)}, \theta^{(8)}, \theta^{(9)}$ and $\theta^{(10)}$, their last 5 coefficents are the same as $\theta^*$  but their first five coefficients are changed to $0.5-0.3a$, $0.5-0.2a$, $0.5-0.1a$ and $0.5+0.1a$, respectively, where $a \in \{1,2,3,4,5\}$ controls the separation between the majority sites and non-majority sites. %When there are 8 majority sites, we set the first five coefficients of $\theta^{(9)}$ and $\theta^{(10)}$ to $0.5-0.3a$ and $0.5-0.1a$, respectively, and the remaining five coefficients are the same as those of $\theta^*$. We again increase $a$ from 1 to 5. 
Define $\beta\supl = \theta^{(l)}_1$, and our inference target is $\beta^* = \theta^*_1$. The site-specific estimators $\{\widehat\theta^{(l)}\}_{1\leq l\leq L}$ are obtained via logistic regression. In Figure~\ref{fig:parametric6}, we present the empirical coverage and average length of the 95\% CIs over 500 simulation replications based on various methods.

\vspace{-3mm}
\begin{figure}[H]
    \centering
    \includegraphics[width=0.875\textwidth]{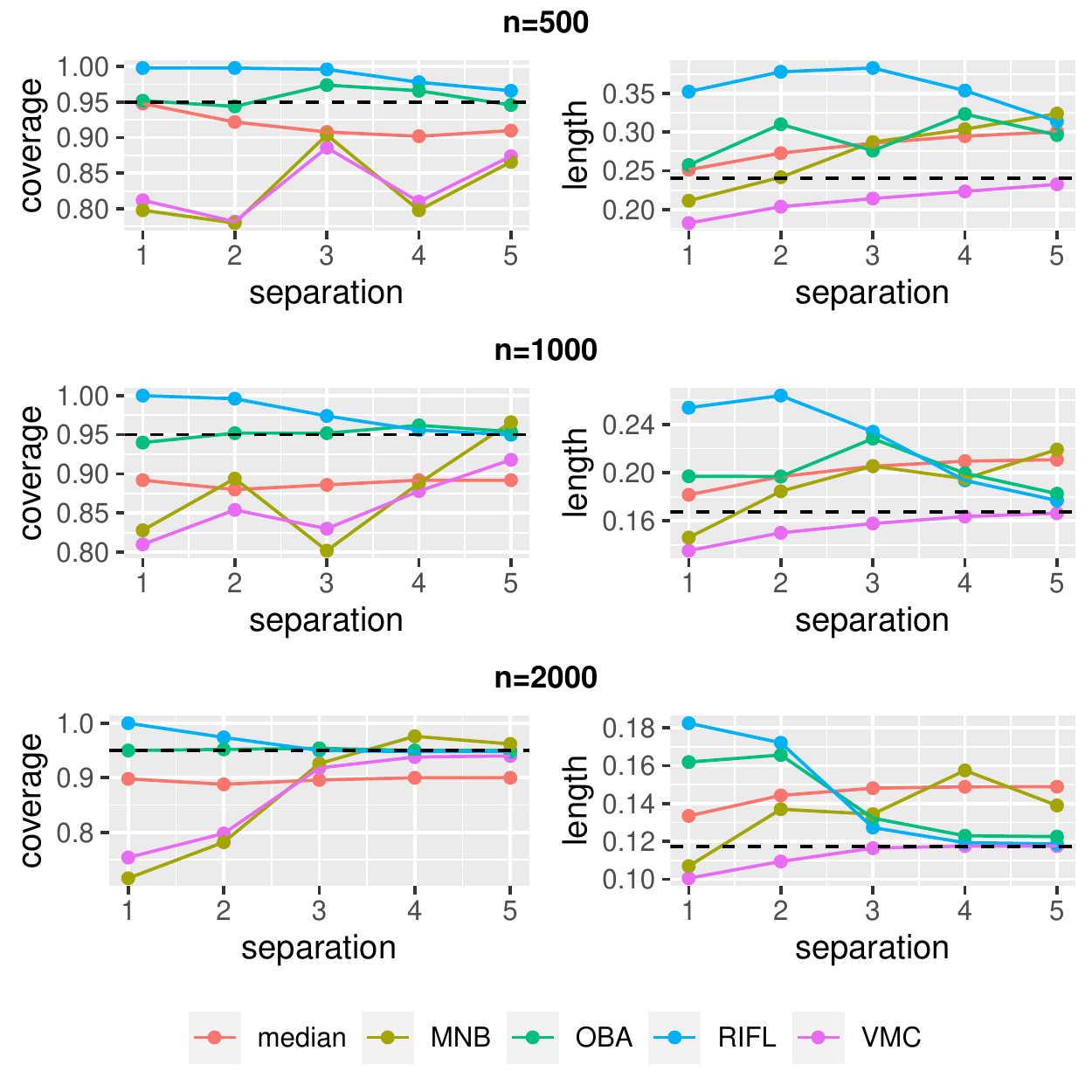}
    \vspace{-5mm}
    \caption{Low-dimensional prediction: coverage and length of $95\%$ CIs for $\beta^* = \theta^*_1$ with 6 majority sites and varying separation levels where $1$ is the lowest (hardest to detect) and $5$ is the highest (easiest to detect). ``median" stands for the CI based on the median estimator, ``MNB" stands for the m-out-of-n bootstrap CI in \eqref{eq: mnb}, ``VMC" stands for the voting with maximum clique estimator and its associated CI in \eqref{eq: post CI}, ``OBA" stands for the oracle bias aware CI in \eqref{eq: bias aware}, and ``RIFL" stands for our proposed CI in \eqref{eq: CI union}. Results are based on 500 simulation replications. Dash lines in the left panels correspond to a nominal coverage level of 0.95; in the right panels correspond to the width of an oracle CI knowing the prevailing set.}
    \label{fig:parametric6}
\end{figure}

 As reported in Figure \ref{fig:parametric6}, the RIFL CIs achieve nominal coverage across all separation levels and have an average length nearly as short as the OBA when the separation level is high. The CI of the VMC estimator shows below nominal coverage when the separation level is low or moderate. Only when the separation level is high, and the sample size is large does the VMC estimator have close to nominal coverage. This happens due to the error in separating the majority group and the remaining sites. As explained in Section \ref{sec: median}, the CI based on the median estimator has coverage below the nominal level due to its bias. In this case, as the parameter values in the non-majority sites fall on both sides of $\beta^*$, the bias is moderate, and the coverage generally only drops to $90\%$. The CIs based on m-out-of-n bootstrap tend to undercover with low or moderate separation levels. This observation matches the discussion in \citet{andrews2000inconsistency, guo2021inference}; the MNB may not provide valid inference in non-regular settings. Recall that we chose $m = n^\upsilon$ with $\upsilon = 0.8$. In Appendix~\ref{sec: additional sim} of the supplementary materials, we present the results of MNB based on various choices of $\upsilon$. Our results indicate that for the range of values of $\upsilon$ we explored, the MNB method tends to undercover when the separation level is low.

% \begin{figure}[H]
%     \centering
%     \includegraphics[width=0.88\textwidth]{graphics/lowD_coverage.png}
%     \caption{Low-dimensional prediction: Coverage of $95\%$ CIs of $\beta^* = \theta_1$ with varying separation levels where $1$ $(a=0.1)$ is the lowest (hardest to detect) and $5$ $(a=0.5)$ is the highest (easiest to detect).}
%     \label{fig:lowD_coverage}
% \end{figure}

% \begin{figure}[H]
%     \centering
%     \includegraphics[width=0.88\textwidth]{graphics/lowD_length.png}
%     \caption{Low-dimensional prediction: Average length of $95\%$ CIs of $\beta^* = \theta_1$ with varying separation levels where $1$ $(a=0.1)$ is the lowest (hardest to detect) and $5$ $(a=0.5)$ is the highest (easiest to detect).}
%     \label{fig:lowD_length}
% \end{figure}

\vspace{-5mm}

\subsection{Multi-source high-dimensional prediction simulation}\label{sims: app 2}
We next examine the performance of RIFL for the high-dimensional multi-source prediction problem described in subsection \ref{sec: app 2} with a continuous outcome and $d=500$ covariates. For $1\leq l\leq L,$ the covariates $\{X_{i}\}_{1\leq i\leq n_l}$ are i.i.d. generated from a multivariate normal distribution with mean $\boldsymbol{0}$ and covariance $\Sigma/2$, where $\Sigma_{j,k} = 0.6^{|j-k|}$ for $1\leq j,k \leq d$. For the $l$-th source site, we generate the outcome $Y^{(l)}$ according to the following model
$
    Y_i^{(l)} = \mu_l + [X_i^{(l)}]^\top \theta^{(l)} + \varepsilon_i^{(l)},$ for $1\leq i\leq n_l,$
with $\varepsilon_i^{(l)} \sim N(0, 1)$.
We let $\mu_l = 0.05$ for $l \in \{1,2,3,7,9\}$ and $0$ otherwise, and set $\theta^{(l)} = \theta^*$ for $l \in \{1,\ldots,6\}$ with $\theta^*_j = (0.1j - 0.6)\mathbf{1}(1 \le j \le 11)$ to represent sparse signal and varying signal strength. For the non-majority sites, we set $\theta_j^{(7)} = \theta_j^{(8)} = \theta_j^* + 0.2 + 0.05a$ and $\theta_j^{(9)} =\theta_j^{(10)} = \theta_j^* + 0.15 + 0.05a$ for $6 \leq j \leq 11$. We define $\beta\supl = \theta_{11}^{(l)}$ and focus on inference for $\beta^* = \theta_{11}^*$ in the following. %\tcomm{would people criticize us for focusing on the signal w highest strength?} 
{Additional results for $\theta_8^*$ are given in Appendix~\ref{sec: additional sim} in the supplementary materials.} We vary $a$ in $\{1,2,3,4,5\}$ to represent varying levels of separation between the majority and non-majority sites. Due to the high computation cost, we do not implement the m-out-of-n bootstrap method.

In Figure~\ref{fig:high6}, we present the empirical coverage and average length of the CIs from different methods. The CI coverage based on the median estimator is even lower compared to the previous example in Section~\ref{sims: app1}, suffering from more severe bias since the parameter values in the non-majority sites are biased in the same direction. More specifically, when there are 6 majority sites, the median estimator corresponds to the average of the largest and second-largest site-specific estimators from the majority sites; see more discussions in Section \ref{sec: median}.

When $n=500$, the VMC CI has consistently low coverage. This is because, with a limited sample size, we do not have enough statistical power to detect even the largest separation level we considered (corresponding to $a=5$.) As a result, some non-majority sites were wrongly included in the estimated prevailing set. When the sample size increases to $n = 1000$ and $2000$, the VMC CI achieves nominal coverage when the separation level is sufficiently large but still undercovers when the separation level is low. The OBA and the RIFL CIs achieve the desired coverage across all settings. However, when $n=500$, the OBA interval gets longer as the separation level grows. As we discussed earlier, some non-majority sites were included in the estimated prevailing set regardless of the separation level due to the limited sample size. In fact, the bias of the VMC estimator resulting from these wrongly selected sites increases as the separation level increases since the non-majority sites are further away from the majority sites. The OBA CI accounts for this increasing bias and becomes longer as the separation level increases. Moreover, when $n=1000$, the OBA interval is much longer than the RIFL interval for small to moderate separation levels. This is again due to the OBA interval correcting for the bias of the VMC estimator, which can be as large as the empirical standard error of the VMC estimator. Across all settings, the RIFL CI has a length comparable to that of the OBA and is often shorter. 
%In the 6 majority sites, we set $\theta_j^{\supl} = 0.2j$ for $1 \le j \le 5$, whereas in the 4 non-majority sites, we set $\theta_j^{(l)}$ to be $0.2j - a$, $0.2+a$, $0.2+2a$, and $0.2+3a$, where $a \in \{0.1,0.2,0.3,0.4,0.5\}$ to represent varying separation levels for $1 \le j \le 5$. 
%We implement LASSO-penalized logistic regression models to obtain the initial estimators. 
%In Figures~ \ref{fig:hd_coverage} and \ref{fig:hd_length}, we show results for the empirical coverage and average length, respectively, of the $95\%$ CIs for $\beta^* = \theta_1$.

\begin{figure}[H]
    \centering
    \includegraphics[width=0.875\textwidth]{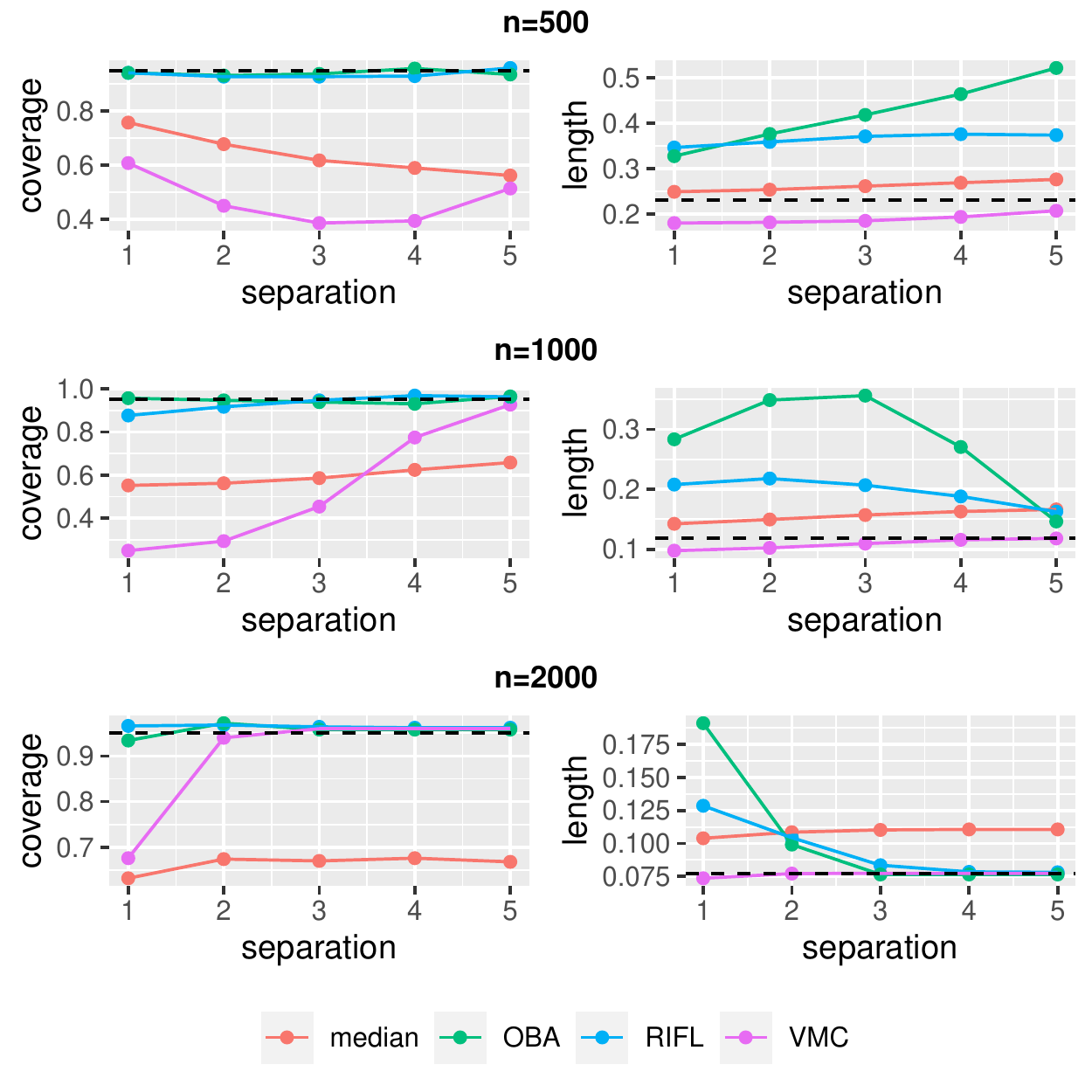}
    \vspace{-5mm}
    \caption{High-dimensional prediction: coverage and length of $95\%$ CIs for $\beta^* = \theta^*_{11}$ with 6 majority sites and varying separation levels where $1$ is the lowest (hardest to detect) and $5$ is the highest (easiest to detect). ``median" stands for the CI based on the median estimator, ``VMC" stands for the voting with maximum clique estimator and its associated CI in \eqref{eq: post CI}, ``OBA" stands for the oracle bias aware CI in \eqref{eq: bias aware}, and ``RIFL" stands for our proposed CI in \eqref{eq: CI union}. Results are based on 500 simulation replications. Dash lines in the left panels correspond to a nominal coverage level of 0.95; in the right panels correspond to the width of an oracle CI knowing the prevailing set.}
    \label{fig:high6}
\end{figure}

% \begin{figure}[H]
%     \centering
%     \includegraphics[width=0.875\textwidth]{graphics/highD_coverage.png}
%     \caption{High-dimensional prediction: Coverage of $95\%$ CIs of $\beta^* = \theta_1$ with varying separation levels where $1$ $(a=0.1)$ is the lowest (hardest to detect) and $5$ $(a=0.5)$ is the highest (easiest to detect).}
%     \label{fig:hd_coverage}
% \end{figure}

% \begin{figure}[H]
%     \centering
%     \includegraphics[width=0.875\textwidth]{graphics/highD_length.png}
%     \caption{High-dimensional prediction: Avereage length of $95\%$ CIs of $\beta^* = \theta_1$ with varying separation levels where $1$ $(a=0.1)$ is the lowest (hardest to detect) and $5$ $(a=0.5)$ is the highest (easiest to detect).}
%     \label{fig:hd_length}
% \end{figure}

%%%%%%%%%%%%%%%%%%%%%%%%%%%%%%%%%%%%%%%%%%%%%%%%%%%%%%%%%%%%

\subsection{Multi-source causal simulation}\label{sims: app 3}
We evaluate the performance of RIFL for the multi-source causal modeling problem where the interest lies in making inferences for the ATE of the target population, as described in subsection \ref{sec: app 3}. %We again consider $L=10$ sites and $15$ settings, where we vary the separation levels between the majority and non-majority sites (5 levels), and the sample size $n_l=n$ (500, 1000 or 2000), as described below. Again, we focus on the case where $|\VV| = 6$.
We consider a 10-dimensional covariate vector, that is, $p = 10$. For $l \in \{1,\ldots, L\}$, the covariates $\{X_i^{(l)}\}_{1\leq i\leq n_l}$ are i.i.d. generated following a multivariate normal distribution with mean $\mu_X^{(l)}$ and covariance matrix $\Sigma$ where $\Sigma_{jk} = 0.6^{|j-k|}$ for $1 \leq j,k \leq p$. For $l \in \{1,2,3,7,9\}$, we set $\mu_X^{(l)}$ to $\boldsymbol{0}$; and for $l \in \{4,5,6,8,10\}$, we set $\mu_X^{(l)}$ to $(0.5,0.5,0,0,\ldots,0)$. For $1\leq i \leq n_l$, given $X_i^{(l)}$, the treatment assignment is generated according to the conditional distribution $A_i^{(l)}|X_i^{(l)} \sim \textnormal{Bernoulli}\{\textnormal{expit}(\alpha_1^{(l)}X^{(l)}_{i1} + \alpha_2^{(l)}X^{(l)}_{i2} + \alpha_{12}^{(l)}X^{(l)}_{i1}X^{(l)}_{i2} )\}$, with $\alpha_1^{(l)} = 0.5$, $\alpha_2^{(l)} = -0.5$ and $\alpha_{12}^{(l)} = 0.1$. The outcome for the $l$-th site is generated according to
$Y_i^{(l)} = \mu_l + [X_i^{(l)}]^\top \zeta^{(l)} + \beta\supl A_i\supl+ \varepsilon_i\supl,$ with $\varepsilon_i\supl \sim N(0, 1)$ for $1 \leq i \leq n_l.$
Here, the regression coefficient $\zeta^{(l)}$ is set to $(0.5,0.5,0.5,0.5,0.5,0.1,0.1,0.1,0,0)$ across all sites and the site-specific intercept $(\mu_1,\mu_2,\ldots,\mu_{10})$ is set to $(0.05,-0.05,0.1,-0.1,0.05,-0.05,0.1,-0.1,0,0)$. The coefficient $\beta\supl$ measures the treatment effect. For the majority sites, that is, $l \in \{1,\ldots,6\}$, $\beta\supl = -1$. We set $\beta^{(7)} = \beta^{(8)} = -1 -0.2a$ and $\beta^{(9)} = \beta^{(10)} = -1 -0.1a$, where $a \in \{1,2,3,4,5\}$ controls the degree of separation. %When there are 8 majority sites, we set $\beta_9 = -1 -0.2a$, $\beta_{10} = -1 -0.1a$ and again vary $a$ from 1 to 5. 
The covariate in the target population is generated from multivariate normal with mean $\boldsymbol{0}$ and covariance matrix $\Sigma$. We generated $N = 10,000$ realizations of $X$ from the target population when evaluating the ATE.

%\Zijian{Xiudi, this section might need a double check; see my questions in the following.}
To estimate the outcome regressions, we fit a linear model with the main effects of each covariate within the treatment arm in each site. The propensity score is estimated via a logistic model with main effects of $X_{\cdot,1}^{(l)}$ and $X_{\cdot,2}^{(l)}$ only within each site. Finally, we assume the following model for the density ratio, $\omega_l(X_i^{(l)};\bge^{(l)}) = \exp([\bge^{(l)}]^\top \widetilde{X}_i^{(l)})$, where $\widetilde{X}_i^{(l)} = (1,(X_i^{(l)})^\top)^\top$. We estimate $\bge^{(l)}$ by solving the estimating equation $\sum_{i=1}^n \exp([\bge^{(l)}]^\top \widetilde{X}_i^{(l)})\widetilde{X}_i^{(l)}/n = \sum_{j=1}^N \widetilde{X}_j^{\mathcal{T}}/N$, where $\widetilde{X}_j^{\mathcal{T}} = (1,(X_j^{\mathcal{T}})^\top)^\top$ and $X_j^{\mathcal{T}}$ denotes the $j$-th observation in the target dataset, for $j \in \{1,\ldots,N\}$. It is worth noting that, although the propensity score model is misspecified, the double robustness property guarantees that the resulting estimator is still consistent and asymptotically normal given that the outcome regression model is correctly specified.
%\tcomm{Xiudi plz check above to make sure that the descriptions are accurate: we generate $Y$ from linear regression? if so we will not fit logistc reg?}

Plots of empirical coverage and average length of the CIs produced via various methods based on 500 simulation replications are provided in Figure~\ref{fig:TATE6} for 6 majority sites. We observe similar patterns as in the low-dimensional prediction example in Section~\ref{sims: app1}. In particular, the coverage of VMC CI approaches the nominal level as the level of separation and sample size increase, but is generally below the nominal level when separation is small to moderate. The MNB CI also undercovers when separation is not sufficiently large. Both RIFL and OBA achieve the nominal coverage across all settings and their length are generally comparable. In this case, the coverage of the CI based on the median estimator is low because the average treatment effects in the non-majority sites are all smaller than that in the majority sites, and therefore the median estimator suffers from severe bias. %\tcomm{plz comment on how the patterns change over $n$ and degree of separation} \Xiudi{Tianxi, this paragraph has been updated}

\subsection{Robustness to different choices of tuning parameters}% and the majority rule}
\label{sec: tuning parameter}
%%%%%%%%%%%%%%%%%%%%%%%%%%%%%%%%%%%%%%%%%%%%%%%%%%%%%%%%%%%%%%%%%%%%%%%%%%%%%%%%%%%%%%%%%%%%%%%%%%%%%%%%%%%%%%%%%%%%%%

We empirically assess the sensitivity of coverage and precision of the RIFL CIs to the choices of the tuning parameter $\rho(M)$ and the resampling size $M.$ We continue with the low-dimensional multi-source prediction problem in Section~\ref{sims: app1} where the majority rule is satisfied with 6 majority sites and $n_l=n=1000$ for $1 \leq l \leq 10$. Moreover, we set the separation level $a=3$. As reported in Figure \ref{fig:rho}, the RIFL CIs vary slightly with different choices of resampling size $M$ (500, 1,000, or 5,000) in terms of both empirical coverage and average length. As discussed in Section \ref{sec: tuning selection}, we choose the smallest $\rho(M)$ such that more than a proportion ({\rm prop}) of the resampled sets satisfy the majority rule. We test sensitivity to the choice of $\rho(M)$ by varying the proportion ({\rm prop}) from $5\%$ to $50\%$ with a $5\%$ increment. In Figure \ref{fig:rho}, we observe that the RIFL CIs with different choices of $\rho(M)$ achieve the desired coverage. The average lengths of the RIFL CIs increase with the proportion. 

\begin{figure}[H]
    \centering
    \includegraphics[width=0.875\textwidth]{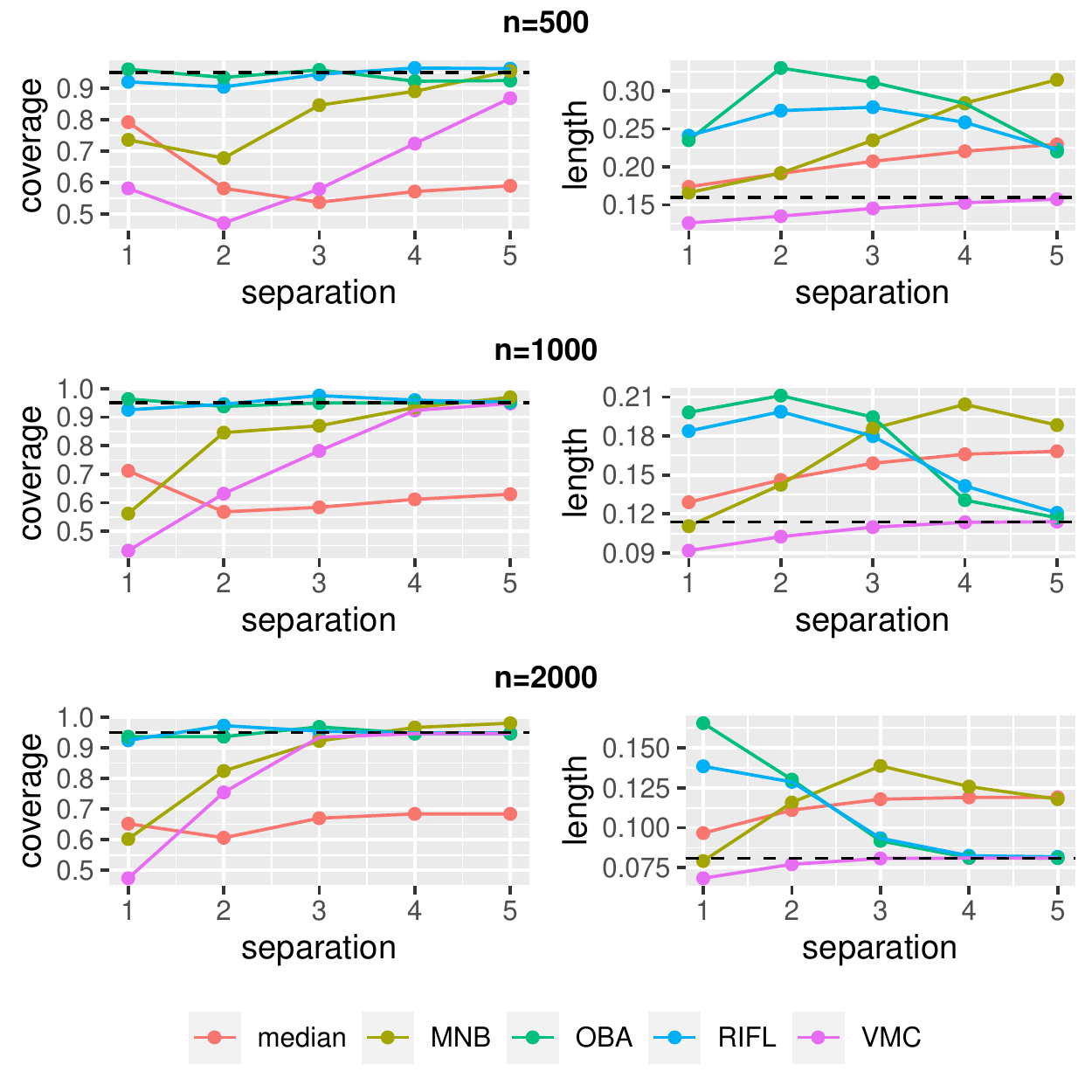}
    \caption{Causal inference: coverage and length of $95\%$ CIs for target ATE of the prevailing sites with 6 majority sites and varying separation levels where $1$ is the lowest (hardest to detect) and $5$ is the highest (easiest to detect). ``median" stands for the CI based on the median estimator, ``MNB" stands for the m-out-of-n bootstrap CI in \eqref{eq: mnb}, ``VMC" stands for the voting with maximum clique estimator and its associated CI in \eqref{eq: post CI}, ``OBA" stands for the oracle bias aware CI in \eqref{eq: bias aware}, and ``RIFL" stands for our proposed CI in \eqref{eq: CI union}.  Results are based on 500 simulation replications. Dash lines in the left panels correspond to a nominal coverage level of 0.95; in the right panels correspond to the width of an oracle CI knowing the prevailing set.}
    \label{fig:TATE6}
\end{figure}
%We empirically assess the sensitivity of coverage and precision of the RIFL CIs to the choices of the tuning parameter $\rho(M)$ and the resampling size $M.$ We continue with Example \ref{exm: demonstration} where the majority rule is satisfied with 6 majority sites and $n_l=n=1000$ for $1 \leq l \leq 10$. Moreover, we set the separation level $a=3$. As reported in Figure \ref{fig:rho}, the RIFL CIs vary slightly with different choices of resampling size $M$ (500, 1,000, or 5,000) in terms of both empirical coverage and average length. As discussed in Section \ref{sec: tuning selection}, we choose the smallest $\rho(M)$ such that more than a proportion ({\rm prop}) of the resampled sets satisfy the majority rule. We test sensitivity to the choice of $\rho(M)$ by varying the proportion ({\rm prop}) across $5\%$, $10\%$, $20\%$ $30\%$, $40\%$, and $50\%$. The chosen $\rho(M)$ corresponding to these proportions are $0.3,0.325,0.35,0.375,0.4,0.425$, and these values are consistent for different choices of $M$ (500, 1,000, and 5,000). In Figure \ref{fig:rho}, we observe that the RIFL CIs with different choices of $\rho(M)$ achieve the desired coverage. The average lengths of the RIFL CIs increase with the proportion. 

\begin{figure}
    \centering
    \includegraphics[width=0.875\textwidth]{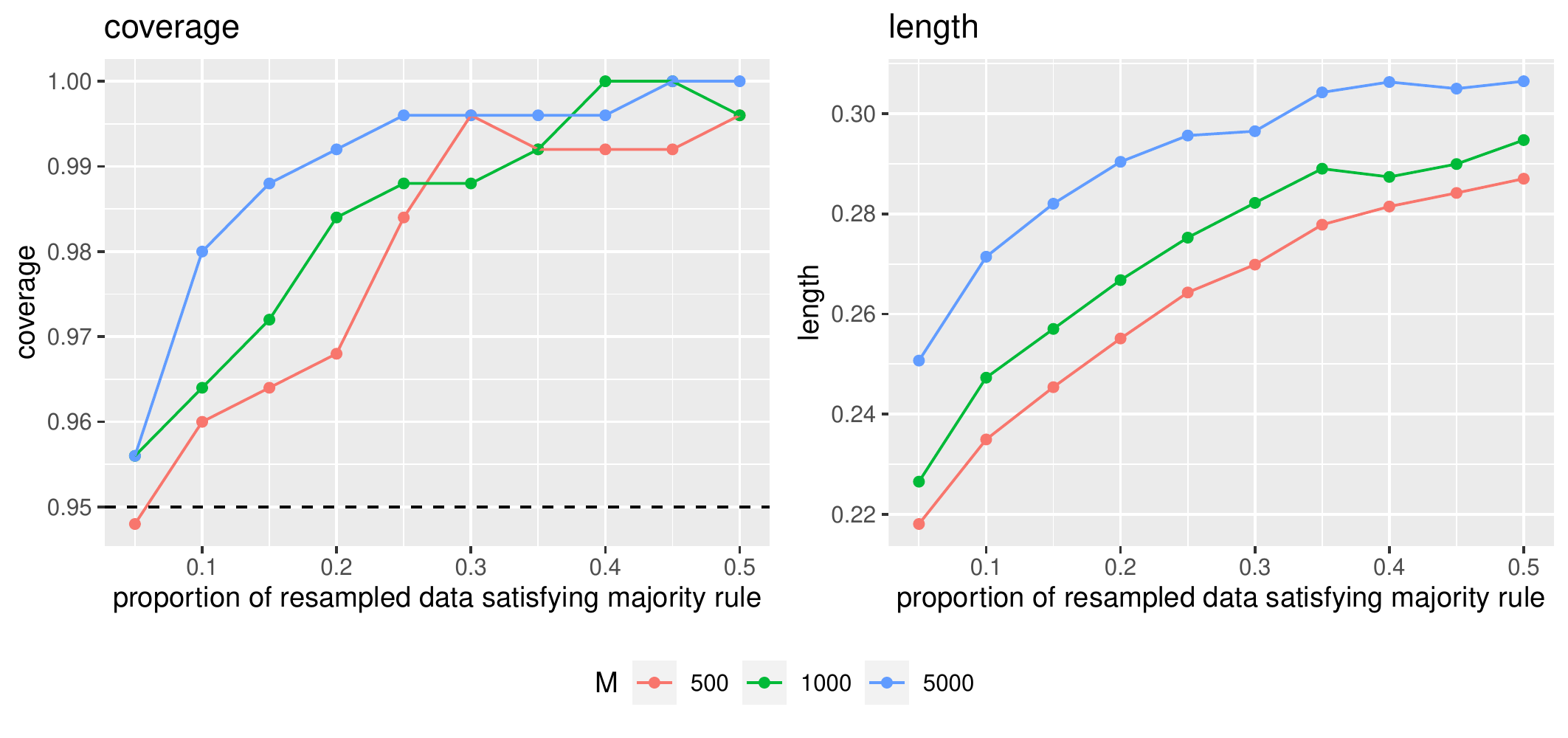}
    \caption{Sensitivity of coverage and the average length of the RIFL CI to different $M$ and $\rho(M)$. Results are based on 200 simulation replications. The x-axis represents the value $\texttt{prop}$, where the tuning parameter $\rho(M)$ is chosen as the smallest value such that more than $\texttt{prop}$ of the resampled dissimilarity measures produce maximum clique sets satisfying the majority rule. }
    \label{fig:rho}
\end{figure}

%In the following, we provide an empirical assessment of the majority rule, which is the critical assumption for our proposal. Recall that equation \eqref{eq: tuning parameter condition} implies that $\mathcal{M}$ is non-empty if we choose $\rho(M)=c_*\left({\log n }/{M}\right)^{1/L(L-1)}$ for some positive constant $c_*.$ With a large resampling size $M$, the tuning parameter $\rho(M)$ is close to zero. However, if we are not able to find any $\rho(M)\in (0,1/2)$ such that $|\mathcal{M}|> 10\% \cdot M$, then we claim that the majority rule is likely to be violated. We empirically assess how RIFL performs when the majority rule does not hold. Specifically, we consider a setting with $L = 10$ sites, where three sites have an ATE of 2.5, four sites have an ATE of 3, and three sites have an ATE of 3.5. We find that over 500 simulations, where we set the number of resamples to be $M = 500$ in each simulation, no choice of $\rho(M)\in (0,1/2)$ leads to $|\mathcal{M}|>0.1*500$ in any of the simulations. Thus, we correctly conclude that the majority rule does not hold in this setting.
% Figure \ref{fig:no_majority} illustrates this phenomenon for the first ten resamples, $m=1,...,10$.

% \begin{figure}[H]
%     \centering
%     \includegraphics[scale=0.28]{graphics/no_majority.png}
%     \caption{For each resample $m$, $m=1,...,10$, the number of blocks in purple is $\leq L/2$, indicating that RIFL can detect when the majority rule does not hold. \Zijian{I do not quite understand this figure.}}
%     \label{fig:no_majority}
% \end{figure}

\section{RIFL Integrative Analysis of EHR Studies on COVID-19 Mortality Prediction}
\label{sec: real data}
%%%%%%%%%%%%%%%%%%%%%%%%%%%%%%%%%%%%%%%%%%%%%%%%%%%%%%%%%%%%
Severe acute respiratory syndrome coronavirus 2 (SARS-CoV-2) has
led to millions of COVID-19 infections and death. Although most individuals present with only a mild form of viral pneumonia, a fraction of individuals develop severe disease. COVID-19 remains a deadly disease for some, including elderly and those with compromised immune systems \citep{wu2020characteristics, goyal2020clinical, wynants2020prediction}. Identifying patients at high risk for mortality can save lives and improve the allocation of resources in resource-scarce health systems.  

To obtain a better understanding of mortality risk for patients hospitalized with COVID-19, we implement our RIFL algorithm using data assembled by $L = 16$ healthcare centers from four countries, representing $275$ hospitals as part of the multi-institutional Consortium for the Clinical Characterization of COVID-19 by EHR (4CE) \citep{brat2020international}. To be included in the study, patients were required to have a positive SARS-CoV-2 reverse transcription polymerase chain reaction (PCR) test, a hospital admission with a positive PCR test between March 1, 2020 and January 31, 2021, and the hospital admission to have occurred no more than seven days before and no later than 14 days after the date of their first positive PCR test. Patients who died on the day of admission were excluded. In the RIFL analysis, we focused on the subset of sites whose summary level data including the covariance matrices of the mortality regression models were available, resulting in a total of 42,655 patients for analyses. Due to patient privacy constraints, individual-level data remained within the firewalls of the institutions, while summary-level data have been generated for a variety of research projects \citep[e.g.,][]{weber2022international}. 
% \begin{table}[H]
% \caption{Healthcare center descriptions, locations, and sample sizes of training data.  \label{tab:ss}}
% \begin{center}
% \begin{tabular}{lllr}
% Center & Description & Location & Size \\\hline
% APHP & Assistance Publique - Hopitaux de Paris & Paris, France & 12,957   \\
% BIDMC & Beth Israel Deaconess Medical Center & Boston, MA, USA & 772\\
% FRBDX & Bordeaux University Hospital & Bordeaux, France & 750 \\
% MGB & Massachusetts General Brigham & Boston, MA, USA & 4,419 \\
% NWU & Northwestern University & Chicago, IL, USA & 6,587 \\
% UCLA & Univ. of California, Los Angles & Los Angeles, CA, USA & 542 \\
% UKFR & University of Freiburg & Freiburg, Germany & 250 \\
% UMICH & Univ. of Michigan & Ann Arbor, MI, USA & 892 \\
% UPENN & Univ. of Pennsylvania  & Philadelphia, PA, USA & 2,695 \\
% UPITT & Univ. of Pittsburgh & Pittsburgh, PA, USA & 1,331 \\
% VA NA & Veterans Affairs, North Atlantic & North Atlantic states, USA & 2,477 \\
% VA SW & Veterans Affairs, Southwestern & Southwest states, USA & 3,094 \\
% VA MW & Veterans Affairs, Midwest & Midwest states, USA & 2,869 \\
% VA CO & Veterans Affairs, Continental & Continental states, USA &  2,499 \\
% VA PAC & Veterans Affairs, Pacific & Pacific states, USA & 1,818 \\
% \end{tabular}
% \end{center}
% \end{table}

% This resulted in a total of 83,178 patients for federated analyses.

We focus on baseline risk factors that were measured in all 16 sites. These baseline risk factors included age group (18-25, 26-49, 50-69, 70-79, 80+), sex, pre-admission Charlson comorbidity score, and nine laboratory test values at admission: albumin, aspartate aminotransferase (AST), AST to alanine aminotransferase ratio (AST/ALT), bilirubin, creatinine, C-reactive protein (CRP), lymphocyte count, neutrophil count, and white blood cell (WBC) count. Laboratory test data had relatively low missingness $(<30\%)$, and missing values were imputed via multivariate imputation by chained equations and averaged over five imputed sets.

We aim to identify and estimate a prevailing mortality risk prediction model to better understand the effect of baseline risk factors on mortality. Specifically, for each healthcare center $l$, $1 \leq l \leq 16$, we let $\theta^{(l)} \in \mathbb{R}^{15}$ denote the vector of log hazard ratios associated with the 15 risk factors, that is, the vector of regression coefficients in a multivariate Cox model. We make statistical inferences for the log hazard ratio associated with each risk factor and let $\beta\supl = \theta^{(l)}_j$ while varying $j$ from 1 to 15. We fit Cox models with adaptive LASSO penalties to obtain $\widehat\theta\supl$, the estimated log hazard ratios of the $15$ baseline risk factors on all-cause mortality within 30-days of hospitalization. Since the sample size is relatively large compared to $p=15$, we expect the oracle property of the adaptive LASSO estimator to hold and hence can rely on asymptotic normality for these local estimators along with consistent estimators of their variances \citep{zou2006adaptive}. 

We implement in Figure \ref{fig:allcovariates} both RIFL and VMC to generate $95\%$ CIs for each baseline risk factor. % \tcomm{i suggest that you also show the point estimates and/or CI from individual sites} 
For RIFL, we set $M = 500$ and choose the value of $\rho(M)$ according to Section~\ref{sec: tuning selection}. That is, we start with a small value (e.g., $1/12$) and set $\rho(M)$ to be the smallest value below 1 such that more than $10\%$ of the resampled dissimilarity measures produce maximum clique sets satisfying the majority rule. %\Zijian{Xiudi, may we unify the discussion on how to choose $\rho(M)$? Please check my discussion in Section 3.3 and you may modify it if needed. One question: what is the reasonable upper bound for $\rho$? 0.6 or 1? In the discussion section, we set $\rho$ to be smaller than 1.} 
The generalizability measures $\{\widehat{p}_l\}_{1\leq l\leq 16}$, defined as $\widehat{p}_l={\sum_{m\in \mathcal{M}} {\bf 1}(l\in \widetilde{\mathcal{V}}^{[m]})}/{|\mathcal{M}|}$, represent the proportion of times that healthcare center $l$ was included in the majority group. Note that we can calculate a set of generalizability measures for the 16 sites for each  risk factor. A visualization of the full generalizability measure is given in Figure \ref{fig:gm} in the supplement.   %We then take the average generalizability measure for each site, averaged over all risk factors. 

As reported in Figure \ref{fig:allcovariates}, the RIFL and VMC intervals are mostly comparable, although for some risk factors, the RIFL CI is notably longer than the VMC CI. However, as we demonstrated in the simulation studies, VMC CI may suffer from under-coverage due to site selection errors. Indeed, some sites have low generalizability scores when we focus on a particular risk factor. For example, site 4 has a generalizability measure of 0.020 when we focus on the variable of age group 18-25; and site 2 has a generalizability measure of 0.016 when we focus on the variable of age $>80$. %; and site 9 has a generalizability measure of 0.118 when we focus on the risk factor AST/ALT. This suggests that \Xiudi{I will add this}. 
This suggests that these sites are not aligned well with other sites when we study the effect of the corresponding risk factor. However, VMC included these sites in the estimated prevailing sets, and the resulting CIs can be misleading. In general, across all 15 risk factors, most sites except for sites 2 and 4 have high generalizability measures,  as shown in Figure \ref{fig:gm} of the supplement.

Our results indicate that older age, male sex, higher Charlson comorbidity score, lower serum albumin level, and higher creatinine, CRP and AST levels are associated with higher mortality risk for patients hospitalized with a positive PCR COVID-19 test. The laboratory tests predictive of mortality represent a mix of general health status (e.g., serum albumin), renal function (e.g., creatinine), hepatic function (e.g., AST), and acute inflammatory response (e.g., CRP). These results are consistent with previous findings in related studies of laboratory measurements to predict COVID-19 mortality \citep{weber2022international}. Recent literature suggests that measuring serum albumin can identify COVID-19 infected patients who are more likely to progress to severe disease and that serum albumin can serve as a useful marker for disease progression \citep{turcato2022severity}. Our results corroborate this finding. The use of regularly collected laboratory tests, in addition to baseline demographic and
comorbidity information, can aid in the development of a clinically useful prediction
tool for mortality risk following hospital admission with COVID-19. Patients identified as having a high risk for mortality could then be prioritized for closer monitoring and potentially more aggressive interventions when deemed appropriate by the physician. 
\vspace{-3mm}

\begin{figure}[H]
    \centering    \includegraphics[width=0.8\textwidth]{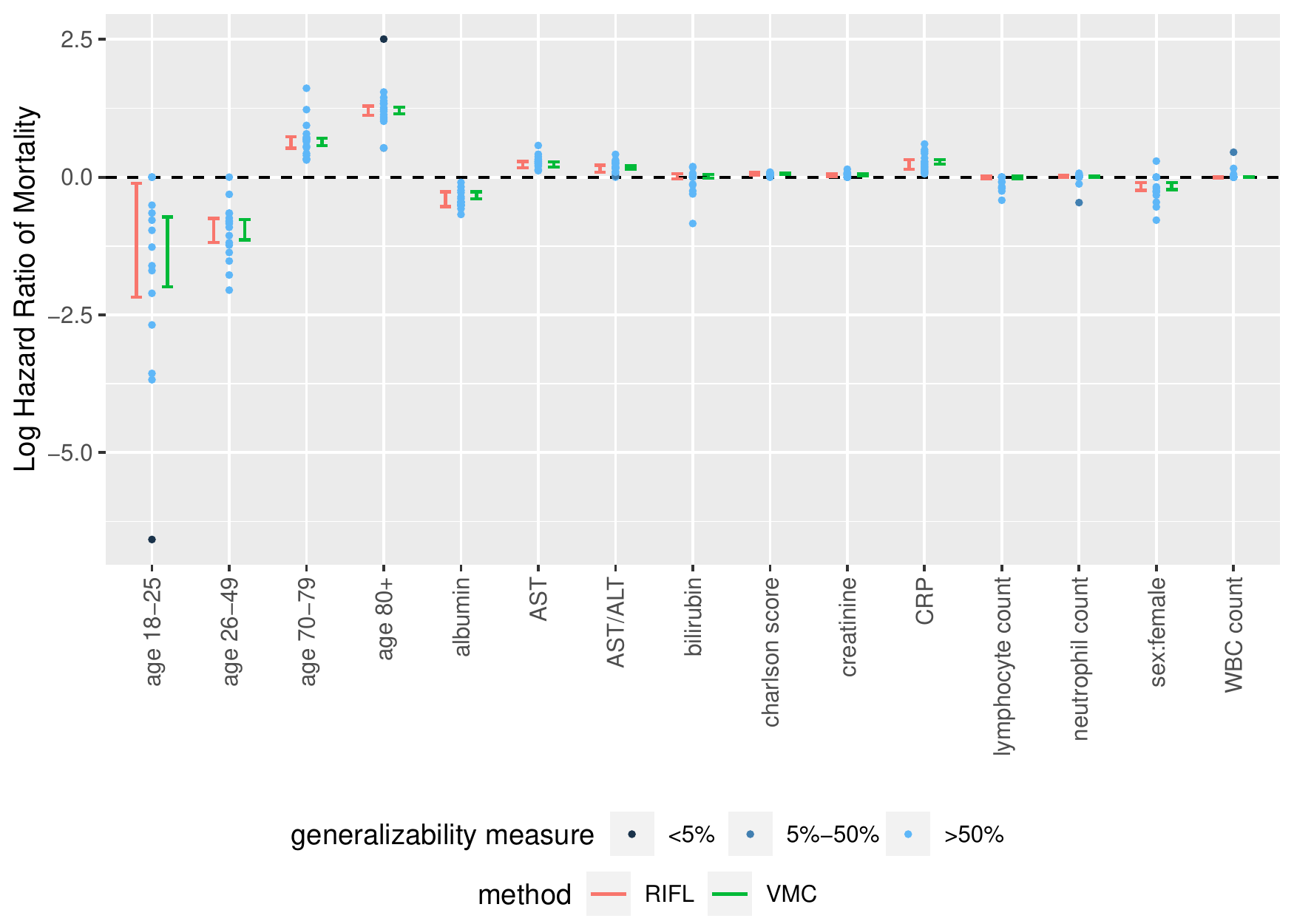}
    \vspace{-3mm}
    \caption{RIFL and VMC $95\%$ CIs for log hazard ratios of mortality within 14 days of hospitalization with COVID-19 for each of the $15$ baseline risk factors. Dots represent the point estimates from individual sites, and darker color of the dots corresponds to lower generalizability measure of individual sites.}
    \label{fig:allcovariates}
\end{figure}

% Our results indicate that older age, black or Hispanic race/ethnicity, low serum albumin levels, and high CRP and AST levels are associated with higher mortality risk for patients hospitalized with a positive PCR COVID-19 test. Our findings have the potential to inform clinical practice. The laboratory tests predictive of mortality represent a mix of general health status (e.g., serum albumin), hepatic function (e.g., AST), and acute inflammatory response (e.g., CRP). These results are consistent with previous findings in related studies of laboratory measurements to predict COVID-19 mortality \citep{weber2022international}. Recent literature  suggests that measuring serum albumin can identify COVID-19 infected patients who are more likely to progress to severe disease and that serum albumin can serve as a useful marker for disease progression \citep{turcato2022severity}. The use of these regularly collected laboratory tests, in addition to baseline demographic and
% comorbidity information, can aid in the development of a clinically useful prediction
% tool for mortality risk following hospital admission with COVID-19. Patients who are identified as having high risk for mortality could then be prioritized for closer monitoring and potentially more aggressive interventions when deemed appropriate by the physician.

%%%%%%%%%%%%%%%%%%%%%%%%%%%%%%%%%%%%%%%%%%%%%%%%%%%
\section{Conclusion and Discussion}
\label{sec: discussion}
%%%%%%%%%%%%%%%%%%%%%%%%%%%%%%%%%%%%%%%%%%%%%%%%%
%\Zijian{Tianxi, double check.}
The RIFL confidence interval,  robust to the errors in separating the non-majority sites from the majority sites, guarantees uniformly valid coverage of the prevailing model. When the majority of sites are well separated from the other sites, the RIFL CI performs similarly to the oracle CI assuming the knowledge of the majority group. 

The majority rule is a crucial assumption for our proposal, and an empirical assessment of the majority rule is vital to validate our proposed procedure. Our proposed RIFL method might provide a heuristic assessment of the majority rule. Suppose the index set $\mathcal{M}$ defined in \eqref{eq: index set} has the cardinality below $10\%\cdot M$ even for $\rho$ approaching $1$. In that case, we claim that the majority rule may fail since it is difficult to verify the majority rule with most of the resampled data.
As a relaxation of the majority rule, we might relax the strict equality of the definition of $\mathcal{V}(\theta)$ in \eqref{eq: equal in distribution} to approximate equality, 
\begin{equation}
\mathcal{V}_{\kappa}(\theta) \coloneqq \{1\leq l\leq L: \|\theta^{(l)}-\theta\|_2\leq \kappa\}, \quad \text{for some} \quad \kappa>0.
\label{eq: relaxed set}
\end{equation}
When $\kappa$ is sufficiently close to zero, our main results may be generalized to hold under the majority rule (Assumption \ref{assump: majority rule}) defined with $\mathcal{V}_{\kappa}(\theta)$.

In practice, domain experts might believe that more than $50\%$ (e.g. $80\%$) of the total sites are similar. Our proposed RIFL can be directly extended to accommodate additional prior information. For example, we may consider the 80\% rule; that is, the domain experts expect more than 80\% of the sites to share the parameters of interest. We can modify our construction by replacing $L/2$ in \eqref{eq: index set} and \eqref{eq: enlarged resampling} with $80\%\cdot L.$ Importantly, the RIFL methods leveraging either the majority rule or the 80\% rule will guarantee uniformly valid CIs for the parameter of the prevailing model, provided that {there are indeed more than $80\%$ of the sites in the prevailing set}. However, the RIFL leveraging the 80\% rule will have a shorter length due to the additional prior information, {as shown in Section \ref{sec: 80 perc rule} of the supplement where we compare the coverage and lengths of RIFL CIs under different prior knowledge. It is also possible to consider alternative strategies for choosing $\rho$ to incorporate further data-adaptive assessment of prevailing set size based on $\{\widehat{p}_l\}_{ 1\leq l\leq L}$. Theoretically justified approaches to adaptively choose $\rho(M)$ warrant further research.} %\tcomm{zijian: plz edit} %We have further explored the numerical studies in Section~\ref{sec: additional sim} in the supplement. 

We discuss two possible directions for robust information aggregation when the majority rule or its relaxed version does not hold. Firstly, the target prediction model can be defined as the one shared by the largest number of sites. If we cluster $\{\theta^{(l)}\}_{1\leq l\leq L}$ into subgroups according to their similarity, the group of the largest size naturally defines a target model, even though this largest group does not contain more than $L/2$ sites. Secondly, we can define the set $\mathcal{V}_{\kappa}(\theta)$ in \eqref{eq: relaxed set} with a relatively large $\kappa$, that is, $\mathcal{V}_{\kappa}(\theta)$ contains all sites with the parameters similar to but possibly different from $\theta$. We may still pool over the information belonging to $\mathcal{V}_{\kappa}(\theta)$ through defining a group distributionally robust model as in \citet{meinshausen2015maximin,hu2018does,sagawa2019distributionally,guo2020inference}. The post-selection problem persists in both settings. It is arguably vital to account for the uncertainty in identifying useful sites. Our proposal is potentially useful to account for the site selection and make inferences for these target models, which is left to future research.

\section*{Acknowledgement}
The research was partly supported by the NSF grant DMS 2015373 as well as NIH grants R01GM140463, R01HL089778, and R01LM013614. 

\bibliographystyle{plainnat}
\bibliography{Ref}

\newpage
\appendix
\renewcommand{\thefigure}{C\arabic{figure}}
\setcounter{figure}{0}
\setcounter{page}{1}
\newpage

%%%%%%%%%%%%%%%%%%%%%%%%%%%%%%%%%%%%%%%%%%%%%%%%%%%%%%%%%%%%
\section{Extra Method and Theory}
%%%%%%%%%%%%%%%%%%%%%%%%%%%%%%%%%%%%%%%%%%%%%%%%%%%%%%%%%%%%
%%%%%%%%%%%%%%%%%%%%%%%%%%%%%%%%%%%%%%%%%%%%%%%%%%%%%%%%%%%%
\subsection{RIFL Algorithm}
\label{sec: algorithm}
\begin{algorithm}[htp!]
\caption{Proposed sampling method for the multi-source inference}
\begin{flushleft}
\textbf{Input:} Site-specific estimators
$\{\widehat{\beta}\supl,\widehat{\sigma}_l\}_{1\leq l\leq L};$ dissimilarity measures $\{\widehat{\DD}_{l,k}\}_{1\leq l<k\leq L}$ with SE estimates $\{\widehat{\rm SE}(\widehat{\DD}_{l,k})\}_{1\leq l<k\leq L}$;  $M\geq 1$, $\rho(M)\in (0,1)$, and levels $\sig,\alpha>0$. \\
\textbf{Output:} Confidence interval ${\rm CI}$; measure of generalizability for each site \\
\end{flushleft}
\vspace{-2.5mm}
\begin{algorithmic}[1]
\For{$l \gets 1$ to $L-1$}  
\For{$k \gets l+1$ to $L$}                                      
\State Compute $\widehat{\LL}_{l,k}=\betahat\supl-\betahat\supk$  and $\widehat{\rm SE}(\widehat{\LL}_{l,k})=\sqrt{\widehat\sigma_l^2+\widehat\sigma_k^2}$;
\EndFor
\EndFor
\For{$m \gets 1$ to $M$}                    
          \State Resample the dissimilarity measures $\{\widehat{\DD}^{[m]}_{l,k},\widehat{\LL}^{[m]}_{l,k}\}_{1\leq l<k\leq L}$ as in \eqref{eq: resampling};
 \State         Compute the resampled test statistics $\{\widehat{S}^{[m]}_{l,k}\}_{1\leq l<k\leq L}$ as in \eqref{eq: global difference sampling};
         \State Construct the sampled voting matrix
$\widehat{H}^{[m]}$ as in \eqref{eq: voting sampling};
\State Construct the maximum clique $\widehat{\mathcal{V}}^{[m]}$ as in \eqref{eq: Vhat sampling};
\State Construct the aggregation set $\widetilde{\mathcal{V}}^{[m]}$ as in \eqref{eq: enlarged resampling};
\State Construct the confidence interval ${\rm CI}^{\m}$ as in \eqref{eq: CI sampling}; 
    \EndFor
 \State Construct  the index set $\mathcal{M}$ as in \eqref{eq: index set};
 \State Return the ${\rm CI}$ defined in \eqref{eq: CI union};
\For{$l \gets 1$ to $L$}                    
 \State Return the generalizability measure $\widehat{p}_l={\sum_{m\in \mathcal{M}} {\bf 1}(l\in \widetilde{\mathcal{V}}^{[m]})}/{|\mathcal{M}|}$;
 \EndFor
\end{algorithmic}
\label{algo: multi-source sampling}
\end{algorithm}

%%%%%%%%%%%%%%%%%%%%%%%%%%%%%%%%%%%%%%%%%%%%%%%%%%%%%%%%%%%%
\subsection{Theoretical properties of $\widehat{\DD}_{l,k}$ in high dimensions}
\label{sec: high-dim theory}
%%%%%%%%%%%%%%%%%%%%%%%%%%%%%%%%%%%%%%%%%%%%%%%%%%%%%%%%%%%%
We summarize our proposed estimators of high-dimensional distance measures in the following Algorithm \ref{algo: high-dim}. 
\begin{algorithm}[htp!]
\caption{High-dimensional Distance Measures}
\begin{flushleft}
\textbf{Input:} the multi-source data $\{X^{(l)},Y^{(l)}\}_{1\leq l\leq L}$. \\
\textbf{Output:} $\{\widehat{\beta}^{(l)},\widehat{\rm SE}(\widehat{\beta}^{(l)}),\widehat{\DD}_{l,k}, \widehat{\rm SE}(\widehat{\DD}_{l,k})\}_{1\leq l<k\leq L}$ \\
\end{flushleft}
\vspace{-2.5mm}
\begin{algorithmic}[1]
\For{$l \gets 1$ to $L$}  
\State Compute the initial $\widetilde{\mu}_l, \widetilde{\theta}^{(l)}$ as in \eqref{eq: penalized estimator};
\State Compute $\widehat{\beta}^{(l)}$ and ${\rm SE}(\widehat{\beta}^{(l)})$ satisfying \eqref{eq: debiasing}; 
\EndFor
\State Broadcast $\{\widetilde{\mu}_l,\widetilde{\theta}^{(l)}, \widehat{\beta}\supl, \widehat{\rm SE}(\widehat{\beta}\supl)\}_{1\leq l\leq L}$ to all $L$ sites;  
\For{$l \gets 1$ to $L$}  
\State Compute the projection direction $\{\widehat{u}_k\supl\}_{k\neq l}$ as in \eqref{eq: projection direction};
\State Compute $\{\widehat{\delta}^{(l)}_{k}\}_{k\neq l}$ as in \eqref{eq: error estimate} and $\{\widehat{\rm V}^{(l)}_k\}_{k\neq l}$ as in \eqref{eq: var component};
 \EndFor
\State 
Compute $\{\widehat{\DD}_{l,k}\}_{1\leq k<l\leq L}$ as in \eqref{eq: dist high-dim} and $\{\widehat{\rm SE}(\widehat{\DD}_{l,k})\}_{1\leq k<l\leq L}$ as in \eqref{eq: SE high-dim}.
\end{algorithmic}
\label{algo: high-dim}
\end{algorithm}
We shall consider the high-dimensional linear or logistic model and establish the theoretical properties of $\widehat{\DD}_{l,k}.$  We recall the notations from the main paper. For $1\leq l\leq L$, we define $\bge^{(l)}=(\mu_l,[\theta^{(l)}]^{\intercal})^{\intercal}\in \R^{d+1}$ and $\widetilde{\bge}^{(l)}=(\widetilde{\mu}_l,[\widetilde{\theta}^{(l)}]^{\intercal})^{\intercal}\in \R^{d+1}$ and $\tX^{(l)}_{i}=(1,(X^{(l)}_{i})^{\intercal})^{\intercal}$ for  $1\leq i\leq n_l.$ Let $\widetilde{\gamma}_{l,k}\coloneqq(0,[\widetilde{\theta}^{(l)}-\widetilde{\theta}^{(k)}]^{\intercal})^{\intercal}$ and $\gamma_{l,k}\coloneqq (0,[\theta\supl-\theta\supk]^{\intercal})^{\intercal}$ for $1\leq l<k\leq L$. When it is clear from the context, we shall write $\widetilde{\gamma}$ and $\gamma$ for $\widetilde{\gamma}_{l,k}$ and $\gamma_{l,k},$ respectively.

Our theoretical analysis follows from that in \citet{guo2021inference}. We generate the model assumptions in \citet{guo2021inference} to the settings with $L$ high-dimensional regression models.

\begin{enumerate}
\item[(A1)]  For $1\leq l\leq L,$ the rows $\{\tX^{(l)}_{i}\}_{1\leq i\leq n} $ are i.i.d. $d$-dimensional {Sub-gaussian} random vectors with $\Sigma^{(l)}=\E (\tX^{(l)}_{i} [\tX^{(l)}_{i}]^{\intercal})$ where $\Sigma^{(l)}$ satisfies $c_0\leq \lambda_{\min}\left(\Sigma^{(l)}\right) \leq \lambda_{\max}\left(\Sigma^{(l)}\right) \leq C_0$ for some positive constants $C_0\geq c_0>0$; The high-dimensional vector $\eta^{(l)}$ is assumed to be of less than $s$ non-zero entries.  
\end{enumerate}
Condition {\rm (A1)} imposes the tail condition for the high-dimensional covariates and assumes that the population second-order moment matrix is invertible.

For the high-dimensional logistic regression, we impose the following condition,
\begin{enumerate}
\item[(A2)]  With probability larger than $1-d^{-c}$, $\min\{h([\tX^{(l)}_{i}]^{\intercal}{\eta}^{(l)}),1-h([\tX^{(l)}_{i}]^{\intercal}{\eta}^{(l)})\}\geq c_{\min}$
for $1\leq i\leq n$ and some small positive constant $c_{\min}\in(0,1)$.
\end{enumerate}
 Condition ${\rm (A2)}$ is imposed such that the case probability is uniformly bounded away from $0$ and $1$. Condition (A2) holds for the setting with bounded $[\tX^{(l)}_{i}]^{\intercal}\eta^{(l)}$ for $1\leq i\leq n$ with a high probability. 

We assume that the penalized MLE estimator
$\widetilde{\theta}^{(l)}$ satisfies the following property:
\begin{enumerate}
\item[(B)] With probability greater than $1-d^{-c}-\exp(-cn)$ for some constant $c>0$, 
\begin{equation*}
\|\widetilde{\eta}^{(l)}-\eta^{(l)}\|_1\leq C s \left({\log d}/{n}\right)^{1/2} \quad \text{and} \quad \|\widetilde{\eta}^{(l)}_{S_l^{c}}-{\eta}^{(l)}_{S_l^{c}}\|_1 \leq C_0\|\widetilde{\eta}^{(l)}_{S_l}-{\eta}^{(l)}_{S_l}\|_1
\end{equation*}
where $S_l$ denotes the support of $\eta^{(l)}$ and $C>0$ and $C_0>0$ are positive constants.
\end{enumerate}
In the high-dimensional linear model, Theorem 7.2 of \citet{bickel2009simultaneous} established that the Lasso estimator satisfies the condition {\rm (B)}.  
In the high-dimensional logistic regression, see Proposition 1 of \citet{guo2021inference} for an example of establishing that the penalized MLE estimator $\widetilde{\theta}^{(l)}$ defined in \eqref{eq: penalized estimator} satisfies the condition ${\rm (B)}$; see also the references within there. 

\begin{Theorem}
Suppose that Conditions {\rm (A1)} and {\rm (B)} hold for the high-dimensional linear regression or Conditions {\rm (A1)}, {\rm (A2)}, and {\rm (B)} hold for the high-dimensional logistic regression, $\tau_{n}\asymp (\log n)^{1/2}$ defined in \eqref{eq: projection direction} satisfies $\tau_{n}{s \log d}/{\sqrt{n}}\rightarrow 0$. For any constant $0<\alpha<1$, the dissimilarity estimator
$\widehat{\DD}_{l,k}$ defined in \eqref{eq: dist high-dim} and the standard error estimator ${\widehat{\rm SE}(\widehat{\DD}_{l,k})}$ defined in \eqref{eq: SE high-dim} satisfy \eqref{eq: key assumption}  for $1\leq l<k\leq L.$
\label{thm: high-dim dissimilarity}
\end{Theorem}

We present the proof of the above theorem in Section \ref{sec: high-dim dissimilarity}. 

%We assume that the dissimilarity estimator $\widehat{\DD}_{l,k}$ for ${1\leq l<k\leq L}$ satisfies 

%and estimate its standard error by 
%\begin{equation}
%\widehat{\rm SE}(\widehat{\DD}_{l,k})=\sqrt{4\widehat{\rm V}^{(k)}_{k}+4\widehat{\rm V}^{(k)}_{l}+\frac{1}{\min\{n_l,n_k\}}}.
%\label{eq: SE high-dim}
%\end{equation}

\subsection{Influence function of the doubly robust ATE estimator}
\label{sec:TATEinfluence}

%\Zijian{Xiudi, may we put this as a proposition or theorem and put the proof to the later section?}
For an i.i.d. sample from the $l$-th source population of size $n_l$, we use $X_i^{(l)} \in \mathbb{R}^p$ to denote the covariate vector in the $i$-th observation and $X_{ij}^{(l)} \in \mathbb{R}$ to denote the $j$-th covariate in the $i$-th observation.

We consider the case where the outcome regression functions and the propensity score are estimated with generalized linear models (GLMs). Specifically, for the propensity score model, we fit the following GLM:
\begin{equation*}
   \EE\left[ A^{(l)} | X^{(l)}\right] =  h\left(\alpha_{l,0} + \sum_{j=1}^B \alpha_{l,j} \psi_j(X^{(l)})\right) = h\left( \bga_l^\top \widetilde{X}^{(l)}\right),
\end{equation*}
where $\{\psi_1(\cdot),\ldots,\psi_B(\cdot)\}$ is a set of basis functions, $\widetilde{X}^{(l)} = (1,\psi_1(X^{(l)}), \ldots, \psi_B(X^{(l)}))^\top$ and $\bga_l = (\alpha_{l,0},\alpha_{l,1},\ldots,\alpha_{l,B})$.  One simple example is to take $\psi_j(X_i^{(l)}) = X_{ij}^{(l)}$ for $j \in \{1,\ldots,p\}$, and we get a GLM that includes the main effect of each covariate. We estimate the coefficient by solving the following estimating equation:
\begin{equation*}
\frac{1}{n_l}\sum_{i=1}^{n_l} \widetilde{X}_i^{(l)}\left\{A_i^{(l)} - h\left(\bga^\top \widetilde X_i^{(l)}\right)\right\} = 0,
\end{equation*}
and obtain an estimated coefficient which we denote as $\widehat\bga_l$. 

For the outcome regressions, we fit generalized linear models within each treatment arm in each source site:
\begin{equation*}
    \EE(Y^{(l)} \mid A^{(l)}=a, X^{(l)}) = g\left(\bgb_{a,0}^{(l)} + \sum_{j=1}^{B_{\bgb}}\bgb\supl_{a,j} \phi_j(X^{(l)})\right) = g\left([W\supl]^\top \bgb\supl_{a}\right), \quad \textnormal{ for } a \in \{0,1\},
\end{equation*}
where $W\supl = (1,\phi_1(X\supl),\ldots,\phi_{B_{\bgb}}(X\supl))$ for some set of basis functions $\{\phi_1(\cdot),\ldots,\phi_{B_{\bgb}}(\cdot)\}$, and $\bgb\supl_a = (\bgb_{a,0}^{(l)},\bgb\supl_{a,1},\ldots, \bgb\supl_{a,B_{\bgb}})$. Let $I\{\cdot\}$ denote the indicator function. We estimate the coefficients $\bgb\supl_a$ by solving the following estimating equation:
\begin{equation*}
    \frac{1}{n_l}\sum_{i=1}^{n_l}I\left\{A\supl_i = a\right\}W\supl_i\left\{Y\supl_i - g\left([W\supl_i]^\top \bgb\supl_{a}\right)\right\} = 0,
\end{equation*}
and we denote the estimate as $\widehat\bgb\supl_a$.

For the density ratio, we consider an exponential tilt model, $\omega_l(X^{(l)};\bge^{(l)}) = \exp([\bge^{(l)}]^\top \widetilde{W}^{(l)})$, where $\widetilde{W}^{(l)} = (1,\varphi_1(X\supl),\ldots,\varphi_{B_\omega}(X\supl))^\top$for a set of basis functions $\{\varphi_1(\cdot),\ldots,\varphi_{B_\omega}(\cdot)\}$. We estimate the parameter $\bge^{(l)}$ by solving the following estimating equation 
\begin{equation*}
    \frac{1}{n_l}\sum_{i=1}^{n_l} \exp\left([\bge^{(l)}]^\top \widetilde{W}_i^{(l)}\right)\widetilde{W}_i^{(l)} = \frac{1}{N}\sum_{j=1}^N \widetilde{W}_j^{\mathcal{T}}/N
\end{equation*}
where $\widetilde{W}_j^{\mathcal{T}} = (1,\varphi_1(X_j^{\mathcal{T}}),\ldots,\varphi_{B_\omega}(X_j^{\mathcal{T}}))^\top$ and $X_j^{\mathcal{T}}$ denotes the $j$-th observation in the target dataset, for $j \in \{1,\ldots,N\}$. We denote the resulting estimate by $\widehat\bge\supl$. Essentially, we estimate the parameter $\bge\supl$ by matching the sample mean of a set of basis functions in the source dataset and the much larger target dataset.

Let $\bga_l^*$, $\bgb_0^{(l),*}$,$\bgb_1^{(l),*}$ and $\bge^{(l),*}$ denote the probabilistic limits of $\widehat\bga_l$, $\widehat\bgb\supl_0$, $\widehat\bgb\supl_1$ and $\widehat\bge\supl$, respectively. The precise definitions of these population parameters are given in the proof of Theorem~\ref{tateIF}. Define the following matrices:
\begin{align*}
    C_{\bga,l} &= \EE_l\left[\widetilde X^{(l)}h^\prime\left((\bga_l^*)^\top \widetilde X^{(l)}\right)\left(\widetilde X^{(l)}\right)^\top\right]; \\
    C_{\bgb_a,l} &= \EE_l\left[I\left\{A\supl=a\right\}W\supl g^\prime\left([W\supl]^\top \bgb^{(l),*}_a\right)[W\supl]^\top\right], \quad a \in \{0,1\}; \\
    C_{\bge,l} &= - \EE_l\left[\exp\left([\bge^{(l),*}]^\top \widetilde{W}^{(l)}\right)\widetilde{W}^{(l)}[\widetilde{W}^{(l)}]^\top\right],
\end{align*}
where $\EE_l$ denotes the expectation with respect to the joint distribution of $(X^{(l)},A^{(l)}, Y^{(l)})$ in the $l$-th source population.

Recall from Section~\ref{sec: app 3} that the target ATE estimator takes the form $\widehat{\theta}^{(l)}=\widehat{M}^{(l)} + \widehat{\delta}^{(l)}$ where 
\begin{equation*}
    \widehat{M}^{(l)} = \frac{1}{N} \sum_{i=1}^{N}
    \left\{m(1,X_i^{\mathcal{T}}; \widehat{\bgb}^{(l)}_{1})  - m(0,X_i^{\mathcal{T}}; \widehat{\bgb}^{(l)}_{0})\right\}
\end{equation*}
and 
\begin{equation*}
\begin{aligned}
    \widehat{\delta}^{(l)} &= \frac{1}{n_l}\sum_{i=1}^{n_l} \omega_l(X_i^{(l)};\widehat{\bge}^{(l)})
    \left\{\sum_{a=0}^1\frac{(-1)^{a+1}I\{A_i\supl = a\}}{\pi_{l} (a,X^{(l)}_i; \widehat{\bga}_l)} \{Y^{(l)}_i - m(A_i\supl,X^{(l)}_i; \widehat{\bgb}^{(l)}_{a})\} \right\} . \label{aug-target}
    \end{aligned}
\end{equation*}
In this case, we have that
\begin{align*}
    m(a,X_i\supl;\widehat\bgb_a\supl) &= g\left([W\supl_i]^\top \widehat\bgb^{(l)}_a\right); \\
    m(a,X_i^{\mathcal{T}};\widehat\bgb_a\supl) &= g\left([W^{\mathcal{T}}_i]^\top \widehat\bgb^{(l)}_a\right); \\
    \omega_l(X_i^{(l)};\widehat{\bge}^{(l)}) &= \exp\left([\widehat\bge\supl]^\top \widetilde{W}_i\supl\right); \\
    \pi_{l} (a,X^{(l)}_i; \widehat{\bga}_l) &= h\left((\widehat{\bga}_l)^\top \widetilde X_i^{(l)}\right)^a\left\{1-h\left((\widehat{\bga}_l)^\top \widetilde X_i^{(l)}\right)\right\}^{(1-a)}.
\end{align*}
For the ease of notation, we define the functions $\tau_{\bgb_0,\bgb_1}(\cdot)$ and $\xi_{\bge,\bga,\bgb_0,\bgb_1}(\cdot)$ such that
\begin{equation*}
    \tau_{\bgb_0,\bgb_1}(x\supl) = m(1,x\supl;\bgb_1) - m(0,x\supl;\bgb_0),
\end{equation*}
and
\begin{equation*}
    \xi_{\bge,\bga,\bgb_0,\bgb_1}(x\supl, a\supl, y\supl) = \omega_l(x\supl;\bge)
    \left\{\sum_{a=0}^1\frac{(-1)^{a+1}I\{a\supl = a\}}{\pi_{l} (a,x\supl; \bga)} \{y\supl - m(a,x\supl; \bgb_a\} \right\} .
\end{equation*}
These functions will appear frequently in the following derivation and in the influence function of $\widehat\theta\supl$. Also note that the functions $\tau$ and $\xi$ are indexed by the nuisance model parameters. 

Define the following quantities:
\begin{align*}
    \boldsymbol{d}_{\bgb_0} 
    &= -\EE_{\mathcal{T}}\left[g^\prime\left([W^\mathcal{T}]^\top\bgb_0^{(l),*}\right)W^{\mathcal{T}}\right] + \EE_l\left[\exp\left([\bge^{(l),*}]^\top \widetilde{W}\supl\right)\frac{I\{A\supl=0\}}{1-h\left([\widetilde{X}\supl]^\top \bga_l^*\right)}\left\{g^\prime\left([W\supl]^\top\bgb_0^{(l),*}\right)W\supl\right\}\right]; \\
    \boldsymbol{d}_{\bgb_1}
    &= \EE_{\mathcal{T}}\left[g^\prime\left([W^\mathcal{T}]^\top\bgb_1^{(l),*}\right)W^{\mathcal{T}}\right] - \EE_l\left[\exp\left([\bge^{(l),*}]^\top \widetilde{W}\supl\right)\frac{I\{A\supl=1\}}{h\left([\widetilde{X}\supl]^\top \bga_l^*\right)}\left\{g^\prime\left([W\supl]^\top\bgb_1^{(l),*}\right)W\supl\right\}\right]; \\
    \boldsymbol{d}_{\bge} &= \EE_l\left[\exp\left([\bge^{(l),*}]^\top \widetilde{W}\supl\right)\widetilde{W}\supl
    \left\{\sum_{a=0}^1\frac{(-1)^{a+1}I\{A\supl = a\}}{\pi_{l} (a,X^{(l)}; \bga^*_l)} \left\{Y\supl - g\left([W\supl]^\top\bgb_a^{(l),*}\right)\right\} \right\}\right];\\
    \boldsymbol{d}_{\bga}
    &= \EE_l\left[\exp\left([\bge^{(l),*}]^\top \widetilde{W}\supl\right)\left\{\sum_{a=0}^1\frac{-I\{A\supl =a\}\pi_l^\prime(a,X\supl;\bga_l^*)}{\pi_l^2(a,X\supl;\bga_l^*)}\left\{Y\supl - g\left([W\supl]^\top\bgb_a^{(l),*}\right)\right\}\right\}\widetilde{X}\supl\right].
\end{align*}

\begin{Theorem}\label{tateIF}
When $N \gg n_l$, the target ATE estimator $\widehat\theta\supl$ is asymptotically linear with influence function $\tau_\theta$ such that
\begin{align*}
    \tau_\theta(x\supl,a\supl,y\supl) &= \xi_{\bge^{(l),*},\bga_l^*,\bgb_0^{(l),*},\bgb_1^{(l),*}}(x\supl,a\supl,y\supl) \\
    &\quad + \boldsymbol{d}_{\bgb_0}^\top C_{\bgb_0,l}^{-1}I\left\{a\supl = 0\right\}w\supl \left\{y\supl - g\left([w\supl]^\top \bgb^{(l),*}_0\right)\right\} \\
    &\quad + \boldsymbol{d}_{\bgb_1}^\top C_{\bgb_1,l}^{-1}I\left\{a\supl = 1\right\}w\supl \left\{y\supl - g\left([w\supl]^\top \bgb^{(l),*}_1\right)\right\} \\
    &\quad + \boldsymbol{d}_{\bge}^\top C_{\bge,l}^{-1}\left\{\exp\left([\bge^{(l),*}]^\top \widetilde{w}\supl\right)\widetilde{w}\supl- \EE_{\mathcal{T}}\left[\widetilde{W}^\mathcal{T}\right]\right\} \\
    &\quad + \boldsymbol{d}_{\bga}^\top C_{\bga,l}^{-1}\widetilde{x}\supl\left\{a\supl - h\left((\bga_l^*)^\top \widetilde x\supl\right)\right\}.
\end{align*}
Here the vectors $\widetilde{x}\supl$, $w\supl$ and $\widetilde{w}\supl$ denote the vectors of basis functions derived from the covariate vector $x\supl$.
\end{Theorem}
We present the proof of the above theorem in Section \ref{sec: tateinfluence}. 

The variance of $\widehat\theta\supl$ can be consistently estimated by the empirical variance of the function $\widehat\tau_\theta$ on the source data, where $\widehat\tau_\theta$ is an estimate of $\tau_\theta$ obtained by plugging in the estimates for the nuisance model parameters $(\bge^{(l),*},\bga_l^*,\bgb_0^{(l),*},\bgb_1^{(l),*})$, the partial derivatives $(\boldsymbol{d}_{\bgb_0},\boldsymbol{d}_{\bgb_1},\boldsymbol{d}_{\bge},\boldsymbol{d}_{\bga})$, and the matrices $C_{\bgb_0,l}$, $C_{\bgb_1,l},C_{\bge,l}$ and $C_{\bga,l}$. Consistent estimators for these partial derivatives and matrices are given in the proof of Theorem~\ref{tateIF}.

\section{Proofs}
%%%%%%%%%%%%%%%%%%%%%%%%%%%%%%%%%%%%%%%%%%%%%%%%%%%%%%%%%%%%
%%%%%%%%%%%%%%%%%%%%%%%%%%%%%%%%%%%%%
\subsection{Proof of Theorem \ref{thm: sampling}}
%%%%%%%%%%%%%%%%%%%%%%%%%%%%%%%%%%%%%
%We shall establish the following result, which will imply Proposition \ref{prop: sampling}.
%\begin{equation}
%\liminf_{n\rightarrow\infty}\PP\left(\min_{1\leq m\leq M}\left[\max_{\beta\in \mathcal{U}(a)}\max_{j\in \Shat} \frac{|\widehat{\Gamma}^{[m]}_j-\gamma_j-\beta(\widehat{\gamma}^{[m]}_j-\gamma_j)|}{\sqrt{(\widehat{\V}^{\Gamma}_{jj}+\beta^2\widehat{\V}^{\gamma}_{jj}-2\beta \widehat{\C}_{jj})/n}}\right]\leq C \err_n(M,\nu)\right)\geq 1- \sig,
%\label{eq: sample quantile strong}
%\end{equation}
%where $\mathcal{M}_0$ is defined as
%{\small
%\begin{equation}
%\mathcal{M}_0=\left\{1\leq m\leq M: \max_{j\in \widehat{\mathcal{S}}}\max\left\{{\left|\widehat{\gamma}^{\m}_j-\widehat{\gamma}_j\right|}/{\sqrt{\widehat{\V}^{\gamma}_{jj}/n}},{\left|\widehat{\Gamma}^{\m}_j-\widehat{\Gamma}_j\right|}/{\sqrt{\widehat{\V}^{\Gamma}_{jj}/n}}\right\}\leq 1.1 z_{\sig/[L(L-1)]}\right\}.
%\label{eq: screening set}
%\end{equation}
%}
 
Denote the observed data by $\Data$. 
 We define the following event for the data $\Data$,
\begin{equation}
%\mathcal{E}=\left\{\max_{1\leq l<k \leq L}\left|\frac{\widehat{\V}_{l,k}}{\V_{l,k}}-1\right|<c_2\right\}, \quad
\mathcal{E}=\left\{\max_{1\leq l<k \leq L}\max\left\{\frac{\left|\widehat{\DD}_{l,k}-{\DD}_{l,k}\right|}{\widehat{\rm SE}(\widehat{\DD}_{l,k})},\frac{\left|\widehat{\LL}_{l,k}-{\LL}_{l,k}\right|}{\widehat{\rm SE}(\widehat{\LL}_{l,k})}\right\}\leq z_{\sig/[2L(L-1)]}\right\}.
\label{eq: event E1}
\end{equation}
Since the site-specific estimators
$\left\{\widehat{\beta}\supl,\widehat{\sigma}_l\right\}_{1\leq l\leq L}$  satisfy \eqref{eq: limiting distribution}, we establish \begin{equation}
\limsup_{n\rightarrow \infty}\PP\left({\left|\widehat{\LL}_{l,k}-{\LL}_{l,k}\right|}/{\widehat{\rm SE}(\widehat{\LL}_{l,k})}\geq z_{\alpha}\right)\leq \alpha \quad \text{for}\quad 0<\alpha<1.
\label{eq: implied difference prob}
\end{equation}

Since the dissimilarly measures $\{\widehat{\DD}_{l,k},\widehat{\rm SE}(\widehat{\DD}_{l,k})\}_{1\leq l<k \leq L}$ satisfy \eqref{eq: key assumption}, we apply the union bound and establish
\begin{equation}
\liminf_{n\rightarrow \infty} \PP(\mathcal{E})\geq 1-\sig.
\label{eq: high prob general}
\end{equation}

In the following, we study the theoretical analysis for a given sample size $n$ and then take the limit with respect to $n.$ We define the stacked vectors $\widehat{U}, \{{U}^{\m}\}_{1\leq m\leq M}\in \R^{L(L-1)}$ as
$$\widehat{U}=\begin{pmatrix}\left\{\frac{\widehat{\DD}_{l,k}-{\DD}_{l,k}}{\widehat{\rm SE}(\widehat{\DD}_{l,k})}\right\}_{1\leq l<k\leq L}\\
\left\{\frac{\widehat{\LL}_{l,k}-{\LL}_{l,k}}{\widehat{\rm SE}(\widehat{\LL}_{l,k})}\right\}_{1\leq l<k\leq L}
\end{pmatrix} \quad \text{and}\quad {U}^{\m}=\begin{pmatrix}\left\{\frac{\widehat{\DD}_{l,k}-{\DD}^{\m}_{l,k}}{\widehat{\rm SE}(\widehat{\DD}_{l,k})}\right\}_{1\leq l<k\leq L}\\
\left\{\frac{\widehat{\LL}_{l,k}-{\LL}^{\m}_{l,k}}{\widehat{\rm SE}(\widehat{\LL}_{l,k})}\right\}_{1\leq l<k\leq L}
\end{pmatrix}.$$ Recall that $\widehat{U}$ is a function of the observed data $\Data.$ Let
$f(\cdot\mid {\Data})$ denote the conditional density function of $U^{\m}$ given the data $\Data$, that is, \begin{equation*}
f({U}^{\m}=U\mid {\Data})=\prod_{1\leq j \leq L(L-1)} \frac{1}{\sqrt{2\pi}}\exp\left(-\frac{U_{j}^2}{2}\right).
\end{equation*}
On the event $\mathcal{E}$, we have
$$
\sum_{1\leq j \leq L(L-1)}\frac{\widehat{U}_{j}^2}{2}\leq\frac{L(L-1)}{2} \left[z_{\sig/[2L(L-1)]}\right]^2,
$$
and further establish
\begin{equation}
f(U^{\m}=\widehat{U}\mid \Data)\cdot {\bf 1}_{\Data\in \mathcal{E}}\geq c(\nu)\coloneqq\left(\frac{1}{\sqrt{2\pi}}\right)^{L(L-1)}\exp\left(-\frac{L(L-1)}{2}\left[z_{\sig/[2L(L-1)]}\right]^2\right).
\label{eq: event}
\end{equation}

%Note that $$\|U^{\m}-\widehat{U}\|_{\infty} =\sqrt{n}\cdot\max\left\{\left\|\widehat{\gamma}^{[m]}-\gamma\right\|_{\infty},\left\|\widehat{\Gamma}^{[m]}-\gamma\right\|_{\infty}\right\}.$$
%To establish Proposition \ref{prop: sampling}, it is sufficient to control $$\PP\left(\min_{1\leq m\leq M}\|U^{\m}-\widehat{U}\|_{\infty} \leq \err_n(M,\nu)\right).$$
We use $\PP(\cdot \mid \Data)$ to denote the conditional probability with respect to the observed data $\Data$.
Note that
\begin{equation*}
\begin{aligned}
&\PP\left(\min_{1\leq m\leq M}\|U^{\m}-\widehat{U}\|_{\infty} \leq \err_n(M,\nu)\mid \Data\right)\\
&=1-\PP\left(\min_{1\leq m\leq M}\|U^{\m}-\widehat{U}\|_{\infty} \geq \err_n(M,\nu)\mid \Data\right)\\
&=1-\prod_{m=1}^{M}\left[1-\PP\left(\|U^{\m}-\widehat{U}\|_{\infty} \leq \err_n(M,\nu)\mid \Data\right)\right]\\
&\geq 1-\exp\left[-\sum_{m=1}^{M}  \PP\left(\|U^{\m}-\widehat{U}\|_{\infty} \leq \err_n(M,\nu)\mid \Data\right)\right],
\end{aligned}
\label{eq: exp lower}
\end{equation*}
where the second equality follows from the conditional independence of $\{U^{\m}\}_{1\leq m\leq M}$ given the data $\mathcal{O}$ and the last inequality follows from $1-x\leq e^{-x}.$ 
By applying the above inequality, we establish 
\begin{equation}
\begin{aligned}
&\PP\left(\min_{1\leq m\leq M}\|U^{\m}-\widehat{U}\|_{\infty} \leq \err_n(M,\nu)\mid \Data\right)\cdot {\bf 1}_{ \Data\in \mathcal{E}}\\
&\geq \left(1-\exp\left[-\sum_{m=1}^{M}\PP\left(\|U^{\m}-\widehat{U}\|_{\infty} \leq \err_n(M,\nu)\mid \Data\right)\right]\right)\cdot {\bf 1}_{ \Data\in \mathcal{E}}\\
&=1-\exp\left[-\sum_{m=1}^{M}\PP\left(\|U^{\m}-\widehat{U}\|_{\infty} \leq \err_n(M,\nu)\mid \Data\right)\cdot {\bf 1}_{ \Data\in \mathcal{E}}\right].
\end{aligned}
\label{eq: connection}
\end{equation}

For the remainder of the proof, we establish a lower bound for 
\begin{equation}
\PP\left(\|U^{\m}-\widehat{U}\|_{\infty} \leq \err_n(M,\nu)\mid \Data\right)\cdot {\bf 1}_{ \Data\in \mathcal{E}},
\label{eq: target}
\end{equation}
and apply \eqref{eq: connection} to establish a lower bound for $$\PP\left(\min_{1\leq m\leq M}\|U^{\m}-\widehat{U}\|_{\infty} \leq \err_n(M,\nu)\mid \Data\right).$$

We further decompose the targeted probability in \eqref{eq: target} as
\begin{equation}
\begin{aligned}
&\PP\left(\|U^{\m}-\widehat{U}\|_{\infty} \leq \err_n(M,\nu)\mid \Data\right)\cdot {\bf 1}_{ \Data\in \mathcal{E}}
\\
=& \int f(U^{\m}=U\mid \Data) \cdot {\bf 1}_{\left\{\|U-\widehat{U}\|_{\infty} \leq \err_n(M,\nu)\right\}}d U \cdot {\bf 1}_{ \Data\in \mathcal{E}}\\
%\geq &\int h(U) \cdot {\bf 1}_{\left\{\|U-\widehat{U}\|_{\infty} \leq \err_n(M,\nu)\right\}}d U\cdot {\bf 1}_{ \Data\in \mathcal{E}}\\
=& \int f(U^{\m}=\widehat{U}\mid \Data) \cdot {\bf 1}_{\left\{\|U-\widehat{U}\|_{\infty} \leq \err_n(M,\nu)\right\}}d U \cdot {\bf 1}_{ \Data\in \mathcal{E}}\\
&+\int [f(U^{\m}={U}\mid \Data)-f(U^{\m}=\widehat{U}\mid \Data)] \cdot {\bf 1}_{\left\{\|U-\widehat{U}\|_{\infty} \leq \err_n(M,\nu)\right\}}d U\cdot {\bf 1}_{ \Data\in \mathcal{E}}. 
\end{aligned}
\label{eq: decomposition density}
\end{equation}
By \eqref{eq: event}, we establish 
\begin{equation}
\begin{aligned}
&\int f(U^{\m}=\widehat{U}\mid \Data)\cdot {\bf 1}_{\left\{\|U-\widehat{U}\|_{\infty} \leq \err_n(M,\nu)\right\}}d U \cdot {\bf 1}_{ \Data\in \mathcal{E}}\\
&\geq {c(\nu)}\cdot \int {\bf 1}_{\left\{\|U-\widehat{U}\|_{\infty} \leq \err_n(M,\nu)\right\}}d U \cdot {\bf 1}_{ \Data\in \mathcal{E}}\\
&\geq {c(\nu)}\cdot [2\err_n(M,\nu)]^{L(L-1)} \cdot {\bf 1}_{ \Data\in \mathcal{E}}.
\end{aligned}
\label{eq: main density}
\end{equation}
There exists $t\in (0,1)$ such that
$$f(U^{\m}={U}\mid \Data)-f(U^{\m}=\widehat{U}\mid \Data)=[\triangledown f(\widehat{U}+t(U-\widehat{U}))]^{\intercal} (U-\widehat{U}),$$
with $$\triangledown f(u)=\left( \frac{1}{\sqrt{2\pi}}\left[-u\cdot\exp\left(-\frac{u^2}{2}\right)\right]\right)^{L(L-1)}.$$
%\frac{1}{\sqrt{(2\pi)^{L(L-1)} {\rm det}({\V}+c_2 {\bf I}) }} \exp\left(-\frac{1}{2}{u}^{\intercal}({\V}-c_2 {\bf I})^{-1}{u}\right)^{-1}({\V}-c_2 {\bf I})^{-1}{u}.$$ 
Since $\|\triangledown f\|_2$ is upper bounded, there exists a positive constant $C>0$ such that $$\left|f(U^{\m}={U}\mid \Data)-f(U^{\m}=\widehat{U}\mid \Data)\right|\leq C \sqrt{L(L-1)}\|U-\widehat{U}\|_{\infty}.$$ Then we establish  
\begin{equation}
\begin{aligned}
&\left|\int [f(U^{\m}={U}\mid \Data)-f(U^{\m}=\widehat{U}\mid \Data)] \cdot {\bf 1}_{\left\{\|U-\widehat{U}\|_{\infty} \leq \err_n(M,\nu)\right\}}d U\cdot {\bf 1}_{ \Data\in \mathcal{E}}
\right|\\
&\leq C \sqrt{L(L-1)}\cdot \err_n(M,\nu)\cdot \int {\bf 1}_{\left\{\|U-\widehat{U}\|_{\infty} \leq \err_n(M,\nu)\right\}}d U \cdot {\bf 1}_{ \Data\in \mathcal{E}}\\
&= C \sqrt{L(L-1)}\cdot \err_n(M,\nu)\cdot [2\err_n(M,\nu)]^{L(L-1)} \cdot {\bf 1}_{ \Data\in \mathcal{E}}.
\end{aligned}
\label{eq: approx density}
\end{equation}
Since $\err_n(M,\nu)\rightarrow 0$ and $c(\nu)$ is a positive constant, then there exists a positive integer $M_0$ such that 
$$C\sqrt{L(L-1)} \cdot \err_n(M,\nu) \leq \frac{1}{2}c(\nu) \quad \text{for} \quad M\geq M_0.$$ We combine the above inequality, \eqref{eq: decomposition density}, \eqref{eq: main density} and \eqref{eq: approx density} and obtain that for $M\geq M_0,$
\begin{equation*}
\begin{aligned}
\PP\left(\|U-\widehat{U}\|_{\infty} \leq \err_n(M,\nu)\mid \Data\right)\cdot {\bf 1}_{ \Data\in \mathcal{E}} \geq \frac{1}{2}c(\nu)\cdot[2\err_n(M,\nu)]^{L(L-1)} \cdot {\bf 1}_{ \Data\in \mathcal{E}}.
\end{aligned}
\end{equation*}
Together with \eqref{eq: connection}, we establish that for $M\geq M_0,$
\begin{equation}
\begin{aligned}
&\PP\left(\min_{1\leq m\leq M}\|U^{\m}-\widehat{U}\|_{\infty} \leq \err_n(M,\nu)\mid \Data\right)\cdot {\bf 1}_{ \Data\in \mathcal{E}}\\
&\geq 1-\exp\left[-M\cdot\frac{1}{2}c(\nu)\cdot [2\err_n(M,\nu)]^{L(L-1)} \cdot {\bf 1}_{\Data\in \mathcal{E}}\right]\\
&=\left(1-\exp\left[-M\cdot\frac{1}{2}c(\nu)\cdot [2\err_n(M,\nu)]^{L(L-1)} \right]\right)
\cdot {\bf 1}_{ \Data\in \mathcal{E}}.
\end{aligned}
\label{eq: connection 2}
\end{equation}
With $\E_{\Data}$ denoting the expectation taken with respect to the observed data $\Data,$ we further integrate with respect to $\Data$ and establish that for $M\geq M_0,$
\begin{equation*}
\begin{aligned}
&\PP\left(\min_{1\leq m\leq M}\|U^{\m}-\widehat{U}\|_{\infty} \leq \err_n(M,\nu)\right)\\
&=\E_{\Data}\left[\PP\left(\min_{1\leq m\leq M}\|U^{\m}-\widehat{U}\|_{\infty} \leq \err_n(M,\nu)\mid \Data\right)\right]\\
&\geq \E_{\Data}\left[\PP\left(\min_{1\leq m\leq M}\|U^{\m}-\widehat{U}\|_{\infty} \leq \err_n(M,\nu)\mid \Data\right)\cdot {\bf 1}_{ \Data\in \mathcal{E}}\right]\\
&\geq \E_{\Data}\left[\left(1-\exp\left[-M\cdot \frac{1}{2}c(\nu)\cdot [2\err_n(M,\nu)]^{L(L-1)} \right]\right)\cdot {\bf 1}_{ \Data\in \mathcal{E}}\right].
\end{aligned}
\end{equation*}
By the definition $\err_n(M,\nu) =\frac{1}{2}\left[\frac{2 \log n }{c(\nu) M}\right]^{\frac{1}{L(L-1)}},$ we establish that for $M\geq M_0,$ 
$$\PP\left(\min_{1\leq m\leq M}\|U^{\m}-\widehat{U}\|_{\infty} \leq \err_n(M,\nu)\right)\geq (1-n^{-1})\cdot \PP\left(\mathcal{E}\right).$$
We further apply \eqref{eq: high prob general} and establish
\begin{equation*}
\begin{aligned}
\liminf_{n\rightarrow\infty} \lim_{M\rightarrow\infty} \PP\left(\min_{1\leq m\leq M}\|U^{\m}-\widehat{U}\|_{\infty} \leq \err_n(M,\nu)\right)\geq \PP\left(\mathcal{E}\right)\geq 1-\sig.
\end{aligned}
\end{equation*}

%%%%%%%%%%%%%%%%%%%%%%%%%%%%%%%%%%%%%%%%%%%%%%%%%%%%%%%%%%%%
\subsection{Proof of Theorem \ref{thm: coverage general}}
%%%%%%%%%%%%%%%%%%%%%%%%%%%%%%%%%%%%%%%%%%%%%%%%%%%%%%%%%%%%
In the following, we first conduct the analysis by fixing the sample size $n$. Define the separation 
\begin{equation*}
\mathcal{L}_{\min}(n)=\min_{l\in \mathcal{V}^{c}}|\beta\supl-\beta^*|.
\end{equation*}
Note that $\mathcal{L}_{\min}(n)$ is a function of the sample size $n$ and might decrease to zero with a growing sample size $n.$
We define the event $$\mathcal{E}_1=\left\{\min_{1\leq m\leq M}\max_{1\leq l<k\leq L}\max\left\{\left|\frac{\widehat{\DD}^{[m]}_{l,k}-\DD_{l,k}}{\widehat{\rm SE}(\widehat{\DD}_{l,k})}\right|,\left|\frac{\widehat{\LL}^{[m]}_{l,k}-\LL_{l,k}}{\widehat{\rm SE}(\widehat{\LL}_{l,k})}\right|\right\}  \leq \err_n(M,\nu)\right\}.$$
On the event $\mathcal{E}_1$, we use $m^*$ to denote the index such that
\begin{equation} 
\max_{1\leq l<k\leq L}\max\left\{\left|\frac{\widehat{\DD}^{[m^*]}_{l,k}-\DD_{l,k}}{\widehat{\rm SE}(\widehat{\DD}_{l,k})}\right|,\left|\frac{\widehat{\LL}^{[m^*]}_{l,k}-\LL_{l,k}}{\widehat{\rm SE}(\widehat{\LL}_{l,k})}\right|\right\}  \leq \err_n(M,\nu).
\label{eq: small distance}
\end{equation}
Recall that $\Thre=z_{\sig/[2L(L-1)]}.$ 
Note that 
\begin{equation}
\PP\left(\theta^*\in {\rm CI}\right)\geq \PP\left(\left\{\theta^*\in {\rm CI}\right\}\cap \mathcal{E}_1\right)\geq \PP\left(\left\{\theta^*\in {\rm CI}^{[m^*]}\right\}\cap \mathcal{E}_1\right).
\label{eq: lower}
\end{equation}
 We first investigate the properties of both $\widehat{\mathcal{V}}^{[m^*]}$ and $\widetilde{\mathcal{V}}^{[m^*]},$ which are useful for the following proof. If ${\rho(M)}\cdot \Thre\geq \err_n(M,\nu),$ then $\mathcal{V}(\theta^*)$ forms a clique in the graph $\mathcal{G}([L],\widehat{H}^{[m^*]})$ and  the maximum clique $\widehat{\mathcal{V}}^{[m^*]}$ satisfies 
\begin{equation*}
\left|\widehat{\mathcal{V}}^{[m^*]}\right|\geq \left|\mathcal{V}(\theta^*)\right|>L/2.
\end{equation*}
The above inequality implies that $m^*\in \mathcal{M},$ and there exists $1\leq l\leq L$ such that  
\begin{equation}
l\in \mathcal{V} \cap \widehat{\mathcal{V}}^{[m^*]}.
\label{eq: non-empty intersection}
\end{equation}

We introduce the following lemma to quantify the set $\widetilde{\mathcal{V}}^{[m^*]}$ defined in \eqref{eq: enlarged resampling}. The proof of the following lemma is postponed to Section \ref{sec: key lemma proof}.
\begin{Lemma}
Assume that the event $\mathcal{E}_1$ holds and the index $m^*$ satisfies \eqref{eq: small distance}.
If the indexes $l,k$ satisfy $\widehat{H}^{[m^*]}_{l,k}=1,$ then we have 
\begin{equation}
\left|\frac{\LL_{l,k}}{\widehat{\rm SE}(\widehat{\LL}_{l,k})}\right|\leq{\rho(M)}\cdot \Thre +\err_n(M,\nu).
\label{eq: key result 1}
\end{equation} 
%\begin{equation}
%\max_{l,k\in \widehat{\mathcal{V}}^{[m^*]}}\left|\frac{\LL_{l,k}}{\widehat{\rm SE}(\widehat{\LL}_{l,k})}\right|\leq{\rho(M)}\cdot z_{\alpha_0/[2L(L-1)]} +\err_n(M,\nu),  
%\label{eq: key result 1}
%\end{equation}
%For any $k\in \widetilde{\mathcal{V}}^{[m^*]},$ there exists $l\in \mathcal{V}$ such that 
%\begin{equation}
%\left|{\LL_{l,k}}\right|\leq 2\left[{\rho(M)}\cdot \Thre +\err_n(M,\nu)\right]\cdot \max_{j_1,j_2 \in \widetilde{\mathcal{V}}^{[m^*]}}\widehat{\rm SE}(\widehat{\LL}_{j_1,j_2}).
%\label{eq: key result 3}
%\end{equation}
In addition, if ${\rho(M)}\cdot  \Thre\geq \err_n(M,\nu)$ and 
\begin{equation}
2{\rho(M)}\cdot \Thre\cdot \max_{j_1\in \mathcal{V}, j_2\in \mathcal{V}^c }\widehat{\rm SE}(\widehat{\LL}_{j_1,j_2})< \mathcal{L}_{\min}(n)
\label{eq: sep fixed n}
\end{equation}
then
\begin{equation}
\widetilde{\mathcal{V}}^{[m^*]}=\mathcal{V}(\theta^*).
\label{eq: key result 2}
\end{equation}
\label{lem: key results}
\end{Lemma}

We now analyze $\PP\left(\left\{\theta^*\in {\rm CI}^{[m^*]}\right\}\cap \mathcal{E}_1\right)$ and then establish the coverage property by applying \eqref{eq: lower}. For a given $n$, since $\rho(M)\rightarrow 0,$ there exists $M_n$ such that if $M\geq M_n$, ${\rho(M)}\cdot  \Thre\geq \err_n(M,\nu)$ and \eqref{eq: sep fixed n} holds.
%Then, for on the event $\mathcal{E}_1$, $\widetilde{\mathcal{V}}^{[m^*]}=\mathcal{V}(\theta^*).$
Hence, for $M\geq M_n$, we have 
\begin{equation*}
\begin{aligned}
\PP\left(\left\{\theta^*\in {\rm CI}^{[m^*]}\right\}\cap \mathcal{E}_1\right)
&= \PP\left(\left\{\left|\frac{\sum_{l\in \mathcal{V}}{\left(\widehat{\beta}\supl-\beta^*\right)
}/{\widehat{\sigma}_{l}^2}}{\sqrt{{\sum_{l\in {\mathcal{V}}}1/{\widehat{\sigma}_{l}^2}}}}\right|
\leq z_{\alpha_1/2}\right\}\cap \mathcal{E}_1\right)\\
&\geq \PP\left(\left|\frac{\sum_{l\in \mathcal{V}}{\left(\widehat{\beta}\supl-\beta^*\right)
}/{\widehat{\sigma}_{l}^2}}{\sqrt{{\sum_{l\in {\mathcal{V}}}1/{\widehat{\sigma}_{l}^2}}}}\right|
\leq z_{\alpha_1/2}\right)- \left[1-\PP\left(\mathcal{E}_1\right)\right].
\end{aligned}
\end{equation*}
By taking limit with respect to $M$, we have 
$$\liminf_{M\rightarrow \infty}\PP\left(\left\{\theta^*\in {\rm CI}^{[m^*]}\right\}\cap \mathcal{E}_1\right)
\geq \PP\left(\left|\frac{\sum_{l\in \mathcal{V}}{\left(\widehat{\beta}\supl-\beta^*\right)
}/{\widehat{\sigma}_{l}^2}}{\sqrt{{\sum_{l\in {\mathcal{V}}}1/{\widehat{\sigma}_{l}^2}}}}\right|
\leq z_{\alpha_1/2}\right)-1+\liminf_{M\rightarrow \infty}\PP\left(\mathcal{E}_1\right).$$
Together with Theorem \ref{thm: sampling} and
$\left\{\widehat{\beta}\supl,\widehat{\sigma}_l\right\}_{1\leq l\leq L}$  satisfying \eqref{eq: limiting distribution}, we establish Theorem \ref{thm: coverage general}.

\subsubsection{Proof of Lemma \ref{lem: key results}}
\label{sec: key lemma proof}
%%%%%%%%%%%%%%%%%%%%%%%%%%%%% 
\noindent{\bf Proof of \eqref{eq: key result 1}}.  For $\widehat{H}^{[m^*]}_{l,k}=1,$ we apply \eqref{eq: voting sampling} and establish 
\begin{equation}
\left|\frac{\widehat{\LL}^{[m^*]}_{l,k}}{\widehat{\rm SE}(\widehat{\LL}_{l,k})}\right|\leq {\rho(M)}\cdot \Thre.
\label{eq: upper 1}
\end{equation}
We apply \eqref{eq: small distance} and establish  
\begin{equation}
\left|\frac{\widehat{\LL}^{[m^*]}_{l,k}-\LL_{l,k}}{\widehat{\rm SE}(\widehat{\LL}_{l,k})}\right|\leq \err_n(M,\nu).
\label{eq: upper 2}
\end{equation}
We establish \eqref{eq: key result 1} by applying \eqref{eq: upper 1} and \eqref{eq: upper 2} and the triangle inequality $$
\left|\frac{\LL_{l,k}}{\widehat{\rm SE}(\widehat{\LL}_{l,k})}\right|\leq \left|\frac{\widehat{\LL}^{[m^*]}_{l,k}-\LL_{l,k}}{\widehat{\rm SE}(\widehat{\LL}_{l,k})}\right|+\left|\frac{\widehat{\LL}^{[m^*]}_{l,k}}{\widehat{\rm SE}(\widehat{\LL}_{l,k})}\right|.$$

%By the definition of $\widehat{\mathcal{V}}^{[m^*]}, $
%we have $\widehat{H}^{[m^*]}_{l,k}=1$ for any $l,k\in \widehat{\mathcal{V}}^{[m^*]}.$ We then directly apply \eqref{eq: key result 1} to establish \eqref{eq: key result 1}.

\vspace{3mm}
\begin{comment}
\noindent{\bf Proof of \eqref{eq: key result 3}}. Recall that $m^*\in \mathcal{M}$ on the event $\mathcal{E}_1.$  Any index $k\in \widetilde{\mathcal{V}}^{[m^*]}$ is connected to more half of all $L$ indexes. Since $|\widehat{\mathcal{V}}^{[m^*]}|>L/2,$ there exists $j\in \widehat{\mathcal{V}}^{[m^*]}$ such that 
$
\widehat{H}^{[m^*]}_{k,j}=1.
$
By applying \eqref{eq: key result 1},  we have 
\begin{equation}
\left|\frac{{\LL}_{k,j}}{\widehat{\rm SE}(\widehat{\LL}_{k,j})}\right|\leq {\rho(M)}\cdot \Thre +\err_n(M,\nu).
\label{eq: intermediate 1}
\end{equation}
For the above $j\in \widehat{\mathcal{V}}^{[m^*]}$, since there exists $l\in \mathcal{V}\cap \widehat{\mathcal{V}}^{[m^*]},$ we have $\widehat{H}^{[m^*]}_{j,l}=1.$ By applying \eqref{eq: key result 1},  we have 
\begin{equation}
\left|\frac{{\LL}_{j,l}}{\widehat{\rm SE}(\widehat{\LL}_{j,l})}\right|\leq {\rho(M)}\cdot \Thre +\err_n(M,\nu).
\label{eq: intermediate 2}
\end{equation}
We establish \eqref{eq: key result 3} by applying \eqref{eq: intermediate 1}m \eqref{eq: intermediate 2}, and the triangle inequality $\left|{\LL}_{k,l}\right|\leq \left|{\LL}_{k,j}\right|+\left|{\LL}_{j,l}\right|.$
\end{comment}
\vspace{3mm}

\noindent{\bf Proof of \eqref{eq: key result 2}}. For $k\in\mathcal{V},$ we apply \eqref{eq: small distance} together with the condition ${\rho(M)}\cdot  \Thre\geq \err_n(M,\nu)$ and establish $\widehat{H}^{[m^*]}_{k,j}=1$ for any $j\in \mathcal{V}.$ By the majority rule, we have $\left\|\widehat{H}^{[m^*]}_{k,\cdot}\right\|_0>L/2$ for $k\in \mathcal{V}.$ 

For $k\not\in \mathcal{V},$ we apply \eqref{eq: key result 1} together with the condition \eqref{eq: sep fixed n} and establish $\widehat{H}^{[m^*]}_{k,j}=0$ for $j\in \mathcal{V}.$ By the majority rule, we have $\left\|\widehat{H}^{[m^*]}_{k,\cdot}\right\|_0<L/2$ for $k\not \in \mathcal{V}.$ Hence, $\widetilde{\mathcal{V}}^{[m^*]}=\mathcal{V}.$
%%%%%%%%%%%%%%%%%%%%%%%%%%%%%%%%%%%%%%%%%%%%%%%%%%%%%%%%%%%%%%%%%%%%%%%%%%%%%%%%%%%%%%%%%%%%%%%%%%%%%%%%%%%%%%%%%%%%%%%%%%%
\subsection{Proof of Theorem \ref{thm: efficiency}}
%%%%%%%%%%%%%%%%%%%%%%%%%%%%%%%%%%%%%%%%%%%%%%%%%%%%%%%%%%%%%%%%%%%%%%%%%%%%%%%%%%%%%%%%%%%%%%%%%%%%%%%%%%%%%%%%%%%%%%%%%%%
Recall $\mathcal{V}^*=\mathcal{V}(\theta^*).$ We define the events 
\begin{equation*}
\begin{aligned}
\mathcal{E}_2&=\left\{\left|\widehat{\LL}^{[m]}_{l,k}-\widehat{\LL}_{l,k}\right|/{\widehat{\rm SE}(\widehat{\LL}_{l,k})}\leq \sqrt{2\log n+2\log M}\right\} \\
\mathcal{E}_3&=\left\{\left|{\LL}_{l,k}-\widehat{\LL}_{l,k}\right|/{\widehat{\rm SE}(\widehat{\LL}_{l,k})}\leq \sqrt{2\log n}\right\}. \\
\end{aligned}
\end{equation*}
By \eqref{eq: resampling} and \eqref{eq: implied difference prob}, we apply the union bound and establish
\begin{equation}
\lim_{n\rightarrow \infty}\lim_{M\rightarrow \infty} \PP(\mathcal{E}_2\cap \mathcal{E}_3)=1.
\label{eq: event prob high}
\end{equation}
By the triangle inequality, we have 
\begin{equation}
\left|\widehat{\LL}^{[m]}_{l,k}/{\widehat{\rm SE}(\widehat{\LL}_{l,k})}-{\LL}_{l,k}/{\widehat{\rm SE}(\widehat{\LL}_{l,k})}\right|\leq \left|\widehat{\LL}^{[m]}_{l,k}-\widehat{\LL}_{l,k}\right|/{\widehat{\rm SE}(\widehat{\LL}_{l,k})}+\left|{\LL}_{l,k}-\widehat{\LL}_{l,k}\right|/{\widehat{\rm SE}(\widehat{\LL}_{l,k})}
\end{equation}
On the event $\mathcal{E}_2\cap \mathcal{E}_3,$ we have 
\begin{equation}\left|\widehat{\LL}^{[m]}_{l,k}/{\widehat{\rm SE}(\widehat{\LL}_{l,k})}-{\LL}_{l,k}/{\widehat{\rm SE}(\widehat{\LL}_{l,k})}\right|\leq 2\sqrt{2\log n+2\log M}.
\label{eq: separation bound}
\end{equation}
The well-separation condition \eqref{eq: well separation} implies that, for $l\in \mathcal{V}$ and $k\in \mathcal{V}^c,$ 
$$\left|{\LL}_{l,k}/{\widehat{\rm SE}(\widehat{\LL}_{l,k})}\right|>2\sqrt{2\log n+2\log M}+\rho(M)\cdot \Thre.$$
On the event $\mathcal{E}_2\cap \mathcal{E}_3,$ if $l\in \mathcal{V}$ and $k\in \mathcal{V}^c$ satisfies \eqref{eq: separation bound}, then 
\begin{equation*}
\left|\widehat{\LL}^{[m]}_{l,k}/{\widehat{\rm SE}(\widehat{\LL}_{l,k})}\right|> \rho(M)\cdot \Thre.
\end{equation*}
which is equivalent to $\widehat{H}^{[m]}_{l,k}=0.$
That is, $k\not \in \mathcal{V}^{[m]}$ for $m\in \mathcal{M}.$

The above derivation shows that if all indexes $k\in \mathcal{V}^c$ satisfy the well separation condition \eqref{eq: separation bound}, we have 
$\mathcal{V}^{[m]}\subset\mathcal{V}^*$ for $m\in \mathcal{M}.$ Since $|\mathcal{V}^{[m]}|> L/2$ and $|\mathcal{V}^*|=\lfloor L/2\rfloor+1$, we have $\mathcal{V}^{[m]}=\mathcal{V}^*$ for $m\in \mathcal{M}.$ This implies $\PP\left({\rm CI}={\rm CI}_{\rm ora}\right)\geq \PP(\mathcal{E}_2\cap \mathcal{E}_3).$ Then the theorem follows from \eqref{eq: event prob high}.

\subsection{Proof of Theorem \ref{thm: low-dim dissimilarity}}
Define $${\rm SE}(\widehat{\DD}_{l,k})=\sqrt{4{\gamma}^{\intercal}{C}^{(l)}{\gamma}/n_l+4{\gamma}^{\intercal}{C}^{(k)}{\gamma}/n_k+1/\min\{n_l,n_k\}} $$ 
and 
$${\rm SE}_0(\widehat{\DD}_{l,k})=\sqrt{4{\gamma}^{\intercal}{C}^{(l)}{\gamma}/n_l+4{\gamma}^{\intercal}{C}^{(k)}{\gamma}/n_k}.$$
Since $\widehat{\theta}^{(l)}$ and $\widehat{C}^{(l)}$ are consistent estimators of ${\theta}^{(l)}$ and ${C}^{(l)},$ respectively, we have 
\begin{equation}
\widehat{\rm SE}(\widehat{\DD}_{l,k})/
{\rm SE}(\widehat{\DD}_{l,k})\cid 1.
\label{eq: consistent SE}
\end{equation}
Define $c_n=(1/\min\{n_l,n_k\})^{1/4}.$
It follows from 
\eqref{eq: limiting ex1 low} that \begin{equation}
\limsup_{n\rightarrow \infty}\PP\left(\| \widehat{\gamma}-\gamma\|_2^2/{\widehat{\rm SE}(\widehat{\DD}_{l,k})}
\geq c_n z_{\alpha}\right)=0.
\label{eq: small prob}
\end{equation}

By the decomposition \eqref{eq: error decomposition}, we have 
\begin{equation}
\begin{aligned}
&\PP\left({\left|\widehat{\DD}_{l,k}-{\DD}_{l,k}\right|}/{\widehat{\rm SE}(\widehat{\DD}_{l,k})}
\geq z_{\alpha}\right)\\
&\leq \PP\left({\left|2\langle \widehat{\gamma}-\gamma,\gamma\rangle\right|}/{\widehat{\rm SE}(\widehat{\DD}_{l,k})}
\geq (1-c_n)z_{\alpha}\right)+\PP\left(\| \widehat{\gamma}-\gamma\|_2^2/{\widehat{\rm SE}(\widehat{\DD}_{l,k})}
\geq c_n z_{\alpha}\right) \\
&\leq \PP\left({\left|2\langle \widehat{\gamma}-\gamma,\gamma\rangle\right|}/{\rm SE}_0(\widehat{\DD}_{l,k})
\geq (1-c_n)z_{\alpha}\cdot \frac{\widehat{\rm SE}(\widehat{\DD}_{l,k})}{
{\rm SE}(\widehat{\DD}_{l,k})}\right)+\PP\left(\| \widehat{\gamma}-\gamma\|_2^2/{\widehat{\rm SE}(\widehat{\DD}_{l,k})}
\geq c_n z_{\alpha}\right),
\end{aligned}
\end{equation}
where the first inequality follows from the union bound and the second inequality follows from the relation ${\rm SE}(\widehat{\DD}_{l,k})\geq {\rm SE}_0(\widehat{\DD}_{l,k}).$ 
We establish \eqref{eq: key assumption} by combing the above decomposition, \eqref{eq: consistent SE}, \eqref{eq: small prob}, and \begin{equation*}
\frac{\langle \widehat{\gamma}-\gamma, \gamma\rangle}{\sqrt{4\gamma^{\intercal}C^{(l)}\gamma/n_l+4\gamma^{\intercal}C^{(k)}\gamma/n_l}}\cid N(0,1).
\end{equation*}

%%%%%%%%%%%%%%%%%%%%%%%%%%%%%%%%%%%%%%%%%%%%%%%%%%%%%%%%%%%%
\subsection{Proof of Theorem \ref{thm: high-dim dissimilarity}}
\label{sec: high-dim dissimilarity}
%%%%%%%%%%%%%%%%%%%%%%%%%%%%%%%%%%%%%%%%%%%%%%%%%%%%%%%%%%%%

The following decomposition is crucial to constructing a consistent estimator of
 the error component: for any vector $u\in \R^{d},$
 \begin{equation}
\begin{aligned}
&[\widehat{u}_k\supl]^{\intercal}\frac{1}{|\mathcal{S}^{(l)}_2|}\sum_{i\in \mathcal{S}^{(l)}_2}W^{(l)}_{i}\tX^{(l)}_{i}(Y_i-h([\tX^{(l)}_{i}]^{\intercal} \widetilde{\eta}^{(l)}))-\langle \eta^{(l)}-\widetilde{\eta}^{(l)}, \widetilde{\gamma}\rangle\\
=&\left(\widehat{\Sigma}^{(l)}\widehat{u}_k\supl-\widetilde{\gamma}\right)^{\intercal}(\eta^{(l)}-\widetilde{\eta}^{(l)})+[\widehat{u}_k\supl]^{\intercal}\frac{1}{|\mathcal{S}^{(l)}_2|}\sum_{i\in \mathcal{S}^{(l)}_2}W^{(l)}_i\epsilon^{(l)}_i\tX^{(l)}_{i}+[\widehat{u}_k\supl]^{\intercal}\frac{1}{|\mathcal{S}^{(l)}_2|}\sum_{i\in \mathcal{S}^{(l)}_2}\Delta^{(l)}_i \tX^{(l)}_{i}, 
\end{aligned}
\label{eq: high-dim bias est}
\end{equation}
with $\widehat{\Sigma}^{(l)}=\frac{1}{|\mathcal{S}^{(l)}_2|}\sum_{i\in \mathcal{S}^{(l)}_2} \tX^{(l)}_{i}\left[\tX^{(l)}_{i}\right]^{\intercal}$ and the approximation error $\Delta^{(l)}_i$ defined as
\begin{equation}\Delta^{(l)}_i=W^{(l)}_i\cdot \int_{0}^{1} (1-t)h''([\tX^{(l)}_{i}]^{\intercal} \widetilde{\eta}^{(l)}+t [\tX^{(l)}_{i}]^{\intercal} [\eta^{(l)}-\widetilde{\eta}^{(l)}]) dt \cdot ([\tX^{(l)}_{i}]^{\intercal} [\eta^{(l)}-\widetilde{\eta}^{(l)}])^2.
\label{eq: approximation error}
\end{equation} 
Note that for the linear outcome model with $h(x)=x$, the approximation error $\Delta^{(l)}_i=0.$ 
Define
\begin{equation*}
\widetilde{\DD}_{l,k}=\|\widetilde{\theta}^{(l)}-\widetilde{\theta}^{(k)}\|_2^2+2\widehat{\delta}^{(l)}_{k}-2\widehat{\delta}^{(k)}_{l}.
\end{equation*}
Since $\DD_{l,k}\geq 0,$ we have 
\begin{equation*}
\left|\widehat{\DD}_{l,k}-\DD_{l,k}\right|\leq \left|\widetilde{\DD}_{l,k}-\DD_{l,k}\right|.
\end{equation*}
To establish \eqref{eq: key assumption}, it is sufficient to establish 
\begin{equation}
\limsup_{n\rightarrow \infty}\PP\left({\left|\widetilde{\DD}_{l,k}-{\DD}_{l,k}\right|}/{\widehat{\rm SE}(\widehat{\DD}_{l,k})}\geq z_{\alpha}\right)\leq \alpha \quad \text{for}\quad 0<\alpha<1
\label{eq: key assumption stronger}
\end{equation}
where $z_{\alpha}$ denotes the upper quantile of a standard normal distribution.

In the following, we establish \eqref{eq: key assumption stronger}. 
We apply \eqref{eq: error decomposition high} and \eqref{eq: high-dim bias est} and obtain 
\begin{equation}
\begin{aligned}
&\widetilde{\DD}_{l,k}-\DD_{l,k}=-\| \widetilde{\gamma}-\gamma\|_2^2
\\
&+\left(\widehat{\Sigma}^{(l)}\widehat{u}_k\supl-\widetilde{\gamma}\right)^{\intercal}(\eta^{(l)}-\widetilde{\eta}^{(l)})+[\widehat{u}_k\supl]^{\intercal}\frac{1}{|\mathcal{S}^{(l)}_2|}\sum_{i\in \mathcal{S}^{(l)}_2}W^{(l)}_i\epsilon^{(l)}_i\tX^{(l)}_{i}+[\widehat{u}_k\supl]^{\intercal}\frac{1}{|\mathcal{S}^{(l)}_2|}\sum_{i\in \mathcal{S}^{(l)}_2}\Delta^{(l)}_i \tX^{(l)}_{i}\\
&+\left(\widehat{\Sigma}^{(k)}\widehat{u}^{(k)}_{l}-\widetilde{\gamma}\right)^{\intercal}(\eta^{(k)}-\widetilde{\eta}^{(k)})+[\widehat{u}^{(k)}_{l}]^{\intercal}\frac{1}{|\mathcal{S}^{(k)}_2|}\sum_{i\in \mathcal{S}^{(k)}_2}W^{(k)}_i\epsilon^{(k)}_i\tX^{(k)}_{i}+[\widehat{u}^{(k)}_l]^{\intercal}\frac{1}{|\mathcal{S}^{(k)}_2|}\sum_{i\in \mathcal{S}^{(k)}_2}\Delta^{(k)}_i \tX^{(k)}_{i}.
\end{aligned}
\label{eq: decomp Dtilde}
\end{equation}
We apply Lemma 5 in \citet{guo2021inference} and establish 
\begin{equation}
\frac{[\widehat{u}_k\supl]^{\intercal}\frac{1}{|\mathcal{S}^{(l)}_2|}\sum_{i\in \mathcal{S}^{(l)}_2}W^{(l)}_i\epsilon^{(l)}_i\tX^{(l)}_{i}+[\widehat{u}^{(k)}_{l}]^{\intercal}\frac{1}{|\mathcal{S}^{(k)}_2|}\sum_{i\in \mathcal{S}^{(k)}_2}W^{(k)}_i\epsilon^{(k)}_i\tX^{(k)}_{i}}{\sqrt{\widehat{\rm V}^{(l)}_k+\widehat{\rm V}^{(k)}_l}}\cid N(0,1)
\label{eq: normal limit}
\end{equation}
with 
\begin{equation*}
\widehat{\rm V}^{(l)}_k=\left(\widehat{u}_k\supl\right)^{\intercal}\left[\frac{1}{|\mathcal{S}^{(l)}_2|^2}\sum_{i\in \mathcal{S}^{(l)}_2}W^{(l)}_i\tX^{(l)}_{i}[\tX^{(l)}_{i}]^{\intercal}\right]\widehat{u}_k\supl,
\end{equation*}
and 
\begin{equation*}
\widehat{\rm V}^{(k)}_l=\left(\widehat{u}^{(k)}_{l}\right)^{\intercal}\left[\frac{1}{|\mathcal{S}^{(k)}_2|^2}\sum_{i\in \mathcal{S}^{(k)}_2}W^{(k)}_i\tX^{(k)}_{i}[\tX^{(k)}_{i}]^{\intercal}\right]\widehat{u}^{(k)}_{l}.
\end{equation*}

By condition (B), we have 
\begin{equation}
\| \widetilde{\gamma}-\gamma\|_2^2\leq C\frac{s \log d}{\min\{n_l,n_k\}}.
\label{eq: bound 1}
\end{equation}
We apply (22) in \citet{guo2021inference} and establish 
\begin{equation}
\begin{aligned}
\left|\left(\widehat{\Sigma}^{(l)}\widehat{u}_k\supl-\widetilde{\gamma}\right)^{\intercal}(\eta^{(l)}-\widetilde{\eta}^{(l)})\right|&\leq C \frac{s \log d}{n_l},\quad 
\left|\left(\widehat{\Sigma}^{(k)}\widehat{u}^{(k)}_{l}-\widetilde{\gamma}\right)^{\intercal}(\eta^{(k)}-\widetilde{\eta}^{(k)})\right|& \leq C \frac{s \log d}{n_k}.
\end{aligned}
\label{eq: bound 2}
\end{equation}
We apply (23) in \citet{guo2021inference} and establish 
\begin{equation}
\begin{aligned}
\left|[\widehat{u}_k\supl]^{\intercal}\frac{1}{|\mathcal{S}^{(l)}_2|}\sum_{i\in \mathcal{S}^{(l)}_2}\Delta^{(l)}_i \tX^{(l)}_{i}\right|&\leq C \tau_n\frac{s \log d}{n_l}\\
\left|[\widehat{u}^{(k)}_{l}]^{\intercal}\frac{1}{|\mathcal{S}^{(k)}_2|}\sum_{i\in \mathcal{S}^{(k)}_2}\Delta^{(k)}_i \tX^{(k)}_{i}\right|&\leq C \tau_n\frac{s \log d}{n_k}
\end{aligned}
\label{eq: bound 3}
\end{equation}

We establish \eqref{eq: key assumption stronger} by combining the decomposition \eqref{eq: decomp Dtilde}, the asymptotic limit \eqref{eq: normal limit}, the error bounds \eqref{eq: bound 1}, \eqref{eq: bound 2}, \eqref{eq: bound 3}, and the condition $\tau_{n}{s \log d}/{\sqrt{n}}\rightarrow 0.$

\subsection{Proof of Theorem~\ref{tateIF}}\label{sec: tateinfluence}
We let $\bga_l^*$ denote the probabilistic limit of $\widehat\bga_l$, which satisfies the following equation:
\begin{equation*}
\EE_l \left[ \widetilde{X}^{(l)}\left\{A^{(l)} - h\left((\bga_l^*)^\top \widetilde X^{(l)}\right)\right\} \right] = 0,
\end{equation*}
where $\EE_l$ denotes the expectation with respect to the joint distribution of $(X^{(l)},A^{(l)}, Y^{(l)})$ in the $l$-th source population.
By the theory of estimating equations, $\widehat\bga_l$ admits the following asymptotic linear expansion:
\begin{equation*}
    \widehat\bga_l - \bga_l^* = \frac{1}{n_l}\sum_{i=1}^{n_l}C_{\bga,l}^{-1}\widetilde{X}_i^{(l)}\left\{A_i^{(l)} - h\left((\bga_l^*)^\top \widetilde X_i^{(l)}\right)\right\} + o_P(n_l^{-1/2}),
\end{equation*}
where the matrix $C_{\bga,l}$ is defined as
\begin{equation*}
    C_{\bga,l} = \EE_l\left[\widetilde X^{(l)}h^\prime\left((\bga_l^*)^\top \widetilde X^{(l)}\right)\left(\widetilde X^{(l)}\right)^\top\right].
\end{equation*}
Note that $C_{\bga,l}$ can be consistently estimated by \begin{equation*}
    \widehat{C}_{\bga,l} = \frac{1}{n_l}\sum_{i=1}^{n_l} \widetilde X_i^{(l)}h^\prime\left((\widehat{\bga}_l)^\top \widetilde X_i^{(l)}\right)\left(\widetilde X_i^{(l)}\right)^\top.
\end{equation*}
Let $\bgb_a^{(l),*}$ denote the probabilistic limit of $\widehat\bgb\supl_a$, which solves the following equation:
\begin{equation*}
    \EE_l\left[I\left\{A\supl = a\right\} W\supl \left\{Y\supl - g\left([W\supl]^\top\bgb_a^{(l),*}\right)\right\}\right] = 0.
\end{equation*}
The estimator $\widehat\bgb\supl_a$ admits the following asymptotic linear expansion:
\begin{equation*}
    \widehat\bgb\supl_a - \bgb_a^{(l),*} = \frac{1}{n_l}\sum_{i=1}^{n_l}C_{\bgb_a,l}^{-1}I\left\{A\supl_i = a\right\}W\supl_i\left\{Y\supl_i - g\left([W\supl_i]^\top \bgb^{(l),*}_a\right)\right\} + o_P(n_l^{-1/2}),
\end{equation*}
where the matrix $C_{\bgb_a,l}$ is defined as
\begin{equation*}
    C_{\bgb_a,l} = \EE_l\left[I\left\{A\supl=a\right\}W\supl g^\prime\left([W\supl]^\top \bgb^{(l),*}_a\right)[W\supl]^\top\right].
\end{equation*}
The matrix $C_{\bgb_a,l}$ can be consistently estimated by an empirical version of it,
\begin{equation*}
    \widehat C_{\bgb_a,l} = \frac{1}{n_l}\sum_{i=1}^{n_l}I\left\{A\supl_i=a\right\}W\supl_i g^\prime\left([W\supl_i]^\top \widehat\bgb^{(l)}_a\right)[W\supl_i]^\top.
\end{equation*}
Let $\bge^{(l),*}$ denote the probabilistic limit of $\widehat\bge\supl$ such that
\begin{equation*}
    \EE_l\left[\exp\left([\bge^{(l),*}]^\top \widetilde{W}^{(l)}\right)\widetilde{W}^{(l)}\right] = \EE_{\mathcal{T}}\left[ \widetilde{W}^{\mathcal{T}}\right],
\end{equation*}
where $\EE_{\mathcal{T}}$ denotes the expectation operator with respect to the covariate distribution in the target population. Again by standard theory of estimating equations, the estimator $\widehat\bge\supl$ is asymptotically linear with the following expansion:
\begin{equation*}
    \widehat\bge\supl - \bge^{(l),*} = C_{\bge,l}^{-1}\left\{\frac{1}{n_l}\sum_{i=1}^{n_l}\exp\left([\bge^{(l),*}]^\top \widetilde{W}_i^{(l)}\right)\widetilde{W}_i^{(l)} - \frac{1}{N}\sum_{j=1}^N \widetilde{W}_j^{\mathcal{T}}\right\} + o_P(n_l^{-1/2}),
\end{equation*}
where the matrix $C_{\bge,l}$ is defined as
\begin{equation*}
    C_{\bge,l} = - \EE_l\left[\exp\left([\bge^{(l),*}]^\top \widetilde{W}^{(l)}\right)\widetilde{W}^{(l)}[\widetilde{W}^{(l)}]^\top\right],
\end{equation*}
and can be consistently estimated by 
\begin{equation*}
    \widehat C_{\bge,l} = -\frac{1}{n_l}\sum_{i=1}^{n_l} \exp\left([\widehat\bge\supl]^\top \widetilde{W}^{(l)}_i\right)\widetilde{W}^{(l)}_i[\widetilde{W}^{(l)}_i]^\top.
\end{equation*}
Note that when $N \gg n_l$, the asymptotic linear expansion of $\widehat\bge\supl$ simplifies to
\begin{equation*}
    \widehat\bge\supl - \bge^{(l),*} = \frac{1}{n_l}\sum_{i=1}^{n_l}C_{\bge,l}^{-1}\left\{\exp\left([\bge^{(l),*}]^\top \widetilde{W}_i^{(l)}\right)\widetilde{W}_i^{(l)} - \EE_{\mathcal{T}}\left[\widetilde{W}^\mathcal{T}\right]\right\}  + o_P(n_l^{-1/2}).
\end{equation*}

Now, we derive an asymptotic linear expansion of the site-specific target ATE estimator. Recall from Section~\ref{sec: app 3} that the target ATE estimator takes the form $\widehat{\theta}^{(l)}=\widehat{M}^{(l)} + \widehat{\delta}^{(l)}$ where 
\begin{equation*}
    \widehat{M}^{(l)} = \frac{1}{N} \sum_{i=1}^{N}
    \left\{m(1,X_i^{\mathcal{T}}; \widehat{\bgb}^{(l)}_{1})  - m(0,X_i^{\mathcal{T}}; \widehat{\bgb}^{(l)}_{0})\right\}
\end{equation*}
and 
\begin{equation*}
\begin{aligned}
    \widehat{\delta}^{(l)} &= \frac{1}{n_l}\sum_{i=1}^{n_l} \omega_l(X_i^{(l)};\widehat{\bge}^{(l)})
    \left\{\sum_{a=0}^1\frac{(-1)^{a+1}I\{A_i\supl = a\}}{\pi_{l} (a,X^{(l)}_i; \widehat{\bga}_l)} \{Y^{(l)}_i - m(A_i\supl,X^{(l)}_i; \widehat{\bgb}^{(l)}_{a})\} \right\} . \label{aug-target}
    \end{aligned}
\end{equation*}
In this case, we have that
\begin{align*}
    m(a,X_i\supl;\widehat\bgb_a\supl) &= g\left([W\supl_i]^\top \widehat\bgb^{(l)}_a\right); \\
    m(a,X_i^{\mathcal{T}};\widehat\bgb_a\supl) &= g\left([W^{\mathcal{T}}_i]^\top \widehat\bgb^{(l)}_a\right); \\
    \omega_l(X_i^{(l)};\widehat{\bge}^{(l)}) &= \exp\left([\widehat\bge\supl]^\top \widetilde{W}_i\supl\right); \\
    \pi_{l} (a,X^{(l)}_i; \widehat{\bga}_l) &= h\left((\widehat{\bga}_l)^\top \widetilde X_i^{(l)}\right)^a\left\{1-h\left((\widehat{\bga}_l)^\top \widetilde X_i^{(l)}\right)\right\}^{(1-a)}.
\end{align*}
Also recall that we define the following functions $\tau_{\bgb_0,\bgb_1}(\cdot)$ and $\xi_{\bge,\bga,\bgb_0,\bgb_1}(\cdot)$ such that
\begin{equation*}
    \tau_{\bgb_0,\bgb_1}(x\supl) = m(1,x\supl;\bgb_1) - m(0,x\supl;\bgb_0),
\end{equation*}
and
\begin{equation*}
    \xi_{\bge,\bga,\bgb_0,\bgb_1}(x\supl, a\supl, y\supl) = \omega_l(x\supl;\bge)
    \left\{\sum_{a=0}^1\frac{(-1)^{a+1}I\{a\supl = a\}}{\pi_{l} (a,x\supl; \bga)} \{y\supl - m(a,x\supl; \bgb_a\} \right\} .
\end{equation*}
We introduce the following notations frequently used when dealing with empirical processes: for generic functions $f_1$ and $f_2$, we define $P_l f_1 = \int f_1 d\PP_l$ where $\PP_l$ denotes the joint distribution of $(X\supl,A\supl,Y\supl)$ in the $l$-th source population, and $P_{\mathcal{T}} f_2 = \int f_2 d\PP_{\mathcal{T}}$ where $\PP_{\mathcal{T}}$ denotes the covariate distribution in the target population. Moreover, let $P_{n,l} f_1 = \sum_{i=1}^{n_l} f_1(X\supl_i,A\supl_i,Y\supl_i)/n_l$ denote the empirical average of $f_1$ on the $l$-th source data, and $P_{n,\mathcal{T}} f_2 = \sum_{i=1}^{N} f_2(X^{\mathcal{T}}_i)/N$ denote the empirical average of $f_2$ on the target data. With these notations, we are now ready to linearize $\widehat\theta\supl$.

To start, we note that the estimator $\widehat\theta\supl$ can be written as
\begin{equation*}
    \widehat\theta\supl = P_{n,\mathcal{T}}\left(\tau_{\widehat\bgb_0\supl,\widehat\bgb_1\supl}\right) + P_{n,l} \left(\xi_{\widehat\bge\supl,\widehat\bga_l,\widehat\bgb_0\supl,\widehat\bgb_1\supl}\right).
\end{equation*}
Thus, the estimator $\widehat\theta^{(l)}$ has the following expansion
\begin{align*}
    \widehat\theta\supl - \theta\supl &= P_{n,\mathcal{T}}\left(\tau_{\widehat\bgb_0\supl,\widehat\bgb_1\supl} \right) - P_{\mathcal{T}}\left(\tau_{\bgb_0^{(l),*},\bgb_1^{(l),*}} \right) \\
    &\quad + P_{n,l} \left(\xi_{\widehat\bge\supl,\widehat\bga_l,\widehat\bgb_0\supl,\widehat\bgb_1\supl}\right) - P_l \left(\xi_{\bge^{(l),*},\bga_l^*,\bgb_0^{(l),*},\bgb_1^{(l),*}}\right) \\
    &= \left(P_{n,\mathcal{T}} - P_{\mathcal{T}}\right)\left(\tau_{\bgb_0^{(l),*},\bgb_1^{(l),*}} \right) + P_{\mathcal{T}}\left(\tau_{\widehat\bgb_0\supl,\widehat\bgb_1\supl} - \tau_{\bgb_0^{(l),*},\bgb_1^{(l),*}} \right) \\
    &\quad + \left(P_{n,l} - P_l\right)\left(\xi_{\bge^{(l),*},\bga_l^*,\bgb_0^{(l),*},\bgb_1^{(l),*}}\right) + P_l \left(\xi_{\widehat\bge\supl,\widehat\bga_l,\widehat\bgb_0\supl,\widehat\bgb_1\supl} - \xi_{\bge^{(l),*},\bga_l^*,\bgb_0^{(l),*},\bgb_1^{(l),*}} \right) + o_P(n_l^{-1/2}) \\
    &= \left(P_{n,\mathcal{T}} - P_{\mathcal{T}}\right)\left(\tau_{\bgb_0^{(l),*},\bgb_1^{(l),*}} \right) + \left(P_{n,l} - P_l\right)\left(\xi_{\bge^{(l),*},\bga_l^*,\bgb_0^{(l),*},\bgb_1^{(l),*}}\right) + o_P(n_l^{-1/2}) \\
    &\quad + \left\{P_{\mathcal{T}}\left(\frac{\partial\tau_{\bgb_0,\bgb_1}}{\partial\bgb_0}\rvert_{(\bgb_0^{(l),*},\bgb_1^{(l),*})}\right) + P_l \left(\frac{\partial\xi_{\bge,\bga,\bgb_0,\bgb_1}}{\partial\bgb_0}\rvert_{(\bge^{(l),*},\bga_l^*,\bgb_0^{(l),*},\bgb_1^{(l),*})}\right)\right\}^\top \left(\widehat\bgb_0\supl - \bgb_0^{(l),*}\right) \\
    &\quad + \left\{P_{\mathcal{T}}\left(\frac{\partial\tau_{\bgb_0,\bgb_1}}{\partial\bgb_1}\rvert_{(\bgb_0^{(l),*},\bgb_1^{(l),*})}\right) + P_l \left(\frac{\partial\xi_{\bge,\bga,\bgb_0,\bgb_1}}{\partial\bgb_1}\rvert_{(\bge^{(l),*},\bga_l^*,\bgb_0^{(l),*},\bgb_1^{(l),*})}\right)\right\}^\top \left(\widehat\bgb_1\supl - \bgb_1^{(l),*}\right) \\
    &\quad + \left\{P_l \left(\frac{\partial\xi_{\bge,\bga,\bgb_0,\bgb_1}}{\partial\bge}\rvert_{(\bge^{(l),*},\bga_l^*,\bgb_0^{(l),*},\bgb_1^{(l),*})}\right)\right\}^\top \left(\widehat\bge\supl - \bge^{(l),*}\right) \\
    &\quad + \left\{P_l \left(\frac{\partial\xi_{\bge,\bga,\bgb_0,\bgb_1}}{\partial\bga}\rvert_{(\bge^{(l),*},\bga_l^*,\bgb_0^{(l),*},\bgb_1^{(l),*})}\right)\right\}^\top \left(\widehat\bga_l - \bga_l^*\right).
\end{align*}
When the sample size in the target data $N$ is such that $N \gg n_l$, the term $(P_{n,\mathcal{T}} - P_{\mathcal{T}})\tau_{\bgb_0^{(l),*},\bgb_1^{(l),*}}$ is of the order $o_P(n_l^{-1/2})$ and hence is negligible. The partial derivatives can be calculated explicitly. Specifically,
\begin{align*}
    \boldsymbol{d}_{\bgb_0} &= P_{\mathcal{T}}\left(\frac{\partial\tau_{\bgb_0,\bgb_1}}{\partial\bgb_0}\rvert_{(\bgb_0^{(l),*},\bgb_1^{(l),*})}\right) + P_l \left(\frac{\partial\xi_{\bge,\bga,\bgb_0,\bgb_1}}{\partial\bgb_0}\rvert_{(\bge^{(l),*},\bga_l^*,\bgb_0^{(l),*},\bgb_1^{(l),*})}\right) \\
    &= -\EE_{\mathcal{T}}\left[\frac{\partial m}{\partial \bgb_0}(0,X^\mathcal{T};\bgb_0^{(l),*})\right] + \EE_l\left[\omega_l(X\supl;\bge^{(l),*})\frac{I\{A\supl=0\}}{\pi_l(0,X\supl;\bga_l^*)}\left\{\frac{\partial m}{\partial \bgb_0}(0,X\supl;\bgb_0^{(l),*})\right\}\right] \\
    &= -\EE_{\mathcal{T}}\left[g^\prime\left([W^\mathcal{T}]^\top\bgb_0^{(l),*}\right)W^{\mathcal{T}}\right] + \EE_l\left[\exp\left([\bge^{(l),*}]^\top \widetilde{W}\supl\right)\frac{I\{A\supl=0\}}{1-h\left([\widetilde{X}\supl]^\top \bga_l^*\right)}\left\{g^\prime\left([W\supl]^\top\bgb_0^{(l),*}\right)W\supl\right\}\right].
\end{align*}

\begin{align*}
    \boldsymbol{d}_{\bgb_1} &= P_{\mathcal{T}}\left(\frac{\partial\tau_{\bgb_0,\bgb_1}}{\partial\bgb_1}\rvert_{(\bgb_0^{(l),*},\bgb_1^{(l),*})}\right) + P_l \left(\frac{\partial\xi_{\bge,\bga,\bgb_0,\bgb_1}}{\partial\bgb_1}\rvert_{(\bge^{(l),*},\bga_l^*,\bgb_0^{(l),*},\bgb_1^{(l),*})}\right) \\
    &= \EE_{\mathcal{T}}\left[\frac{\partial m}{\partial \bgb_1}(1,X^\mathcal{T};\bgb_1^{(l),*})\right] - \EE_l\left[\omega_l(X\supl;\bge^{(l),*})\frac{I\{A\supl=1\}}{\pi_l(1,X\supl;\bga_l^*)}\left\{\frac{\partial m}{\partial \bgb_1}(1,X\supl;\bgb_1^{(l),*})\right\}\right] \\
    &= \EE_{\mathcal{T}}\left[g^\prime\left([W^\mathcal{T}]^\top\bgb_1^{(l),*}\right)W^{\mathcal{T}}\right] - \EE_l\left[\exp\left([\bge^{(l),*}]^\top \widetilde{W}\supl\right)\frac{I\{A\supl=1\}}{h\left([\widetilde{X}\supl]^\top \bga_l^*\right)}\left\{g^\prime\left([W\supl]^\top\bgb_1^{(l),*}\right)W\supl\right\}\right].
\end{align*}

\begin{align*}
    \boldsymbol{d}_{\bge} &= P_l \left(\frac{\partial\xi_{\bge,\bga,\bgb_0,\bgb_1}}{\partial\bge}\rvert_{(\bge^{(l),*},\bga_l^*,\bgb_0^{(l),*},\bgb_1^{(l),*})}\right) \\
    &= \EE_l\left[\frac{\partial\omega_l}{\partial \bge}(X\supl;\bge^{(l),*})
    \left\{\sum_{a=0}^1\frac{(-1)^{a+1}I\{A\supl = a\}}{\pi_{l} (a,X^{(l)}; \bga^*_l)} \{Y\supl - m(a,X\supl; \bgb^{(l),*}_{a})\} \right\}\right] \\
    &= \EE_l\left[\exp\left([\bge^{(l),*}]^\top \widetilde{W}\supl\right)\widetilde{W}\supl
    \left\{\sum_{a=0}^1\frac{(-1)^{a+1}I\{A\supl = a\}}{\pi_{l} (a,X^{(l)}; \bga^*_l)} \left\{Y\supl - g\left([W\supl]^\top\bgb_a^{(l),*}\right)\right\} \right\}\right].
\end{align*}

\begin{align*}
    \boldsymbol{d}_{\bga} &= P_l\left(\frac{\partial\xi_{\bge,\bga,\bgb_0,\bgb_1}}{\partial\bga}\rvert_{(\bge^{(l),*},\bga_l^*,\bgb_0^{(l),*},\bgb_1^{(l),*})}\right) \\
    &= \EE_l\left[\omega_l(X\supl;\bge^{(l),*})\left\{\sum_{a=0}^1\frac{-I\{A\supl =a\}\pi^\prime(a,X\supl;\bga_l^*)}{\pi^2(a,X\supl;\bga_l^*)}\left\{Y\supl - m(a,X\supl;\bgb_{a}^{(l),*})\right\}\right\}\right] \\
    &= \EE_l\left[\exp\left([\bge^{(l),*}]^\top \widetilde{W}\supl\right)\left\{\sum_{a=0}^1\frac{-I\{A\supl =a\}\pi_l^\prime(a,X\supl;\bga_l^*)}{\pi_l^2(a,X\supl;\bga_l^*)}\left\{Y\supl - g\left([W\supl]^\top\bgb_a^{(l),*}\right)\right\}\right\}\widetilde{X}\supl\right].
\end{align*}
These partial derivatives can be estimated by replacing the expectations with the corresponding sample averages, and replacing the unknown population parameters $\bge^{(l),*},\bga_l^*,\bgb_0^{(l),*}$ and $\bgb_1^{(l),*}$ with their consistent estimators $\widehat\bge\supl,\widehat\bga_l,\widehat\bgb_0\supl$ and $\widehat\bgb_1\supl$, respectively. 

As $\widehat\bge\supl,\widehat\bga_l,\widehat\bgb_0\supl$ and $\widehat\bgb_1\supl$ are all asymptotically linear, the expansion we derived for $\widehat\theta\supl$ implies that $\widehat\theta\supl$ is also asymptotically linear with influence function $\tau_\theta$ such that
\begin{align*}
    \tau_\theta(x\supl,a\supl,y\supl) &= \xi_{\bge^{(l),*},\bga_l^*,\bgb_0^{(l),*},\bgb_1^{(l),*}}(x\supl,a\supl,y\supl) \\
    &\quad + \boldsymbol{d}_{\bgb_0}^\top C_{\bgb_0,l}^{-1}I\left\{a\supl = 0\right\}w\supl \left\{y\supl - g\left([w\supl]^\top \bgb^{(l),*}_0\right)\right\} \\
    &\quad + \boldsymbol{d}_{\bgb_1}^\top C_{\bgb_1,l}^{-1}I\left\{a\supl = 1\right\}w\supl \left\{y\supl - g\left([w\supl]^\top \bgb^{(l),*}_1\right)\right\} \\
    &\quad + \boldsymbol{d}_{\bge}^\top C_{\bge,l}^{-1}\left\{\exp\left([\bge^{(l),*}]^\top \widetilde{w}\supl\right)\widetilde{w}\supl- \EE_{\mathcal{T}}\left[\widetilde{W}^\mathcal{T}\right]\right\} \\
    &\quad + \boldsymbol{d}_{\bga}^\top C_{\bga,l}^{-1}\widetilde{x}\supl\left\{a\supl - h\left((\bga_l^*)^\top \widetilde x\supl\right)\right\}.
\end{align*}
Here the vectors $\widetilde{x}\supl$, $w\supl$ and $\widetilde{w}\supl$ denote the vectors of basis functions derived from the covariate vector $x\supl$. 

\section{Additional Numerical Results}\label{sec: additional sim}

\subsection{Simulation results for 8 majority sites and RIFL with $80\%$ rule}
\label{sec: 80 perc rule}
We now present simulation results when there are 8 majority sites, and we use an $80\%$-rule in RIFL that utilizes prior information on the number of majority sites. Specifically, we modify the definition of the index set in \eqref{eq: index set} and define
\begin{equation}
\mathcal{M}_{80\%}\coloneqq \left\{1\leq m\leq M: |\widehat{\mathcal{V}}^{[m]}|\geq 0.8 L\right\} .
\label{eq: index set 80}
\end{equation}
RIFL with the $80\%$ rule is defined in the same way as the original RIFL except that we replace $\mathcal{M}$ with $\mathcal{M}_{80\%}$.

First, we present results in the low-dimensional prediction example in Section~\ref{sec: app 1}. For $l \in \{1,2,\ldots,8\}$, we set $\theta^{(l)} = \theta^* =  (0.5,0.5,0.5,0.5,0.5,0.1,0.1,0.1,0,0)$. For $\theta^{(9)}$ and $\theta^{(10)}$, their last 5 coefficients are the same as $\theta^*$  but their first five coefficients are changed to $0.5-0.3a$ and $0.5-0.1a$, respectively, where $a \in \{1,2,3,4,5\}$ controls the separation between the majority sites and non-majority sites. All the other aspects of the simulation setup is the same as in Section~\ref{sims: app1}.

We present the empirical coverage and average length of $95\%$ CIs from various methods in Figure~\ref{fig:parametric8}. We observe that both RIFL and RIFL with the $80\%$ rule achieve the nominal coverage across all settings. In addition, RIFL with the $80\%$ rule results in a CI that is much shorter than the original RIFL CI and comparable to the OBA interval when the separation level is high. The performance of all the other methods show similar pattern as in Section~\ref{sims: app1}.

\begin{figure}
    \centering
    \includegraphics[width=\textwidth]{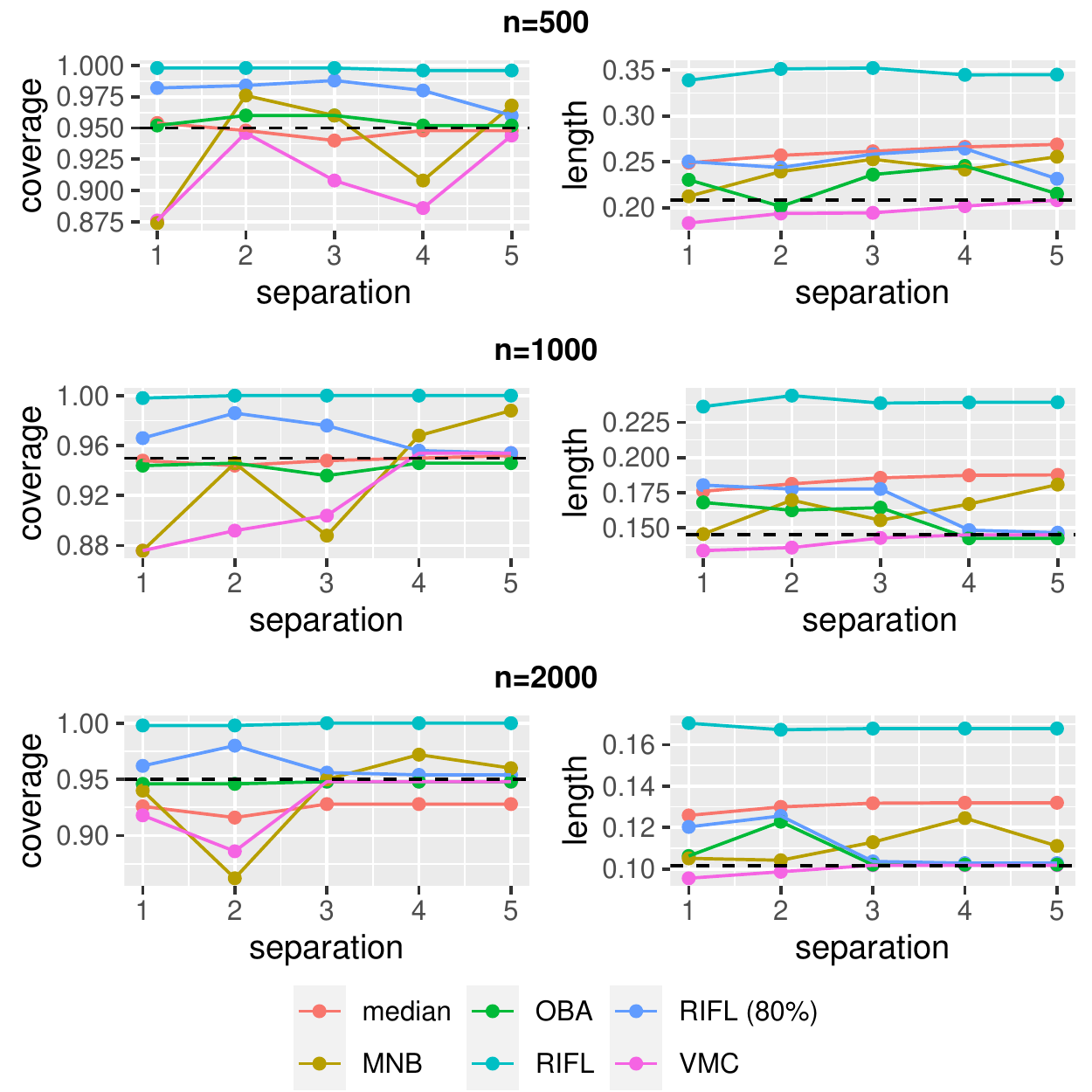}
    \caption{Low-dimensional prediction: coverage and length of $95\%$ CIs for $\beta^* = \theta^*_1$ with 8 majority sites and varying separation levels where $1$ is the lowest (hardest to detect) and $5$ is the highest (easiest to detect). ``median" stands for the CI based on the median estimator, ``MNB" stands for the m-out-of-n bootstrap CI in \eqref{eq: mnb}, ``VMC" stands for the voting with maximum clique estimator and its associated CI in \eqref{eq: post CI}, ``OBA" stands for the oracle bias aware CI in \eqref{eq: bias aware}, ``RIFL" stands for our proposed CI in \eqref{eq: CI union}, and ``RIFL (80\%)" stands for our proposed RIFL CI leveraging 80\% rule. Results are based on 500 simulation replications. Dash lines in the left panels correspond to nominal coverage level 0.95; in the right panels correspond to the width of an oracle CI knowing the prevailing set.}
    \label{fig:parametric8}
\end{figure}

Next, we present additional simulation results for the high-dimensional prediction example in Section~\ref{sec: app 2} for $\theta_{11}^*$ with 8 majority sites. The simulation setup is the same as that in Section~\ref{sims: app 2} except that $\theta^{(l)} = \theta^*$ for $l \in \{1,\ldots,8\}$ and $\theta_j^{(9)} = \theta_j^* + 0.2 + 0.05a$ and $\theta_j^{(10)} = \theta_j^* + 0.15 + 0.05a$ for $6 \leq j \leq 11$ with $a$ varied in $\{1,2,3,4,5\}$ to represent varying levels of separation between the majority and non-majority sites. The results are presented in Figure~\ref{fig:high8}, and we observe similar patterns as in the case with 6 majority sites. Simulation results for $\theta_8^*$ are presented in Figure~\ref{fig:highweak6} and Figure~\ref{fig:highweak8} for 6 and 8 majority sites, respectively. Again, we observe similar patterns as in the case for $\theta_{11}^*$.

\begin{figure}[H]
    \centering
    \includegraphics[width=\textwidth]{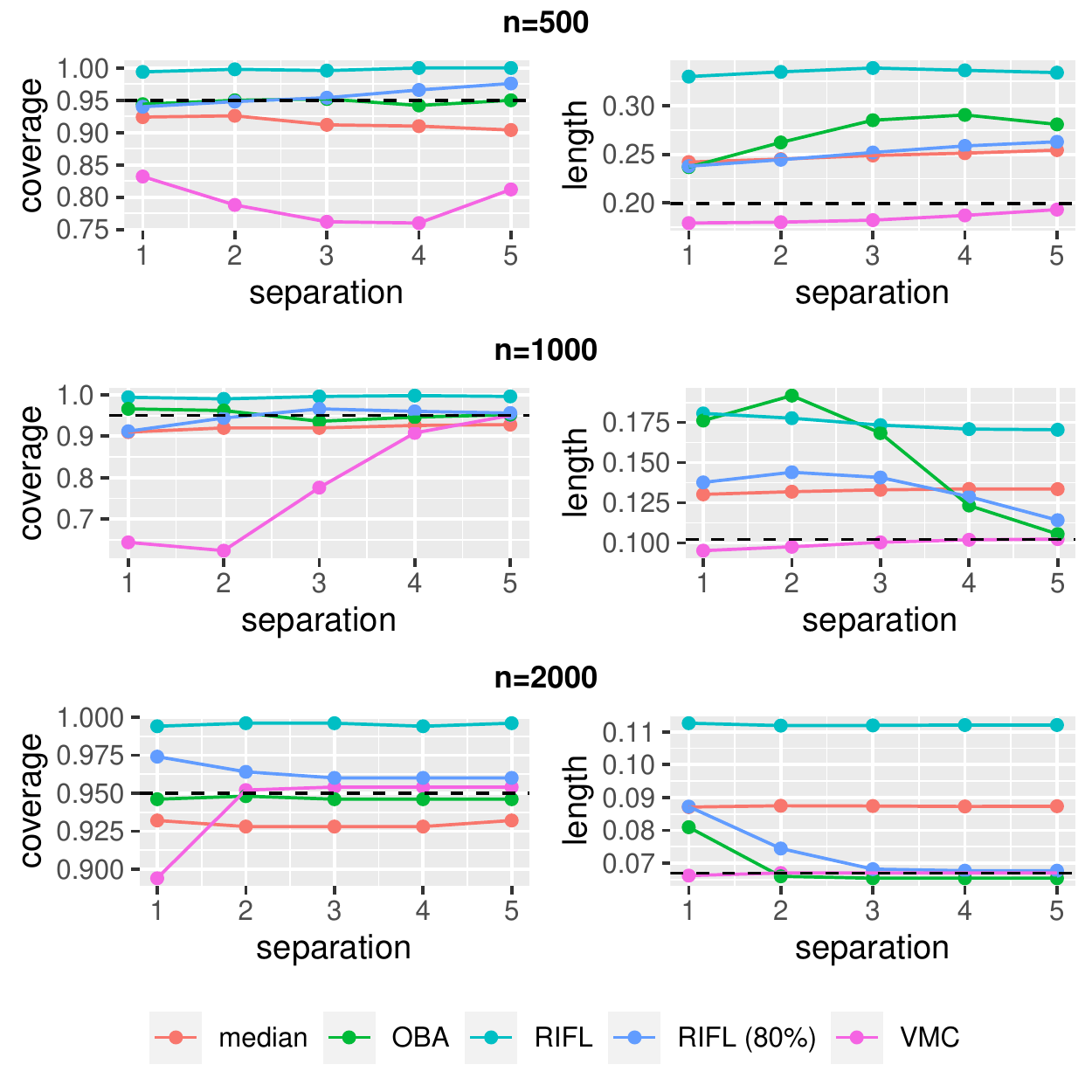}
    \caption{High-dimensional prediction: coverage and length of $95\%$ CIs for $\beta^* = \theta^*_{11}$ with 8 majority sites and varying separation levels where $1$ is the lowest (hardest to detect) and $5$ is the highest (easiest to detect). ``median" stands for the CI based on the median estimator, ``VMC" stands for the voting with maximum clique estimator and its associated CI in \eqref{eq: post CI}, ``OBA" stands for the oracle bias aware CI in \eqref{eq: bias aware}, ``RIFL" stands for our proposed CI in \eqref{eq: CI union}, and ``RIFL (80\%)" stands for our proposed RIFL CI leveraging 80\% rule. Results are based on 500 simulation replications. Dash lines in the left panels correspond to nominal coverage level 0.95; in the right panels correspond to the width of an oracle CI knowing the prevailing set.}
    \label{fig:high8}
\end{figure}

\begin{figure}[H]
    \centering
    \includegraphics[width=\textwidth]{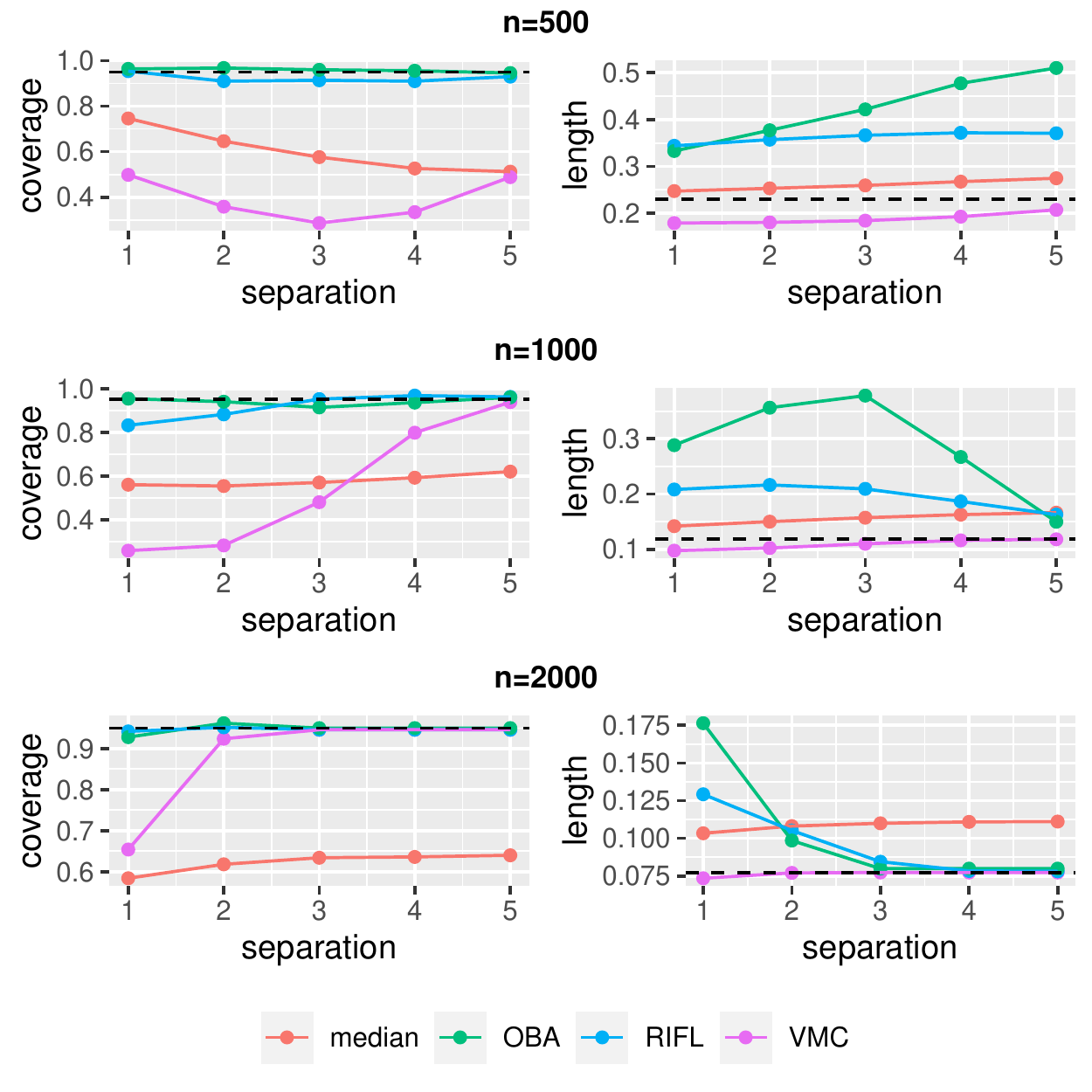}
    \caption{High-dimensional prediction: coverage and length of $95\%$ CIs for $\beta^* = \theta^*_{8}$ with 6 majority sites and varying separation levels where $1$ is the lowest (hardest to detect) and $5$ is the highest (easiest to detect). ``median" stands for the CI based on the median estimator, ``VMC" stands for the voting with maximum clique estimator and its associated CI in \eqref{eq: post CI}, ``OBA" stands for the oracle bias aware CI in \eqref{eq: bias aware}, ``RIFL" stands for our proposed CI in \eqref{eq: CI union}, and ``RIFL (80\%)" stands for our proposed RIFL CI leveraging 80\% rule. Results are based on 500 simulation replications. Dash lines in the left panels correspond to nominal coverage level 0.95; in the right panels correspond to the width of an oracle CI knowing the prevailing set.}
    \label{fig:highweak6}
\end{figure}

\begin{figure}[H]
    \centering
    \includegraphics[width=\textwidth]{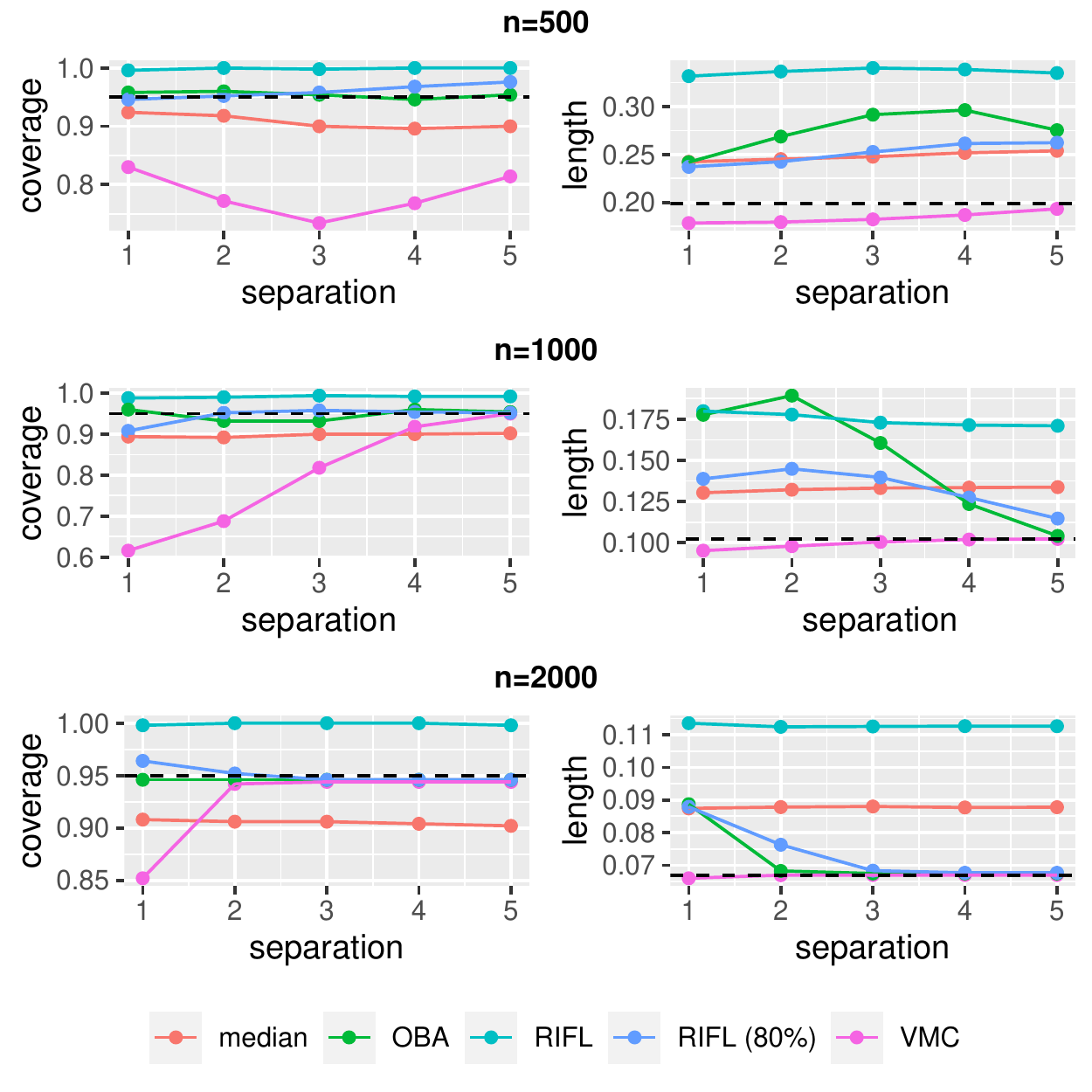}
    \caption{High-dimensional prediction: coverage and length of $95\%$ CIs for $\beta^* = \theta^*_{8}$ with 8 majority sites and varying separation levels where $1$ is the lowest (hardest to detect) and $5$ is the highest (easiest to detect). ``median" stands for the CI based on the median estimator, ``VMC" stands for the voting with maximum clique estimator and its associated CI in \eqref{eq: post CI}, ``OBA" stands for the oracle bias aware CI in \eqref{eq: bias aware}, ``RIFL" stands for our proposed CI in \eqref{eq: CI union}, and ``RIFL (80\%)" stands for our proposed RIFL CI leveraging 80\% rule. Results are based on 500 simulation replications. Dash lines in the left panels correspond to nominal coverage level 0.95; in the right panels correspond to the width of an oracle CI knowing the prevailing set.}
    \label{fig:highweak8}
\end{figure}

Finally, we present results for the multi-source causal inference problem discussed in Sections~\ref{sec: app 3} and \ref{sims: app 3} when there are 8 majority sites. Recall that the outcome for the $l$-th site is generated according to
$$Y_i^{(l)} = \mu_l + [X_i^{(l)}]^\top \zeta^{(l)} + \beta\supl A_i\supl+ \varepsilon_i\supl, \quad \varepsilon_i\supl \sim N(0, 1).$$
For the case with 8 majority sites, we set $\beta^{(l)} = \beta^* = -1$ for $l \in \{1,2,\ldots,8\}$, $\beta^{(9)} = -1 - 0.2a$ and $\beta^{(10)} = -1 - 0.1a$, with $a$ varied in $\{1,2,3,4,5\}$. All the other aspects of the simulation setup is the same as that in Section~\ref{sims: app 3}. The results are summarized in Figure~\ref{fig:TATE8}, and we again observe similar patterns.

\begin{figure}
    \centering
 \includegraphics[width=\textwidth]{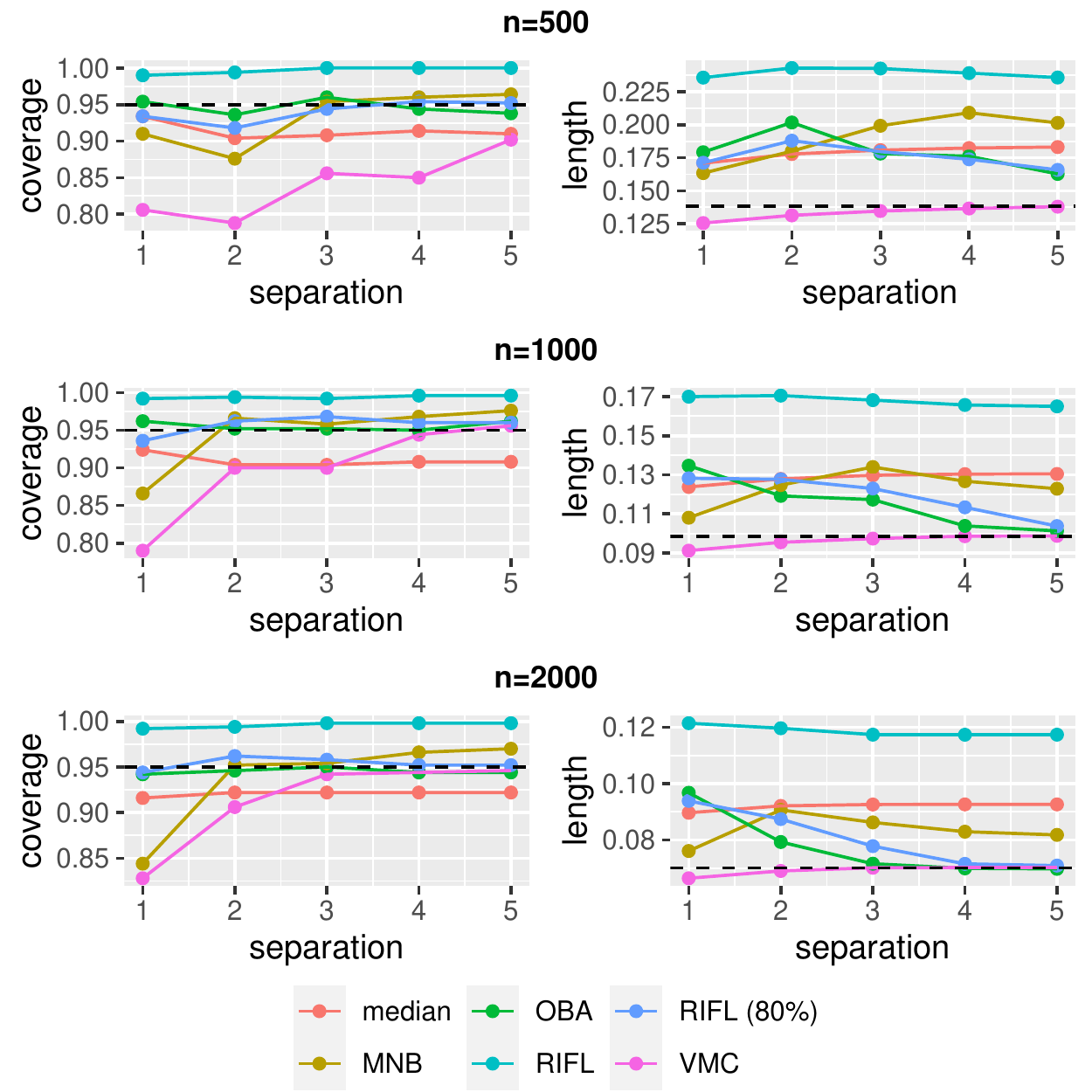}
    \caption{Causal inference: coverage and length of $95\%$ CIs for target ATE of the prevailing sites with 8 majority sites and varying separation levels where $1$ is the lowest (hardest to detect) and $5$ is the highest (easiest to detect). ``median" stands for the CI based on the median estimator, ``MNB" stands for the m-out-of-n bootstrap CI in \eqref{eq: mnb}, ``VMC" stands for the voting with maximum clique estimator and its associated CI in \eqref{eq: post CI}, ``OBA" stands for the oracle bias aware CI in \eqref{eq: bias aware}, ``RIFL" stands for our proposed CI in \eqref{eq: CI union}, and ``RIFL (80\%)" stands for our proposed RIFL CI leveraging 80\% rule. Results are based on 500 simulation replications. Dash lines in the left panels correspond to nominal coverage level 0.95; in the right panels correspond to the width of an oracle CI knowing the prevailing set.}
    \label{fig:TATE8}
\end{figure}

\subsection{Choices of $\nu$ in m-out-of-n bootstrap}
In this section, we present the empirical coverage and average length of the CIs obtained via m-out-of-n bootstrap with different choices of $m$. Specifically, we consider $m = n^\nu$ with $\nu$ varied in $\{0.6, 0.7, 0.8, 0.9, 1\}$. We focus on the low-dimensional prediction example in Section~\ref{sims: app1} with 6 majority sites and $n_l = 1000$ for all 10 sites. All other simulation settings are the same as those described in Section~\ref{sims: app1}.

The results are presented in Figure~\ref{fig:MNB}. We observe similar patterns in coverage for $\nu \in \{0.8,0.9,1\}$. Moreover, out of these 3 values, the choice of $\nu = 0.8$ generally produces the shortest CI. For all choices of $\nu$, the MNB CI has coverage below nominal level when the separation level is low to moderate.

\begin{figure}
    \centering
    \includegraphics[width=\textwidth]{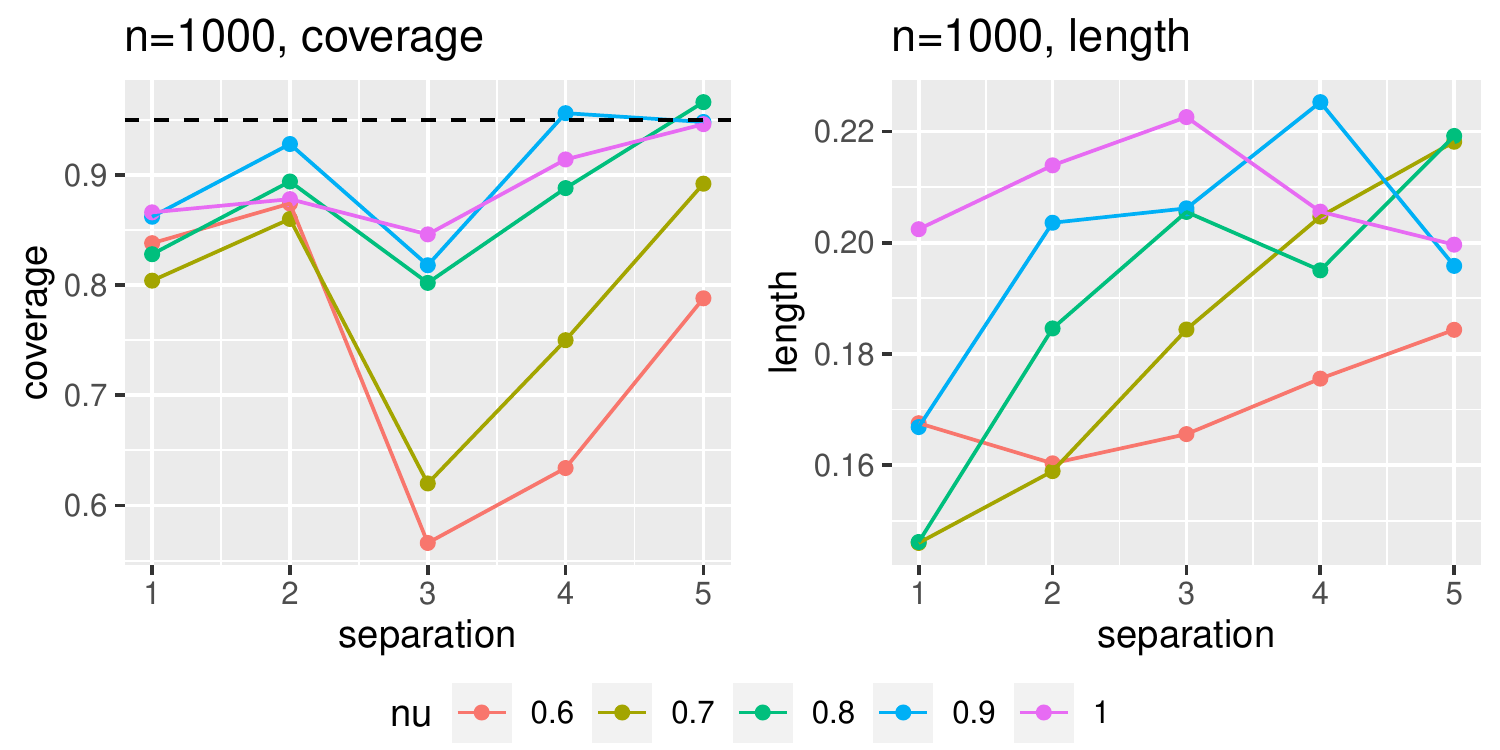}
    \caption{Low-dimensional prediction: coverage and length of $95\%$ CIs for $\beta^* = \theta^*_1$ of m-out-of-n bootstrap (MNB) CI with different choices of $m$. We set $m = n^\nu$ and vary $\nu$ in $\{0.6,0.7,0.8,0.9,1\}$. Sample size in each of the 10 sites is 1,000, and there are 6 majority sites. Results are based on 500 simulation replications. Dash lines in the left panels correspond to nominal coverage level 0.95.}
    \label{fig:MNB}
\end{figure}

\subsection{Generalizability measure for the 16 sites in the real data analysis}
In Figure~\ref{fig:gm}, we present the full set of generalizability measure for all 16 sites in our Covid-19 real data analysis. Site 4 generally has lower generalizability than the other sites. When we focus on one specific risk factor (corresponding to one specific row of the plot in Figure~\ref{fig:gm}), some sites have low generalizability, indicating that they may not belong to the prevailing set.

\begin{figure}
    \centering
    \includegraphics[width=\textwidth]{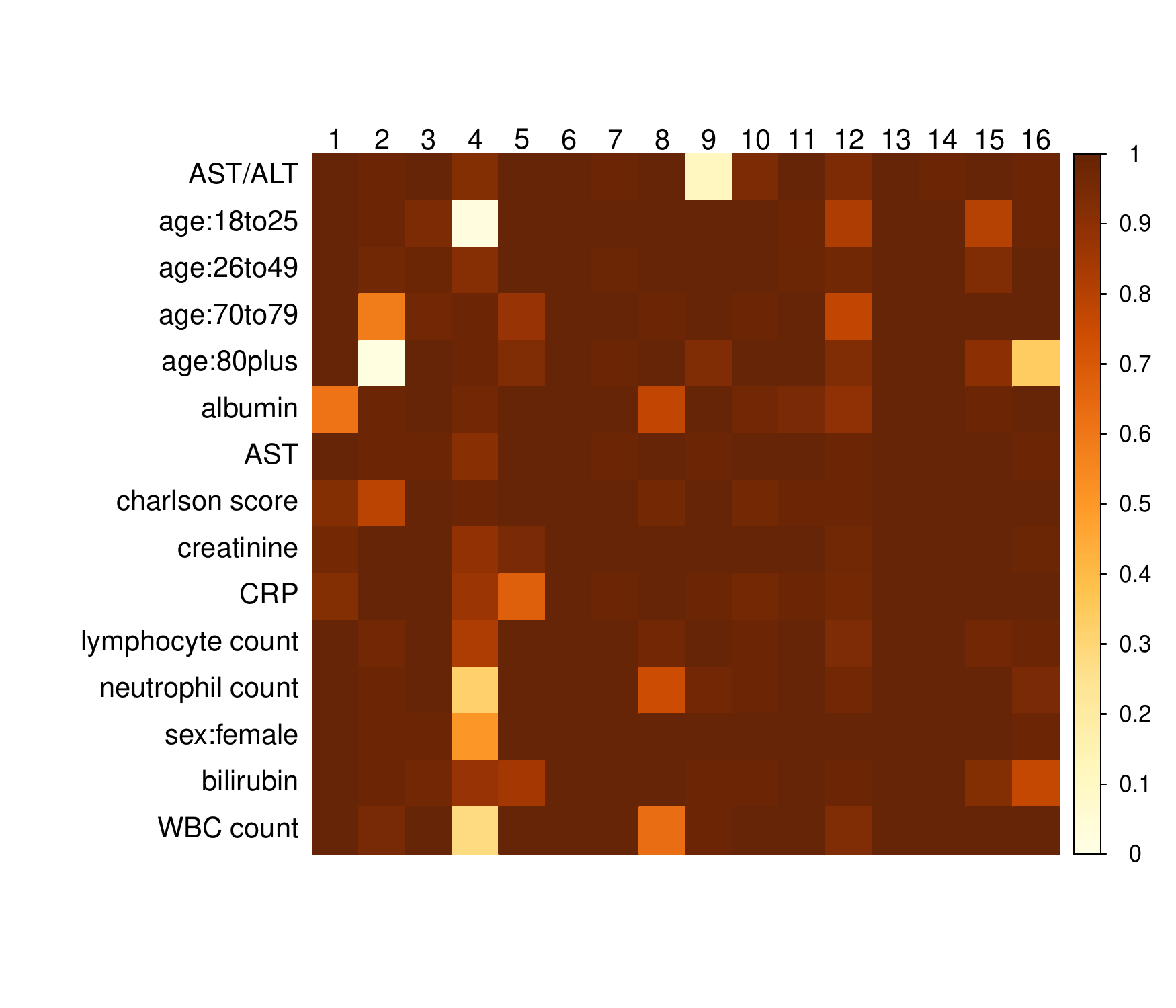}
    \caption{Generalizability measure for all 16 sites for each of the 15 risk factors. Each row corresponds to a risk factor and each column corresponds to a site.}
    \label{fig:gm}
\end{figure}

%%%%%%%%%%%%%%%%%%%%%%%%%%%%%%%%%%%%%%%%%%%%%%%%%%%%%%%%%%%%

%\section{Extra Simulation Results}
%\subsection{Post-selection Problem using $\widehat{\mathcal{V}}$}
%\begin{figure}[H]
%    \centering
%    \includegraphics[scale=0.22]{graphics/Cov1.500.png}
%    \caption{Illustration of under-coverage of $95\%$ CIs of the voting method \ref{eq: voting}, with only $100$ of the $500$ CI's covering the true ATE of $3.0$ (light blue).}
%    \label{fig: undercover1}
%\end{figure}

% \section{Extra Real Data Results}
% \Zijian{Larry, please add some discussions. Please also mention them briefly in the main paper.}

% \begin{figure}[H]
%     \centering
%     \includegraphics[scale=0.3]{graphics/CI.1-10.png}
%     \caption{For each index of the prevailing set, estimates of the $95\%$ CIs for log hazard ratios of mortality within 14 days of hospitalization with COVID-19 for the first 10 baseline risk factors}
%     \label{fig:CI.1-10}
% \end{figure}

% \begin{figure}[H]
%     \centering
%     \includegraphics[scale=0.3]{graphics/CI.11-19.png}
%     \caption{For each index of the prevailing set, estimates of the $95\%$ CIs for log hazard ratios of mortality within 14 days of hospitalization with COVID-19 for the last 9 baseline risk factors}
%     \label{fig:CI.11-19}
% \end{figure}

\end{document}